\documentclass[preprint,showpacs,preprintnumbers,amsmath,amssymb,superscriptaddress]{revtex4}

\usepackage{blindtext}
\usepackage{graphicx}% Include figure files
\usepackage{dcolumn}% Align table columns on decimal point
\usepackage{bm}% bold math
\begin{document}

%\preprint{APS/}

\title{Inherent rhythm of smooth muscle cells in rat mesenteric arterioles: an eigensystem formulation}

\author{I Lin Ho}
\email{sunta.ho@msa.hinet.net}
\affiliation{Department of Physics, National Chung Hsing University, Taichung 402,
Taiwan, R.O.C.}
\affiliation{Dalton Cardiovascular Research Center, University of Missouri, Columbia, MO 65211}
\author{Arash Moshkforoush}
\affiliation{Dept. of Biomedical Engineering, Florida International University, 10555 W. Flagler Str., EC 2674 Miami, FL 33174}
\author{Kwangseok Hong}
\affiliation{Dalton Cardiovascular Research Center, University of Missouri, Columbia, MO 65211}
\author{Gerald A. Meininger}
\affiliation{Dalton Cardiovascular Research Center, University of Missouri, Columbia, MO 65211}
\author{Michael A. Hill}
\affiliation{Dalton Cardiovascular Research Center, University of Missouri, Columbia, MO 65211}
\author{Nikolaos M. Tsoukias}
\affiliation{Dept. of Biomedical Engineering, Florida International University, 10555 W. Flagler Str., EC 2674 Miami, FL 33174}
\author{Watson Kuo}
\email{wkuo@phys.nchu.edu.tw}
\affiliation{Department of Physics, National Chung Hsing University, Taichung 402,
Taiwan, R.O.C.}

\begin{abstract}
On the basis of experimental data and mathematical equations in the literature, we remodel the ionic dynamics of smooth muscle cells (SMCs) as an eigensystem formulation, which is valid for investigating finite variations of variables from the
equilibrium like in common experimental operations.
This algorithm provides an alternate viewpoint from frequency-domain analysis and enables one to probe functionalities of SMC's rhythm by means of a resonance-related mechanism. Numerical results show three types of
calcium oscillations of SMCs in mesenteric arterioles: spontaneous calcium oscillation, agonist-dependent calcium oscillation, and agonist-dependent calcium spike.
For simple single and double SMCs, we demonstrate properties of synchronization among complex signals related to calcium oscillations, and show different correlation relations between calcium and voltage signals for various synchronization and resonance conditions. For practical cell clusters, our analyses indicate that the rhythm of SMCs could (1) benefit enhancements of signal communications among remote cells, (2) respond to a significant calcium peaking against transient stimulations for triggering globally-oscillating modes, and (3) characterize the globally-oscillating modes via frog-leap (non-molecular-diffusion) calcium waves across inhomogeneous SMCs.
\end{abstract}
\maketitle

\section{Introduction}
Rhythmical contractions in smooth muscle have been observed in many different tissues, e.g. in the gastrointestinal tract, urinary tract, and lymphatic vessels \cite{rhy1,rhy2,rhy3}. In blood vessels, this activity, named for vasomotion, is found in larger arteries and in low-resistance vessels in microcirculation \cite{vasoa1,vasoa2}, where vascular rhythmicity is apparently synchronous over considerable lengths of arteries \cite{vasoa3}. While the literature has investigated the underlying mechanism for many years, it has only been recently that, through images of confocal microscopy, the vasomotion is argued as critically depending on calcium waves originating from intracellular stores \cite{vasoa4} and on cell coupling via gap junctions. In addition to these vasomotion phenomena observed in isolated arteries and in some intact mammals (e.g. humans, dogs, rabbits, and rats) \cite{an1,an2,an3,an4}, some operations by in-vitro experiments indicate that vascular rhythmicity can be enhanced with the help of agonists: noradrenaline (NE), acetylcholine (ACh), phenylephrine (PE), neuropeptide Y, and KCl solution \cite{ex1,ex2,ex3,ex4}.
These studies showed that the vasomotion spreads over an increasing distance of the arteriole by raising the dosages of agonists, and tonic contraction can be induced without calcium oscillations at very high concentrations of NE, KCl, and PE; otherwise, the spread of vasomotion is much faster than the movement of molecules by normal diffusion \cite{ex5}, and can be eliminated by clamping the voltage \cite{ex4}.

These physiological reactivities of vascular
rhythmicity corresponding to experimental observations are not fully understood. One such inference is that the functionality of vasomotion can be for low energy-consumption tissue perfusion (1.7 to 8.0 times more efficient than in vessels without vasomotion) \cite{perf1,perf2,perf3} and could be protective of pathological conditions (e.g. hypertension, via regulating vascular resistance) \cite{res1,res2}. In this work we investigate vascular rhythmicity by means of mesenteric microcirculation, which is a region of easy-regulating resistance against blood flow. Due to its accessibility, the rat mesenteric artery is one of the most thoroughly studied vascular beds \cite{mathmodel1,mathmodel2}, bringing forth a vast amount of experimental data. The literature has recognized the need for mathematical models in vasomotion studies and has developed many mathematical models. However, most studies investigated either membrane potential changes or changes in calcium concentrations \cite{study1,study2,study3,study4,study5,study6,study7,study8}, while few works systemically looked at the resonance mechanisms underlying voltage oscillation and the correlated experimental observations with the synchronization of calcium oscillation for smooth muscle cells in rat mesenteric arterioles.

On the basis of experimental data and mathematical equations in the literature, we remodel the ionic dynamics of smooth muscle cells as an eigensystem formulation. By using the first-order Taylor approximation, our approach accurately depicts the characteristic frequencies (eigenvalues) of SMCs and the correlations of signaling pathways (eigenfunctions) under finite variations of model variables, like in common experimental conditions. This algorithm provides an alternate viewpoint on frequency-domain analysis and enables one to probe the functionalities of SMCs' rhythm by means of a resonance-related mechanism. However, the first-order approximation could introduce significant numeric inaccuracy if there exist violent fluctuations of variables. Our work mainly investigates the underlying mechanisms of SMCs' rhythmicity by varying dosages of agonists, i.e. potassium, that can diffuse from muscle fibers at the onset of exercise and the responses to evident changes of vascular rhythm\cite{book_K}.
Our calculations show three types of
calcium oscillations of SMCs in mesenteric arterioles: spontaneous calcium oscillation, agonist-dependent calcium oscillation, and agonist-dependent calcium spike.
For simple single and double SMCs, we demonstrate properties of synchronization among complex signals related to calcium oscillations, and show different correlations between calcium and voltage signals for various synchronization and resonance conditions. For practical cell clusters \cite{study7}, our analyses indicate that the rhythm of SMCs could (1) benefit enhancements of signal communications among remote cells, (2) respond to a significant calcium peaking against transient stimulations for triggering globally-oscillating modes, and (3) characterize the globally-oscillating modes via frog-leap (non-molecular-diffusion) calcium waves across inhomogeneous SMCs.

Our conclusions interpret experimental phenomena in the literature and provide materials for understanding other functionalities of calcium dynamics (e.g. appearance of the significant calcium peaking). Our algorithm also offers preliminary considerations for the inherent rhythm of rat mesenteric arterioles at the cell level, which are proposed to have a relation to efficient energy transports and the heart rate \cite{heart1}.

\section{Mathematical algorithms}

\subsection{Basic mathematical modelings}

\begin{figure*}[ht]
\centering
\includegraphics[scale=0.35]{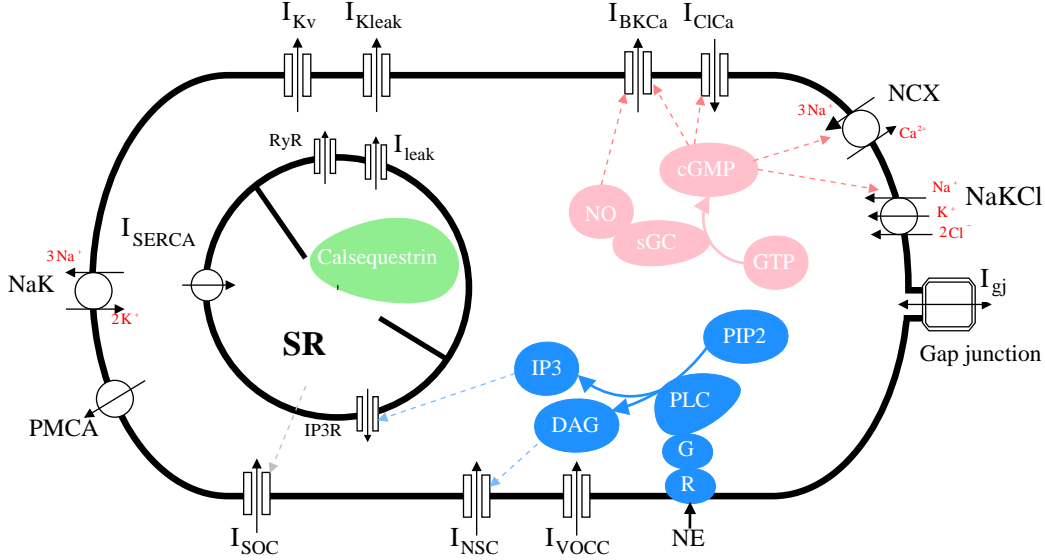}
\caption{ {\protect\small {\ Schematic diagram of model components for
mesenteric smooth muscle cell. $\mathbf{BK_{Ca}}$: large conductance
calcium-activated $K^{+}$ channel; $\mathbf{K_{leak}}$: unspecified $K^{+}$
leak channel; $\mathbf{K_{v}}$: voltage-dependent $K^{+}$ channel; $\mathbf{%
Cl_{Ca}}$: calcium-activated $Cl^{-}$ channel; $\mathbf{NSC}$: non-selective
cation channel; $\mathbf{SOC}$: store-operated calcium-permeable
non-selective cation channel; $\mathbf{VOCC}$: L-type voltage-operated $%
Ca^{2+}$ channel; $\mathbf{NaK}$: $Na^{+}$-$K^{+}$-ATPase; $\mathbf{PMCA}$:
plasma membrane $Ca^{2+}$-ATPase; $\mathbf{NaKCl}$: $Na^{+}$-$K^{+}$-$Cl^{-}$
cotransport; $\mathbf{NCX}$: $Na^{+}$-$Ca^{2+}$ exchange; $\mathbf{SR}$:
sarcoplasmic reticulum; $\mathbf{IP_{3}R}$: $IP_{3}$ receptor; $\mathbf{RyR}$%
: ryanodine receptor; $\mathbf{SERCA}$: SR $Ca^{2+}$-ATPase pumps; $\mathbf{%
CSQN}$: calsequestrin; $\mathbf{CM}$: calmodulin; $\mathbf{R}$: $\protect%
\alpha_{1}$-adrenoceptor; $\mathbf{G}$: G protein; $\mathbf{PLC}$:
phospholipase C; $\mathbf{sGC}$: soluble guanylate cyclase; $\mathbf{GJ}$:
non-selective gap-junction channel. }}}
\label{fig1}
\end{figure*}

The model is composed of three categories: plasma membrane, cytosol,
and intracellular calcium store. Relevant experimental parameters and
mathematical equations are derived on the basis of Tsoukias's previous
developments for rat mesenteric smooth muscle \cite{mathmodel1,mathmodel2}.
Figure (\ref{fig1}) illustrates a schematic diagram of the model: ($%
\mathrm{i}$) The dynamics of plasma membrane include ion channels, pumps,
exchangers, and receptors, for all the major transmembrane currents that
have been identified in SMCs of rat mesenteric arterioles. The ion channels
contain large conductance calcium-activated $K^{+}$ channel ($\mathbf{BKCa}$%
), voltage-dependent $K^{+}$ channel ($\mathbf{K_{v}}$), unspecified leak $%
K^{+}$ channel ($\mathbf{K_{leak}}$), calcium-activated $Cl^{-}$ channel ($%
\mathbf{Cl_{Ca}}$), non-selective cation channel ($\mathbf{NSC}$),
store-operated cation channel ($\mathbf{SOC}$), voltage-operated $Ca^{2+}$
channel ($\mathbf{VOCC}$), the non-selective gap-junction ion channel ($%
\mathbf{GJ}$), and the $IP_{3}$ gap-junction flux. Specific mathematical
descriptions are included for the $Na^{+}$-$K^{+}$-ATPase pump ($\mathbf{NaK}
$), the plasma membrane $Ca^{2+}$-ATPase pump ($\mathbf{PMCA}$), the $Na^{+}$%
-$K^{+}$-$Cl^{-}$ cotransport ($\mathbf{NaKCl}$), and the $Na^{+}$-$Ca^{2+}$
exchanger ($\mathbf{NCX}$). ($\mathrm{ii}$) The intracellular calcium store,
representing the sarcoplasmic reticulum, contains sarcoplasmic reticulum $%
Ca^{2+}$-ATPase pumps ($\mathbf{SERCA}$), $IP_{3}$ receptor $Ca^{2+}$
channel ($\mathbf{IP_{3}R}$) ryanodine receptor $Ca^{2+}$ channel ($\mathbf{%
RyR}$), the leak current ($\mathbf{leak}$), and the $Ca^{2+}$ buffering with
calsequestrin ($\mathbf{CSQN}$). ($\mathrm{iii}$) The the cytosol incldues the
processes of $\alpha_{1}$-adrenoceptor activation and $IP_{3}$ formation,
incorporated with the effects of $NE$. The vasodilatory action
of $NO$ is modeled through a direct effect on the $\mathbf{BK_{Ca}}$ channel
and through the formation of CGMP. The whole dynamics are integrated by
ionic balances for calcium, sodium, potassium, and chloride ions and are
described in the appendix.

\subsection{Eigensystem formulation for a single SMC}

We first consider a single SMC. The 26 transient variables $\nu _{i\in
\{1...26\}}(t)$, having time-dependent evolutions, are re-defined by the
function $\nu _{i}(t)=\bar{\nu}_{i}+\delta _{\nu _{i}}(t)$. Here, $\bar{\nu}%
_{i}$ denotes the time average value for variable $\nu _{i}$, and $\delta
_{\nu _{i}}(t)$ denotes its transient variation. In this work, 26 transient
variables for a single SMC are grouped into a vector, i.e. $\vec{\nu}(t)=($ $%
[Ca]_{i}$ $[Ca]_{r}$ $[Ca]_{u}$ $[Na]_{i}$ $[K]_{i}$ $[Cl]_{i}$ $V_{m}$ $%
d_{L}$ $f_{L}$ $p_{f}$ $p_{s}$ $p_{K}$ $q_{1}$ $q_{2}$ $P_{SOC}$ $R_{10}$ $%
R_{11}$ $R_{01}$ $h_{IP3}$ $[R_{G}^{S}]$ $[R_{P,G}^{S}]$ $[G]$ $[IP_{3}]$ $%
[PIP_{2}]$ $V_{cGMP}$ $[cGMP]$ $)^{T}$. The 20 relevant ionic currents are
also arranged as vector $\vec{I}(t)=($ $I_{\mathbf{VOCC}}$ $I_{\mathbf{BKCa}%
} $ $I_{\mathbf{Kv}}$ $I_{\mathbf{Kleak}}$ $I_{\mathbf{CaNSC}}$ $I_{\mathbf{%
NaNSC}}$ $I_{\mathbf{KNSC}}$ $I_{\mathbf{SOCCa}}$ $I_{\mathbf{SOCNa}}$ $I_{%
\mathbf{ClCa}}$ $I_{\mathbf{PMCA}}$ $I_{\mathbf{NCX}}$ $I_{\mathbf{NaK}}$ $%
I_{\mathbf{NaKCl}}^{Na}$ $I_{\mathbf{NaKCl}}^{K}$ $I_{\mathbf{NaKCl}}^{Cl}$ $%
I_{\mathbf{SERCA}}$ $I_{\mathbf{tr}}$ $I_{\mathbf{rel}}$ $I_{\mathbf{IP3}}$ $%
)^{T}$ for convenience. Considering finite variations such that the
higher-order contributions of variables converge ($\delta _{\nu
_{i}}^{n}\rightarrow 0$ at $n\gg 1$), ionic currents as well as relevant
nonlinear equations can be validly expressed by a Taylor series, e.g. $%
I_{j}(t)=\bar{I}_{j}+\Delta I_{j}(\bar{\nu}_{i},\delta _{\nu i}^{n}(t))$ for
the $j_{th}$ vector component of $\vec{I}(t)$. Here, $\bar{I}_{j}$ denotes
the time average value for component $I_{j}(t)$.

For the quasi-equilibrium conditions, the transient variables $\nu _{i}(t)$ and
ionic currents $I_{j}(t)$ are stable and their time-average terms ($\bar{\nu}%
_{i}$ and $\bar{I}_{j}$) remain constant. All time-dependent behaviors can
be attributed to variant terms $\delta _{\nu _{i}}(t)$, i.e.:
\begin{eqnarray}
I_{j}(t) &=&\bar{I}_{j}+\Delta I_{j}(\bar{\nu}_{i},\delta _{\nu i}(t))
\nonumber \\
&\simeq &\bar{I}_{j}+\sum_{i}\Delta I_{j}^{\nu _{i}}\delta _{\nu i}(t)
\label{ty1}
\end{eqnarray}%
for first-order approximation of ionic currents, and:
\begin{eqnarray}
\frac{d}{dt}\nu _{i}(t) &=&\frac{d}{dt}\left[ \overline{\nu }_{i}+\delta
_{\nu _{i}}(t)\right]   \nonumber \\
&\simeq &\sum_{j}\Delta _{\nu _{i}}^{\nu _{j}}\delta _{\nu
_{j}}(t)+\sum_{j,k}\Delta I_{k}^{\nu _{j}}\delta _{\nu _{j}}(t)  \nonumber \\
\frac{d}{dt}\delta _{\nu i}(t) &=&\sum_{j\in \{1,...,26\}}\left[ \Delta
_{\nu _{i}}^{\nu _{j}}+\sum_{k\in \{1,...,20|\}}\Delta I_{k}^{\nu _{j}}%
\right] \delta _{\nu _{j}}(t)  \label{ty2}
\end{eqnarray}%
for the first-order approximation equations of components $\nu _{i\in
\{1,..,26\}}(t)$, in which relevant variables $\Delta I_{k}^{\nu _{j}}$ and $%
\Delta _{\nu _{i}}^{\nu _{j}}$ are derived in more detail in the appendix
section. Alternatively, Eqs. (\ref{ty1}) and (\ref{ty2}) can be formulated
in matrix forms:
\begin{eqnarray}
\vec{I}(t) &=&\vec{\overline{I}}+\mathbf{\Lambda }\vec{\delta}_{\nu }(t)
\label{ty3} \\
\frac{d}{dt}\vec{\delta}_{\nu }(t) &=&\mathbf{\Omega }\vec{\delta}_{\nu }(t)
\label{ty4}
\end{eqnarray}%
Obviously, Eq. (\ref{ty4}) has an analytical solution:
\begin{equation}
\vec{\delta}_{\nu }(t)=e^{\mathbf{\Omega }(t-t_{0})}\vec{\delta}_{\nu
}(t_{0})  \label{ty5}
\end{equation}%
where $\exp (\mathbf{\Omega }t)$ represents the matrix exponential of $%
\mathbf{\Omega }t$. With the given initial conditions, the transient properties of $\vec{I}(t)$ can otherwise
be straightforwardly obtained by substituting Eq. (\ref{ty5}) into Eq. (\ref%
{ty3}).

With the eigensolution of the matrix $\mathbf{\Omega }$ in Eq.
(\ref{ty4}), the transient characteristics of variables $\vec{\delta}_{\nu
}(t)$ can be conveniently demonstrated from the viewpoint of eigenvalues $%
\omega _{i}$ and eigenfunction $\vec{\theta}_{i}(\delta _{\nu _{j}})$ for eigenmode $i$.
Without loss of generality, the time evolution of $i_{th}$ eigenmode
can be expressed as $\exp (\omega _{i}t)\vec{\theta}_{i}(\delta _{\nu
_{j}})\equiv \exp (\omega _{i,r}t+i\omega _{i,c}t)\vec{\theta}_{i}(\delta
_{\nu _{j}})$ with $j\in \{1,...,26\}$. The positive (negative) real part $%
\omega _{i,r}$ of the eigenvalues denotes the growing (decaying) time by $%
1/\omega _{i,r}$, and the imaginary part $\omega
_{i,c}$ denotes its period of oscillation by $2\pi /\omega _{i,c}$. The
component amplitude of eigenfunction $|\theta _{i,j}|=|\vec{\theta}%
_{i}(\delta _{\nu _{j}})|$ indicates the oscillating amplitude of the $j_{th}
$ variable $\nu _{j}$ in the $i_{th}$ eigenmode, and the phase angle $%
(\vartheta _{i,j})=\mathtt{angle}[\vec{\theta}_{i}(\delta _{\nu _{j}})]$ indicates
its lag phase in the oscillation period. Since $\mathbf{\Omega }$ is a
non-symmetric matrix, eigenfunctions of SMC-systems are not mutually orthogonal in our
case.

\subsection{Eigensystem formulation for multiple SMCs}

We next consider the condition for multiple ($m$) SMCs coupled
via gap junctions. For this case, Eq. (\ref{ty4}) is straightforwardly
extended to be:
\begin{equation}
\frac{d}{dt}\left[
\begin{array}{c}
\vec{\delta}_{\nu ,1}(t) \\
... \\
\vec{\delta}_{\nu ,m}(t)%
\end{array}%
\right] =\left[
\begin{array}{ccc}
\widetilde{\mathbf{\Omega }}_{1} & \mathbf{\Theta }_{1j} & 0 \\
\mathbf{\Theta }_{j1} & \widetilde{\mathbf{\Omega }}_{j} & ... \\
0 & ... & \widetilde{\mathbf{\Omega }}_{m}%
\end{array}%
\right] \left[
\begin{array}{c}
\vec{\delta}_{\nu ,1}(t) \\
... \\
\vec{\delta}_{\nu ,m}(t)%
\end{array}%
\right]   \label{ty6}
\end{equation}%
with matrix components%
\begin{eqnarray}
\widetilde{\mathbf{\Omega }}_{pq,i} &=&\mathbf{\Omega }_{pq,i}+\sum_{j}%
\Delta I_{GJ,j}^{\nu _{q}}\delta _{\nu _{q,i}}  \label{ty7} \\
\mathbf{\Theta }_{pq,iC} &=&\sum_{j}\Delta I_{GJ,j}^{\nu _{q,C}}\delta _{\nu
_{q,C}}  \label{ty8}
\end{eqnarray}%
Here, $\Delta I_{GJ,j}$ corresponds to gap junction currents for ion $%
j$ (or $[IP_{3}]$), and $C$ indexes the nearby SMC coupled to the local
one. The component $\widetilde{\mathbf{\Omega }}_{pq,i}$ in Eq. (\ref{ty7})
includes additional contributions of the gap junction currents by the
terms $\Delta I_{GJ,j}^{\nu _{q}}$, which are associated with
variations of $\delta _{\nu _{q,i}}$ ($q_{th}$ component of $\vec{\delta}%
_{\nu ,i}$) in the local SMC $i$. The component $\mathbf{\Theta }_{pq,iC}$
in Eq. (\ref{ty8}) includes the contributions of the gap junction currents
by the terms $\Delta I_{GJ,j}^{\nu _{q,C}}$, that is associated with
variations of $\delta _{\nu _{q,C}}$ in the nearby SMC $C$. Since the format
of the matrix in Eq. (\ref{ty6}) remains the same as that in Eq. (\ref{ty4}%
), the same process for solving a single SMC is applied for
multiple SMCs.

\section{Numerical results and discussions}
In this section, ($\mathrm{i}$) we first carry out the frequency-domain analysis on a control condition of SMC. We study how the eigenvalue and eigenfunction correspond to time evolutions of transient variables.
($\mathrm{ii}$) Continuing with the frequency-domain approach, we investigate properties of calcium oscillations at different concentrations of potassium; otherwise, we perform the corresponding time-domain analysis to get intuitional ideas of synchronization among complex signals. Another exemplar of two coupled SMCs is prepared to elucidate more realistic (intracellular and intercellular) calcium dynamics and resonance effects.
($\mathrm{iii}$) Lastly, we explore practical finite SMC clusters. With an input delta-function calcium pulse in this resonance medium, we observe physiological functionalities of rhythmic oscillations of SMCs.

Mathematical algorithms were implemented in Visual C++ and were executed on a HP Z800 workstation with 48GB of RAM. Relevant default values of the variables and initial parameters are defined in appendix \ref{appa}. Source codes of C++ language for time-domain and frequency-domain analyses are available online: https://drive.google.com/open?id=0B8l8iii7Z4iqWWJRLTduRl9zMVE. We also refer to JSim computations for time-domain programming: http://www.physiome.org/jsim/.

\subsection{ Frequency-domain analysis on the control set of SMC}
Default parameters for the control condition of SMC are defined in appendix \ref{appa}, except $[K]_e=35.8mM$, $[NE]=2.0\times 10^{-4}mM$, $I_{SERCA,0}=20.4pA$, and $R_{leak}=0.0000535$. We obtain the equilibrium parameters $\vec{\overline{\delta}}_{\nu}$ necessary for frequency-domain analysis after running the time-domain simulation for $10^5s$. Relevant parameters of the SMC model are taken from known experimental measurements \cite{KClexp1}: 25s-period calcium oscillation in rat mesenteric arterioles and the threshold value of $[K^{+}]_e=20nM$ for triggering calcium oscillation for instance.

\begin{figure*}[ht]
\centering
\includegraphics[scale=0.4]{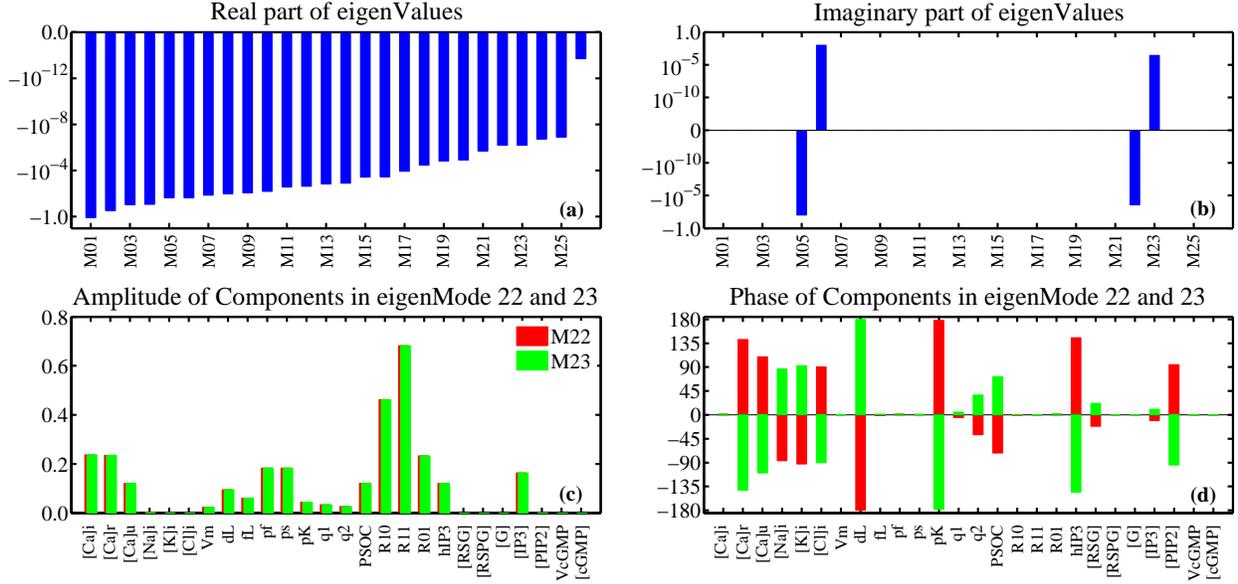}
\caption{ Numerical analysis of eigenvalues and eigenfunctions for SMC on the control condition: (a) real parts of eigenvalues, (b) imaginary parts of eigenvalues, (c) oscillating phase of transient variables in eigenmodes 22 and 23, and (d) oscillating amplitude of transient variables in eigenmodes 22 and 23.}\label{fig2}
\end{figure*}

Figure (\ref{fig2}) shows numerical results of eigenvalues and eigenfunctions for a single SMC on the control condition. The figure is divided into four parts: ($\mathrm{\mathbf{a}}$) Real parts of eigenvalues that decide the growing time or decaying time of eigenmodes, corresponding to positive or negative values, respectively; ($\mathrm{\mathbf{b}}$) Imaginary parts of eigenvalues that indicate the oscillation periods of eigenmodes; ($\mathrm{\mathbf{c}}$) Oscillating amplitude of transient variables in eigenmodes $M22$ and $M23$; and ($\mathrm{\mathbf{d}}$) Oscillating phase of transient variables in eigenmodes that tells the phase lags regarding oscillations. Each oscillating amplitude value in Fig (\ref{fig2}c) is divided by its own average to show the percentage of variations and is dimensionless.

In Figs. (\ref{fig2}a-\ref{fig2}b) for a single SMC model, we obtain two sets of mutually complex-conjugate eigenvalues, with values $-0.028\pm 0.011i$ and $-8\cdot10^{-7} \pm 0.000264i$ $ms^{-1}$, corresponding to eigenmodes $(5,6)$ and $(22,23)$, respectively. We select the set of eigenmodes $(22,23)$, which signify a longer decay-time of $1250s$ and oscillation period of $23.8s$, for studying the properties of transient variables.

For the set of eigenmodes $(22,23)$, Figs. (\ref{fig2}c-\ref{fig2}d) illustrate the oscillating amplitude and phase of transient variables. Several quantities are discussed here for the following study: the oscillation amplitude of $[Ca]_i$ is $22\%$, the oscillation amplitude of $[Ca]_u$ is $10\%$, and the phase difference between $[Ca]_i$ and $[Ca]_u$ is $105^{o}$.

\begin{figure*}[ht]
\centering
\includegraphics[scale=0.375]{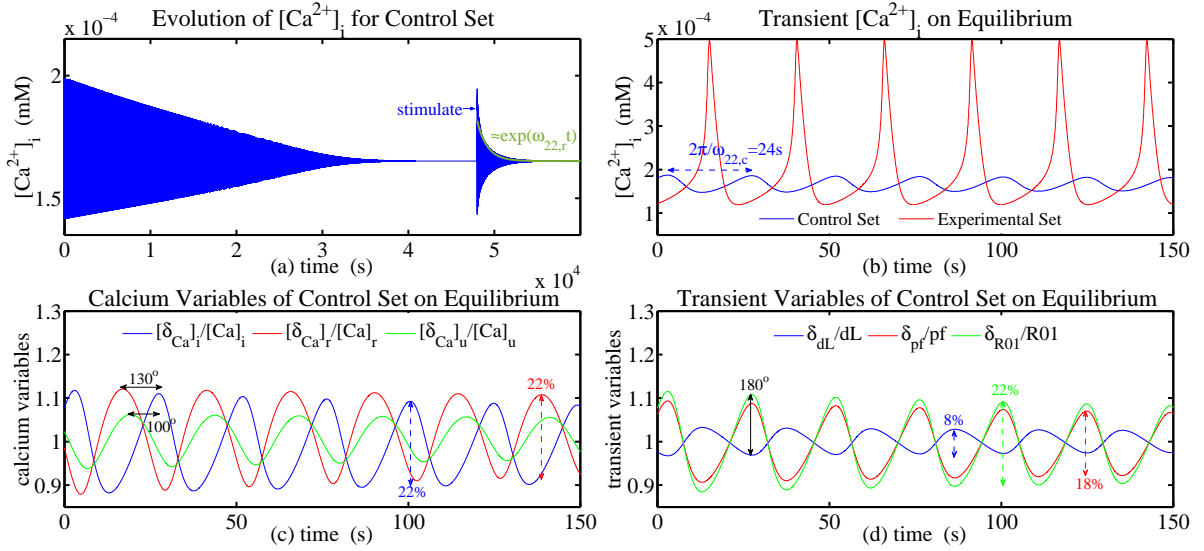}
\caption{ Time evolutions of several transient variables in eigenmodes $(22,23)$ for a single SMC: (a) time evolution of $[Ca^{2+}]_i$ with given delta-function calcium stimulation at equilibrium $t=4.8\times 10^{4}s$, (b) transient variations of $[Ca^{2+}]_i$ after the delta-function calcium stimulation for the control condition $[K^+]_e=35.8mM$ and experimental condition $[K^+]_e=40.0mM$, (c) and (d) transient variations of several variables after the delta-function calcium stimulation.}\label{fig3}
\end{figure*}

To apply the quantities from frequency-domain analysis, we study the time-domain calculations . Figure (\ref{fig3}) shows numerical results for the time evolutions of several transient variables.
Figure (\ref{fig3}a) presents the temporal variation of $[Ca^{2+}]_i$ to arrive at the equilibrium after $t=4\times 10^{4}s$. After equilibrium, we add an input of delta-function calcium stimulation at $t=4.8\times 10^{4}s$. Responses to this stimulation illustrate a calcium oscillation having decaying time $1/\omega_{22,r}=-1250s$ (green curve in Fig. \ref{fig2}a) and oscillation period $2\pi/\omega_{22,c}=24s$ (blue curve in Fig. \ref{fig2}b), which agree with the values from frequency-domain analysis. An extra calculation for SMC on experimental condition ($[K^{+}]_e=40mM$, red curve) is appended here to explain the high dosages of agonists and is in the paragraph below.

Figures (\ref{fig3}c-\ref{fig3}d) characterize oscillating amplitudes and phases for several transient variables in eigenmodes $(22,23)$. Here, each curve is divided by its average of variables to in order to compare the results with Figs. (\ref{fig2}c-\ref{fig2}d): $22\%$ oscillation amplitude of $[Ca]_i$ relatively compared to $10\%$ oscillation amplitude of $[Ca]_u$, and $105^{o}$ phase lags between $[Ca]_i$ and $[Ca]_u$, which agree with those in Figs. (\ref{fig2}c-\ref{fig2}d).

\subsection{Time-domain and frequency-domain analyses for a single SMC}

\begin{figure*}[ht]
\centering
\includegraphics[scale=0.35]{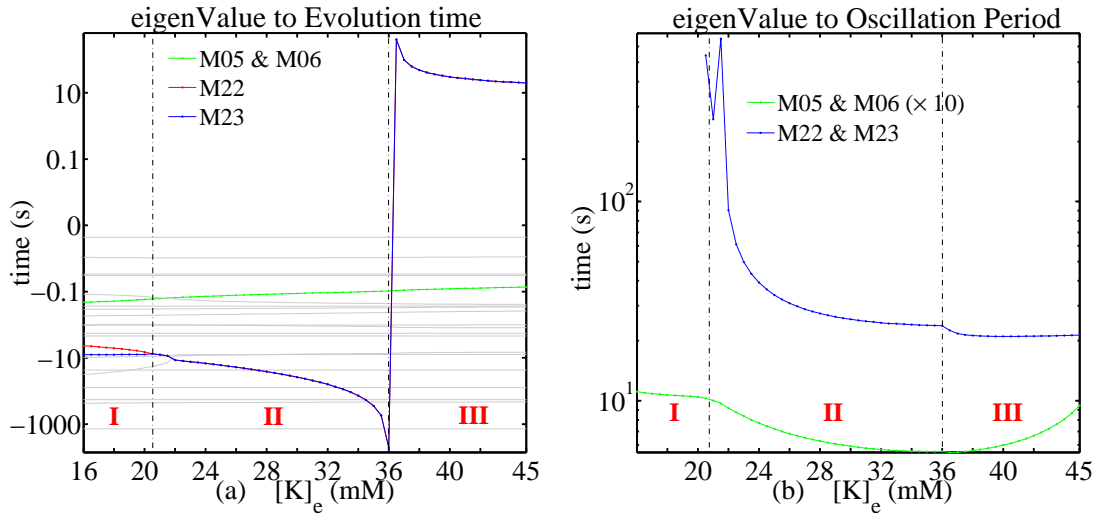}
\caption{ Eigenvalue spectrum of a single SMC in response to extracellular potassium concentrations. (a) real parts of eigenvalues for evolution time $\sim 1/\omega_{22,r}$, and (b) imaginary parts of eigenvalues for oscillation periods $\sim 2\pi/\omega_{22,i}$.}\label{fig4}
\end{figure*}

Continuing with the frequency-domain approach, we investigate properties of calcium oscillations at different concentrations of potassium. Figure (\ref{fig4}) gives the eigenvalue spectrum of a single SMC in response to different extracellular potassium concentrations. The spectrum is depicted in units of time, instead of frequency, for convenient descriptions. For real parts of eigenvalues in Fig. (\ref{fig4}a), the positive (negative) value denotes the growing (decaying) time $\sim (1/\omega_{i,r})$. We find two sets of complex-conjugate solutions corresponding to rhythmic oscillations by frequency-domain analysis: ($\mathrm{\mathbf{i}}$) the set of eigenmode $(5,6)$ (green curves) characterizes a short oscillation period ($\sim1-10s$), fast decay time ($\sim0.1s$), and insensitivity against agonist. This mode exists even under very low dosages of agonists. According to the literature \cite{vasoa3}, rat mesenteric arteries are resistant to spontaneous vasomotion. Moreover, only sparse observations of spontaneous vasomotions with periods $\sim 2s$ are noted in our experiments of mesenteric arterioles. For these reasons, we identify eigenmodes $(5,6)$ as spontaneous vasomotions and observe them infrequently in realistic mesenteric arterioles due to their fast decay. ($\mathrm{\mathbf{ii}}$) The set of eigenmode $(22,23)$ (red and blue curves in Fig. \ref{fig4}), however, exhibits strong dependence on the extracellular potassium concentrations and has different behaviors in regions $\mathrm{\mathbf{I}}$, $\mathrm{\mathbf{II}}$, and $\mathrm{\mathbf{III}}$. In region $\mathrm{\mathbf{I}}$, an oscillation-deactivation section, eigenvalues of modes $22$ and $23$ are real and distinctly separate, and no oscillating actions appear. In region $\mathrm{\mathbf{II}}$, an oscillation-activation section, the decay time of modes $(22,23)$ prolongs exponentially when increasing extracellular potassium concentrations, and is present as a dominant eigenmode at $[K]_e\simeq 36mM$. Around $[K]_e\simeq 36mM$, the set of eigenmode $(22,23)$ can be realized as experimental observations due to the robustness of its life time. For $[K]_e\ll 36mM$, eigenmode $(22,23)$ could be smeared due to the nature of non-orthogonality to other non-oscillating eigenmodes (gray curves). In region $\mathrm{\mathbf{III}}$, as shown in Fig. (\ref{fig4}a), the real part of the eigenvalues turns into a positive value, and hence eigenmode $(22,23)$ transfers to a transient growing type. The oscillating amplitude could rise violently along the temporal curve as the red curve (experimental set) in Fig. (\ref{fig3}b). In this region, the significant deviations from equilibrium values of transient variables invalidate the precision of our algorithm except for qualitative inferences. In summary, numerical results of Fig. (\ref{fig4}) show three types of
calcium oscillations for a single SMC in mesenteric arterioles: spontaneous calcium oscillation (green curves), agonist-dependent calcium oscillation (blue curves in region $\mathrm{\mathbf{II}}$), and agonist-dependent calcium spike (blue curves in region $\mathrm{\mathbf{III}}$) as shown in Fig. (\ref{fig4}).

\begin{figure*}[ht]
\centering
\includegraphics[scale=0.375]{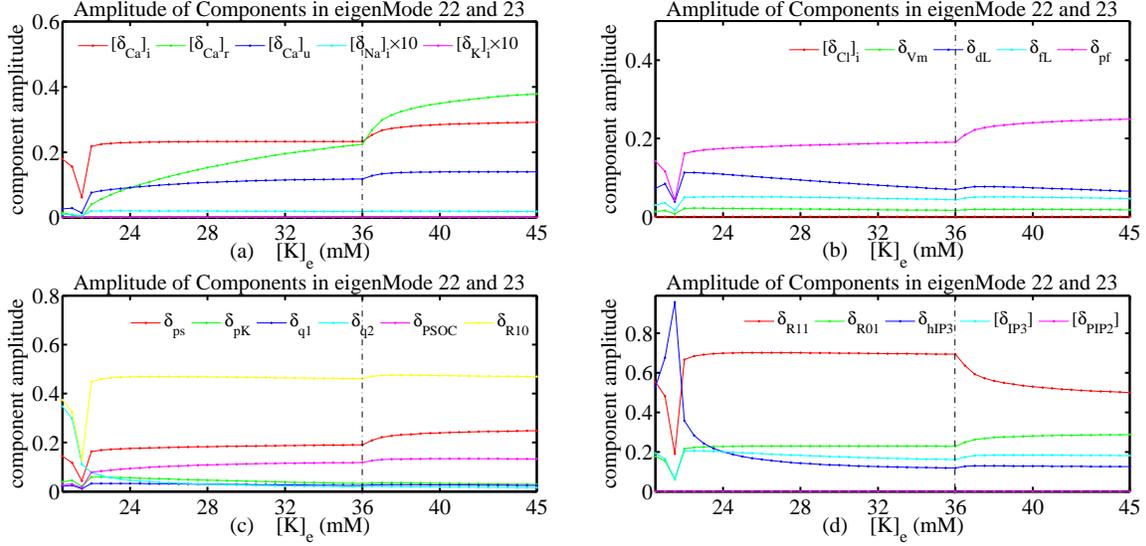}
\caption{ Oscillating amplitude $\hat{\delta}_{\nu i}\equiv |{\delta}_{\nu i}/{\bar{\delta}}_{\nu i}|$ of variables in eigenmode ($22,23$) in response to extracellular potassium concentrations.}\label{fig5}
\end{figure*}

\begin{figure*}[ht]
\centering
\includegraphics[scale=0.375]{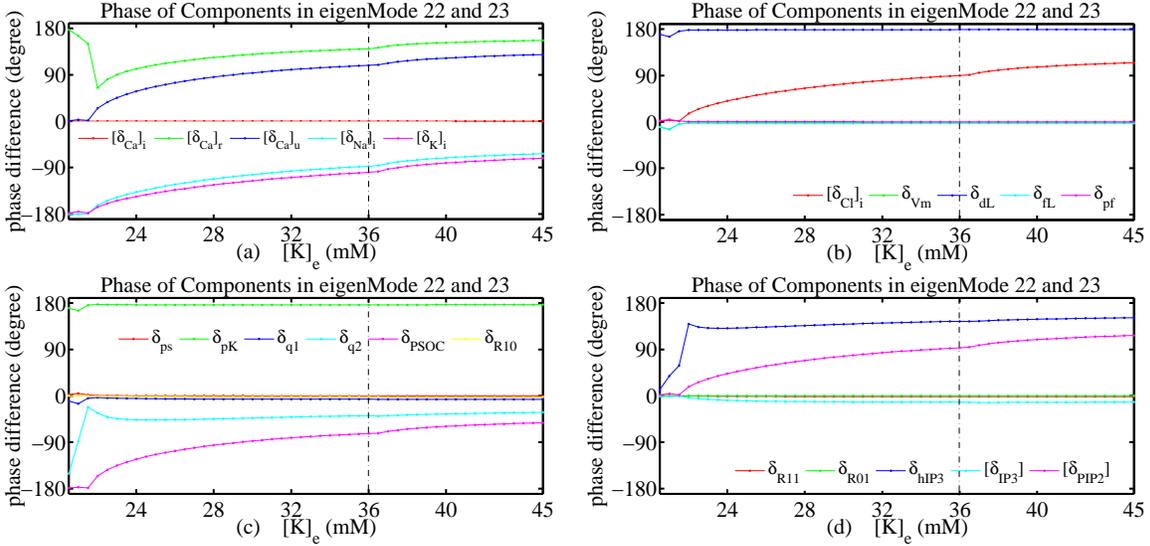}
\caption{ Oscillating phase $\vartheta_{\nu i}$ of variables in eigenmode ($22,23$) in response to extracellular potassium concentrations.}\label{fig6}
\end{figure*}

\begin{figure*}[ht]
\centering
\includegraphics[scale=0.375]{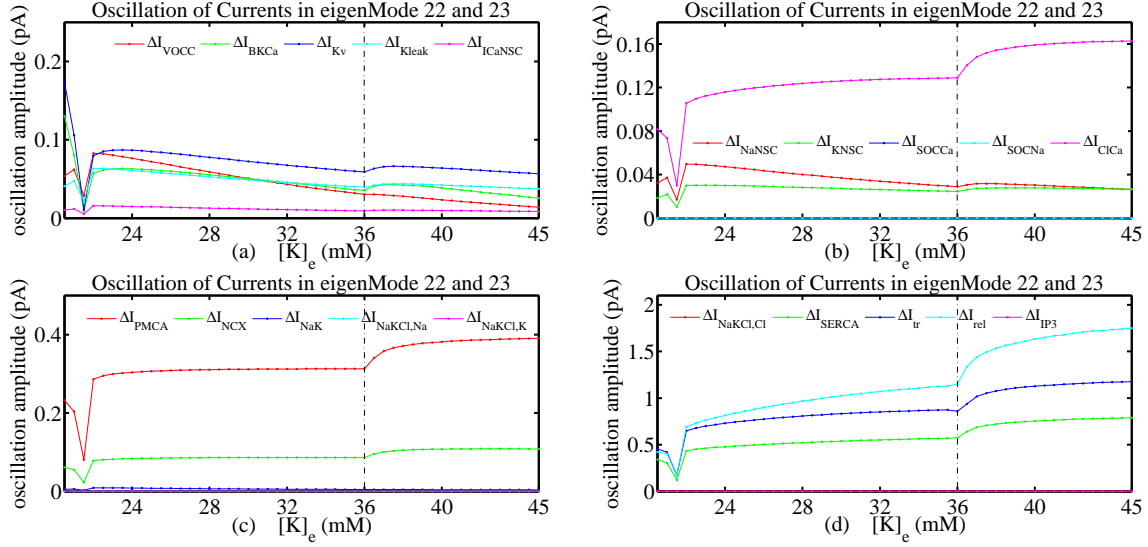}
\caption{ Oscillating amplitude of current $\Delta I_i$ in eigenmode ($22,23$) in response to extracellular potassium concentrations, in which $[\delta_{Ca}]_i=2\times 10^{-5}mM$. }\label{fig7}
\end{figure*}

\begin{figure*}[ht]
\centering
\includegraphics[scale=0.375]{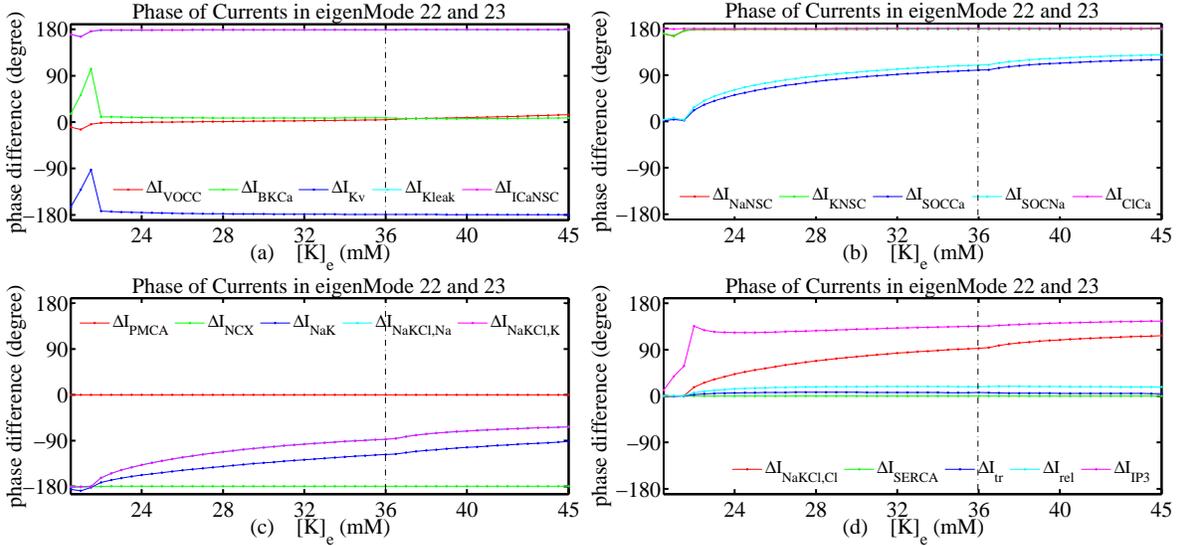}
\caption{ Phase lag regarding oscillations for current $\Delta I_i$ in eigenmode ($22,23$) versus extracellular potassium concentrations, in which $[\delta_{Ca}]_i=2\times 10^{-5}mM$. }\label{fig8}
\end{figure*}

In addition to eigenvalue analysis in Fig. (\ref{fig4}), eigenfunctions also provide explicit information as discussed below. Figures (\ref{fig5}) and (\ref{fig6}) show the oscillating amplitude $\hat{\delta}_{\nu i}\equiv |{\delta}_{\nu i}/{\bar{\delta}}_{\nu i}|$ and oscillating phase $\vartheta_{\nu i}$ for eigenmode ($22,23$) in response to extracellular potassium concentrations, respectively. By Eq. (\ref{ty3}), we transform the eigenfunction associated with $\vec{\delta}_{\nu }$ into an eigenfunction associated with currents $\Delta \vec{I}\sim\mathbf{\Lambda} \vec{\delta}_{\nu }$. Figures (\ref{fig7}) and (\ref{fig8}) illustrate numerical results for the oscillating amplitude and phase of currents in eigenmode ($22,23$) versus extracellular potassium concentrations, respectively. We note that the definitions of phase lag here refer to the calcium oscillation in cytosol.

Previous studies of signal-inhibitions and vascular mechanics have in fact suggested a diverse mechanism (in addition to the known intracellular stores) for rhythmic contractions of SMCs. For instance several mechanical measurements \cite{dyn1,dyn2,dyn3,dyn4,dyn5,dyn6} have demonstrated mandatory or modulatory roles of $K^+$ channels for vasomotion in arteries. The activation of $cGMP$ and the calcium-dependent chloride current for vasomotion have been found in endothelium-denuded mesenteric arteries. In a voltage-dependent coupled oscillator model, the literature has also proposed the depolarization and the involvement of voltage-dependent calcium current to be responsible for agonist-dependent vasomotion in mesenteric arteries \cite{dyn7,dyn8,dyn9}.

To investigate the interplays of a diverse mechanism for rhythmic contractions of SMCs, we study the correlations and synchronizing timings of signal pathways from Figs (\ref{fig5}-\ref{fig8}). We find the following: ($\mathrm{\mathbf{i}}$) By increasing $[K]_e$, electrical oscillations in cytosol gradually change from the cyclic $(K^+_i+Na^+_i\leftrightarrow Ca^{2+}_i+Cl^-_i)$ configuration toward the cyclic $(K^+_i+Na^+_i+Ca^{2+}_i \leftrightarrow Cl^-_i)$ configuration as in Fig. (\ref{fig6}). This is in response to the growing strength of oscillations by alternate positive and negative charge accumulations in cytosol at high $[K]_e$. ($\mathrm{\mathbf{ii}}$) By increasing $[K]_e$, the oscillating amplitude related to transmembrane currents decreases, while the oscillating amplitude related to intracellular stores ($\Delta I_{tr}$,$\Delta I_{rel}$,$\Delta I_{SERCA}$) increase as in Fig. (\ref{fig7}). ($\mathrm{\mathbf{iii}}$) By increasing $[K]_e$, the oscillating amplitude of $[\delta_{Ca}]_r$ rapidly increases while that of $[\delta_{Ca}]_i$ and $[\delta_{Ca}]_u$ almost remain constant as in Fig. (\ref{fig5}). This fact causes the evolvement of the temporal waveform of $[Ca]_i$ from being a sine-like (e.g. red curve in Fig. \ref{fig3}b) to a spike-like (e.g. blue curve in Fig. \ref{fig3}b) function, while keeping the accumulation of $[Ca]_i$ (integral area of waveform) relatively stable. ($\mathrm{\mathbf{iv}}$) By increasing $[K]_e$, the discordance between the increasing amplitudes of $[\delta_{pf}]$ and $[\delta_{ps}]$ (Fig. \ref{fig5}) and the decreasing amplitude $\Delta I_{BKCa}$ (Fig. \ref{fig7}) is symbolized as desynchronization effects for this pathway. This inference can also be deduced from another discordance between $\Delta I_{BKCa}/\bar{I}_{BKCa}\sim1.5\%$ and $\delta_{p_f}/\bar{p}_f=\delta_{p_s}/\bar{p}_s\sim19\%$ at $[K]_e=35.8mM$.

\begin{figure*}[h]
\centering
\includegraphics[scale=0.4]{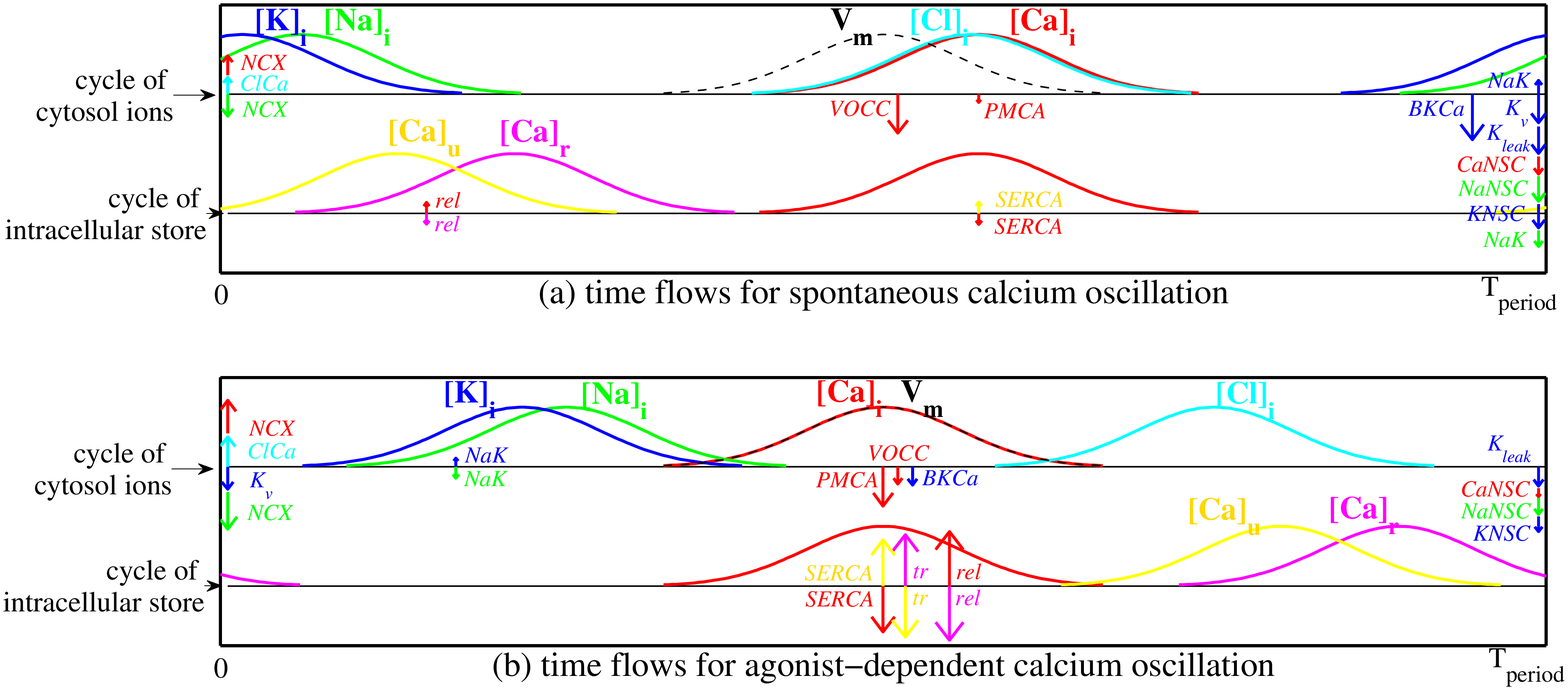}
\caption{ Schematic sketch of coupled time flows of signaling pathways for (a) spontaneous calcium oscillation and (b)agonist-dependent calcium oscillation. The curves show the timings of the maximum values of ionic concentrations $\delta_{\nu i\in\{1..6\}}$ and membrane potential $\delta_{\nu i\in\{7\}}$. The positions of the arrows schedule the time when the maximal amplitudes of current variations occur, and the lengths of the arrows indicate the oscillation amplitudes of channel currents. Arrows are colored as relevant ions. The opposite-direction current variations occur after a time lag $T_{period}/2=\pi/\omega_{22,i}$ and are not shown here. It is emphasized that the positive and negative arrows represent the relatively increasing and decreasing concentration to the values in equilibrium, respectively, and not the absolute values of concentrations.}\label{fig9}
\end{figure*}

On the basis of these eigenfunction analyses, we further investigate how synchronizing timings among signal pathways differentiate the spontaneous and agonist-dependent calcium oscillation.
For simplicity, we categorize time flows of signalling pathways into two streams: one stream for the cycle of cytosol ions, and the other stream for the cycle of intracellular store.
Figure (\ref{fig9}) depicts the schematic sketch of coupled time flows of signaling pathways for (a) spontaneous calcium oscillation and (b)agonist-dependent calcium oscillation on the control condition. The curves show the timings of maximum values of ionic concentrations $\delta_{\nu i\in\{1..6\}}$ and membrane potential $\delta_{\nu i\in\{7\}}$. The positions of the arrows schedule the time when the maximal amplitudes of current variations occur, and the lengths of the arrows indicate the oscillation amplitudes of channel currents. Arrows are colored as relevant ions. The opposite-direction current variations occur after a time lag $T_{period}/2=\pi/\omega_{22,i}$ and are not shown here. For the fast decaying spontaneous calcium and long-lasting agonist-dependent calcium oscillation at $[K]_e=35.8mM$, we find ($\mathrm{\mathbf{i}}$) the cycle of cytosol ions through transmembrane currents is dominant in the former condition, while the cycle of intracellular store is dominant in the latter condition; ($\mathrm{\mathbf{ii}}$) the time lag between calcium and voltage oscillations is finite in the former condition, while the oscillating phases of calcium and voltage oscillations are exactly synchronous in the latter condition; ($\mathrm{\mathbf{iii}}$) Aside from the $Ca^{2+}$ itself, $K^+$ oscillation plays the primary role in the former condition, while $Na^+$ and $Cl^-$ oscillations are relatively intense in the latter condition.
All these findings above interpret the significance of synchronizing timings for vasomotions on different conditions \cite{vasoa4,an2}, and conclude with inevitable involvements of $Na^+,K^+,Cl^-$ ions as well as other relevant channels \cite{dyn7,dyn8,dyn9}.

\begin{figure*}[h]
\centering
\includegraphics[scale=0.375]{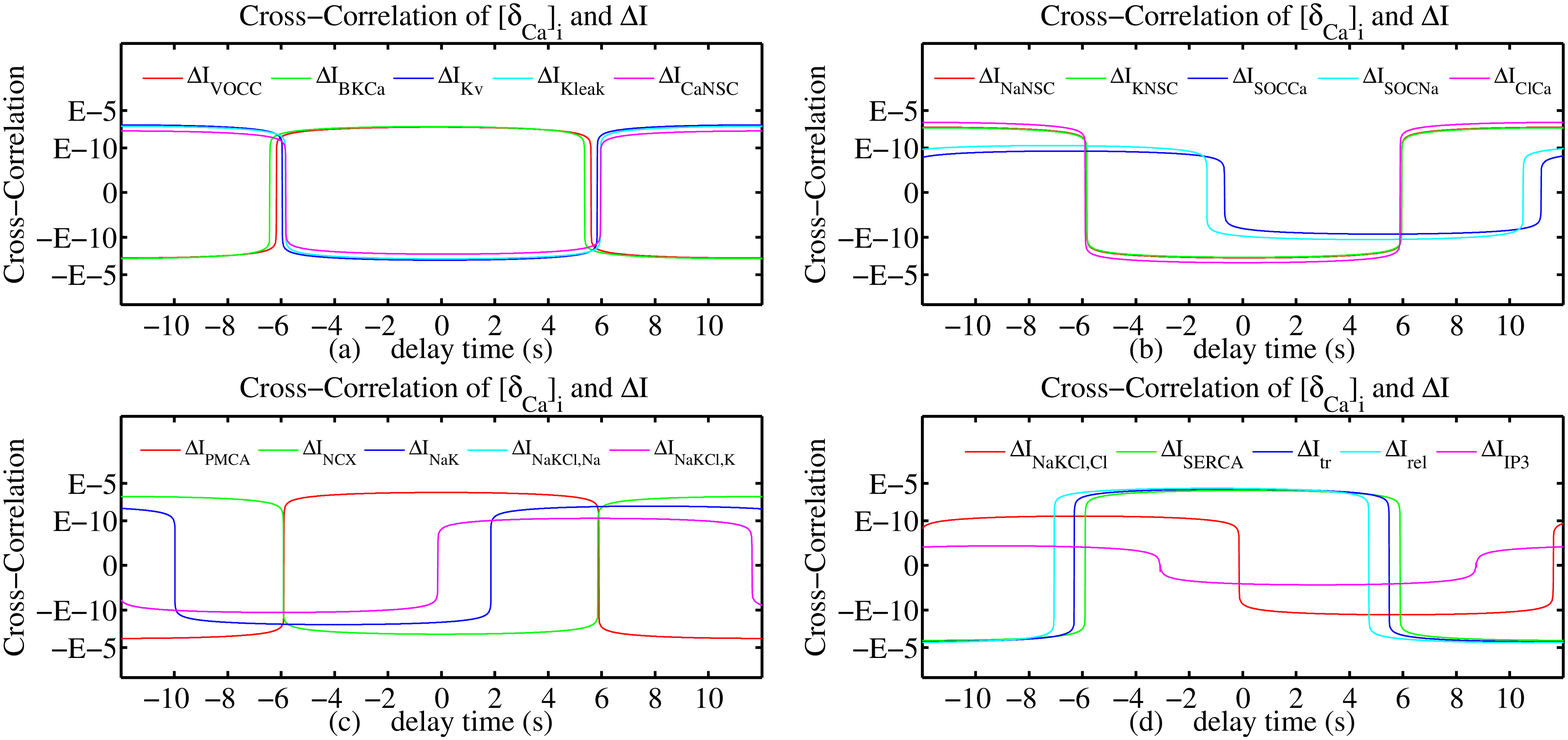}
\caption{ Time-lag cross-correlation between calcium oscillation $[\delta_{Ca}]_i$ and current oscillations $\Delta I_i$.}\label{fig10}
\end{figure*}

\begin{figure*}[h]
\centering
\includegraphics[scale=0.375]{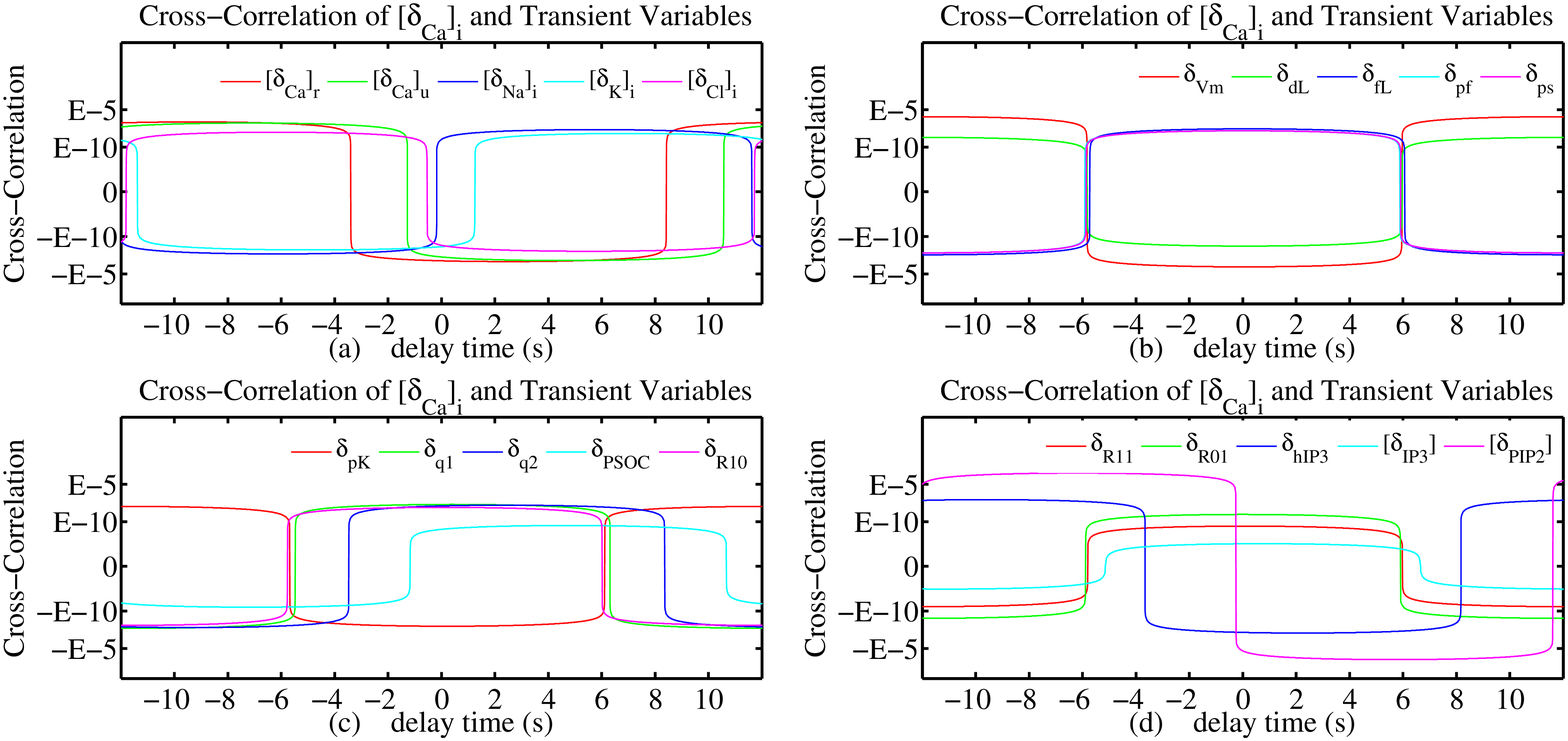}
\caption{ Time-lag cross-correlation between calcium oscillation $[\delta_{Ca}]_i$ and other transient variables $\delta _{\nu i}$.}\label{fig11}
\end{figure*}

We also append time-domain calculations corresponding to analyses in Fig. (\ref{fig9}b) to validate our algorithms. Figure (\ref{fig10}) shows the time-lag cross-correlation between calcium oscillation $[\delta_{Ca}]_i$ and current oscillations $\Delta I_i$. Figure (\ref{fig11}) illustrates the time-lag cross-correlation between calcium oscillation $[\delta_{Ca}]_i$ and other transient variables $\delta _{\nu i}$. With the given oscillation period of $24s$ on the control condition, for instance, the $8s$ time lag for $\Delta I_{NaK}$ in Fig. (\ref{fig10}c) is equivalent to the shift $-(8/24)T_{period}$ of $\Delta I_{NaK}$ from $[Ca]_i$ in Fig. (\ref{fig9}b).

\begin{figure*}[h]
\centering
\includegraphics[scale=0.375]{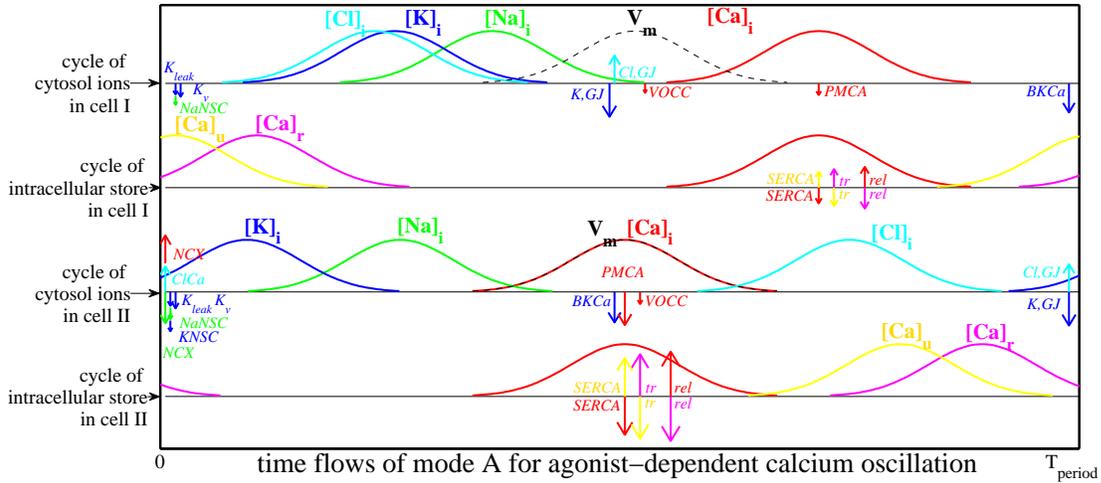}
\caption{ Schematic sketch of coupled time flows of signaling pathways for eigenmode A, with $T_{period}=2\pi/\omega_{A,i}$. Relevant definitions for the curves and arrows are similar to that in Fig. (\ref{fig9}).}\label{fig12}
\end{figure*}

\begin{figure*}[h]
\centering
\includegraphics[scale=0.375]{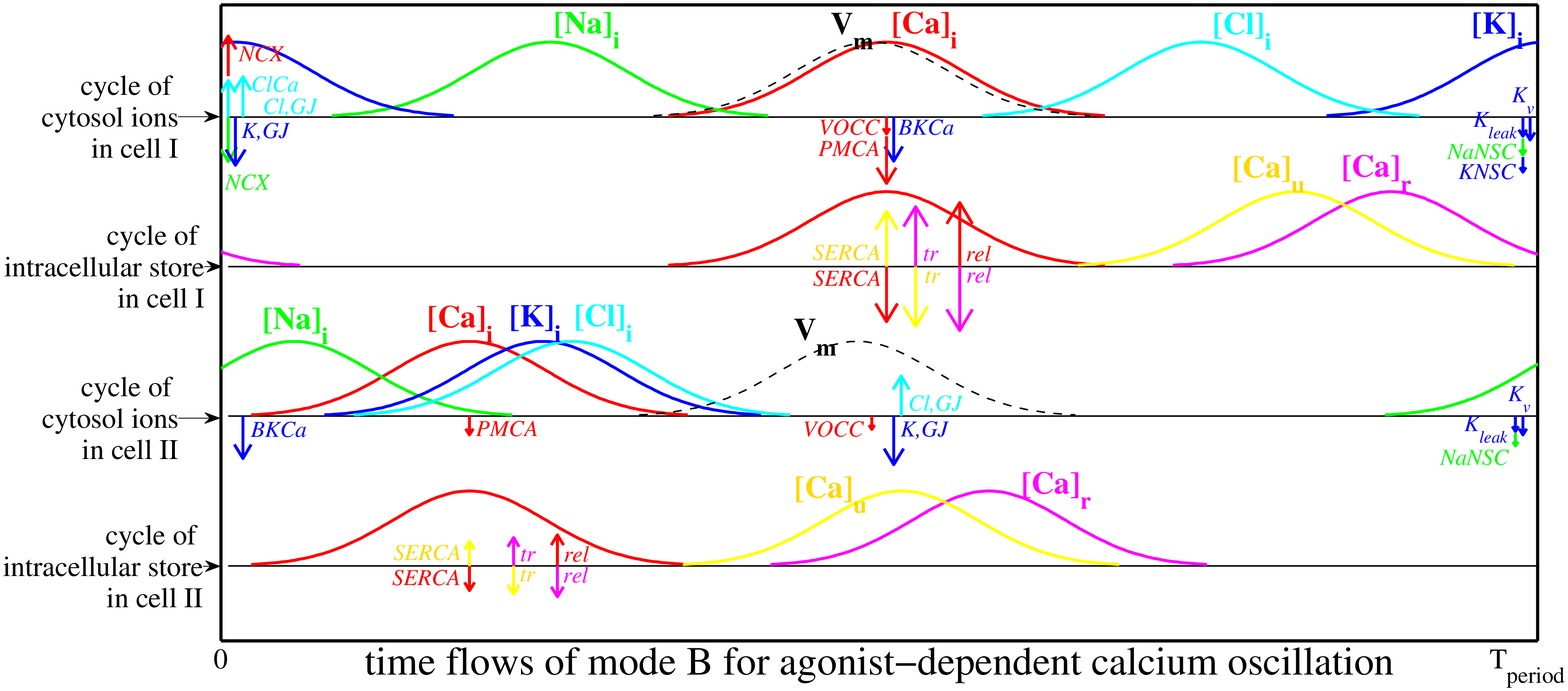}
\caption{ Schematic sketch of coupled time flows of signaling pathways for eigenmode B, with $T_{period}=2\pi/\omega_{B,i}$. Relevant definitions for the curves and arrows are similar to that in Fig. (\ref{fig9}).}\label{fig13}
\end{figure*}

Before studying multiple SMCs, we consider an example of two coupled SMCs (cells $I$ and $II$), including intracellular and intercellular (see the next section) calcium dynamics, for synchronizing and resonance effects. Extra parameters for the 2-SMCs are defined for inhomogeneous cell volumes, with $vol_I=1.6p\ell$ and $vol_{II}=1.1p\ell$. A numerical calculation of frequency-domain algorithm gives three sets of complex-conjugate eigenvalues for the 2-SMCs: ($\mathrm{\mathbf{i}}$) Mode S with $\omega_S=-0.028\pm0.011i$, ($\mathrm{\mathbf{ii}}$) Mode A with $\omega_A=-4.96e-07\pm 2.46e-04i$, and ($\mathrm{\mathbf{iii}}$) Mode B with $\omega_B=-1.49e-05\pm1.98e-04i$. Mode S is responsible for the spontaneous calcium oscillation, and Modes A and B present two kinds of globally agonist-dependent calcium oscillations. Figures (\ref{fig12}) and (\ref{fig13}) show the schematic sketch of time flows of signaling pathways for Mode A and B, respectively. One can refer to Fig. (\ref{fig9}) for the definitions of the curves and arrows. For Mode A in Fig. (\ref{fig12}), we observe that cell $II$ shows more prevailing calcium oscillation than that in cell $I$, while the oscillating phases of calcium and voltage are synchronous in cell $II$ but have a time lag in cell $I$. Oppositely for Mode B in Fig. (\ref{fig13}), we observed that cell $I$ shows more prevailing calcium oscillation than that in cell $II$, while the oscillating phases of calcium and voltage are almost synchronous in cell $I$ but have a time lag in cell $II$. We notice that oscillations of voltage in both cells always remain exactly synchronous in the studied cases.

\begin{figure*}[h]
\centering
\includegraphics[scale=0.375]{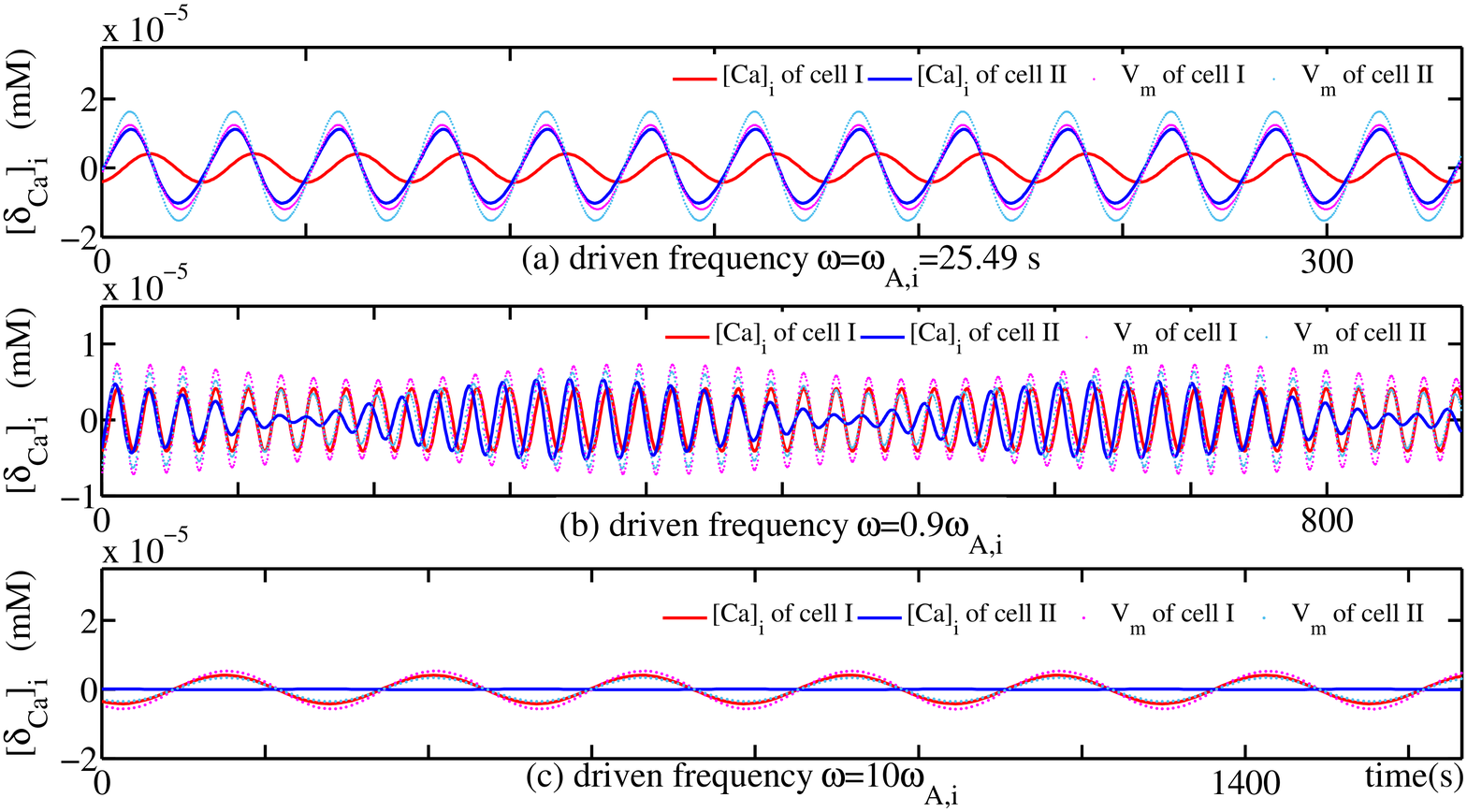}
\caption{ Time evolutions of $[Ca]_i$ and $V_m$ for two coupled SMCs ($I$ and $II$) at (a) on-resonance, (b) near-resonance, and (c) off-resonance conditions. Note that all voltage values are scaled and shifted to be comparable with calcium values.}\label{fig14}
\end{figure*}

To explore the influences of synchronizing timings, we further carry out time-domain calculations to study 2-SMCs's responses to the external signal, delivered from nerve activity or blood flow, for example \cite{book_K}. With Mode A, we consider three kinds of cyclic calcium stimulations (with frequency $\omega$) to cell $I$: $\mathrm{(a)}$ on-resonance condition with $\omega=\omega_{A,i}$, $\mathrm{(b)}$ near-resonance condition with $\omega=0.9\omega_{A,i}$, and $\mathrm{(c)}$ off-resonance condition with $\omega=10\omega_{A,i}$ as in Fig. (\ref{fig14}). Figure (\ref{fig14}) shows time evolutions of $[Ca]_i$ and $V_m$ for cells $I$ and $II$ in these conditions. For the on-resonance condition, the stimulation signals in cell $I$ transfer through gap junction to cell $II$, and bring both cells into calcium oscillation at Mode A; otherwise, the oscillating amplitudes in cell $II$ are more intense than that in cell $I$, and the oscillating phase of $[Ca]_i$ in cell $I$ has a time lag ($\sim T_{period}/5$) to $V_m$, agreeing with that in Fig. (\ref{fig12}). For the near-resonance condition, the stimulation signals from cell $I$ dissipate during transference, and cell $II$ exhibits incomplete synchronization with cell $I$. During the period of in-phase oscillations, the oscillating amplitude of $[Ca]_i$ in cell $II$ is strong. During the period of out-phase oscillations, however, the oscillating amplitude of $[Ca]_i$ in cell $II$ is relatively weak. The time lags among $[Ca]_i$ of cell $II$ and the other three variables vary with time. For the off-resonance condition, the stimulation signals from cell $I$ are mostly blocked from cell $II$. Calcium and voltage oscillations are not observed in cell $II$, and are significantly suppressed in cell $I$. With this case of 2-SMCs, we conclude that two factors are essential for efficient signalling communications among cells: $\mathrm{(i)}$ correct synchronizing timings among signal pathways occur in SMCs, i.e. the existence of definite eigenmodes, and $\mathrm{(ii)}$ stimulation signals having similar frequency to the eigenmode. On the basis of synchronizing and resonance concepts, we investigate functionalities of SMC's rhythm for practical cell clusters in the next section.

\subsection{Frequency-domain analysis for finite SMC clusters}
 We now explore physiological functionalities of SMCs' rhythmic oscillations, especially for signaling communications among cells. For intercellular communication, we include the electro-diffusion coupling \cite{mathmodel2}, which uses the Goldman-Hodgkin-Katz
(GHK) equation for ionic currents through the gap
junctions. These ionic currents were added to the membrane
potential equation and corresponding ionic flux equations as in appendices \ref{appb12}, \ref{appd}, and \ref{appf}.
We assume permeability to be the same for all ions. Gap
junction resistance values from experiments were
used to calculate the permeability \cite{gjvalue1}; otherwise, $15\%$ variations of SMC volumes, as indicated in Ref. \cite{sizedep1}, were introduced to realize the inhomogeneity of cells.

\begin{figure*}[h]
\centering
\includegraphics[scale=0.35]{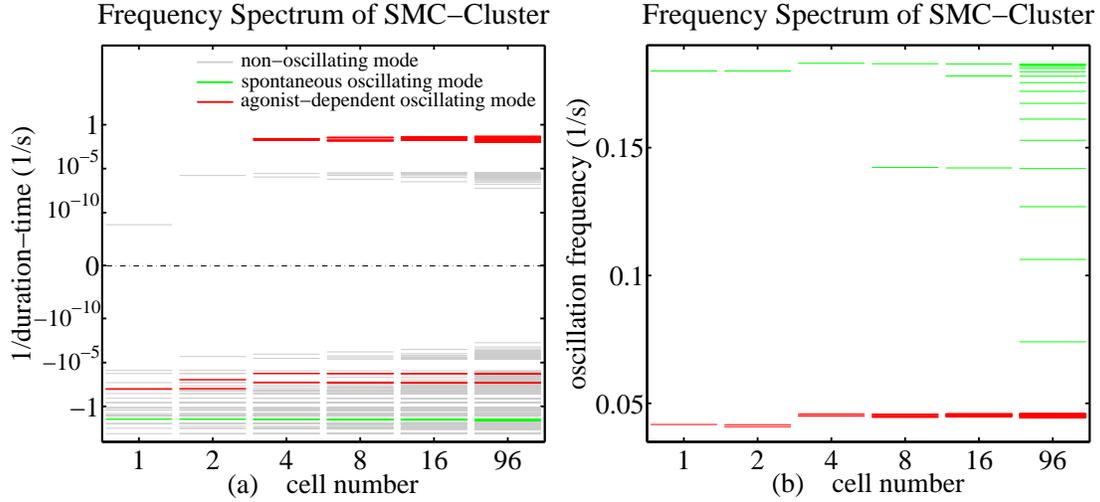}
\caption{ Frequency spectrum of homogeneous SMC clusters at different cell numbers: (a) real parts of eigenvalues related to growing or decaying time,
(b) imaginary parts of eigenvalues related to oscillation periods. }\label{fig15}
\end{figure*}

Figure (\ref{fig15}) shows the frequency spectrum for homogeneous 1D clusters at varying cell numbers. Every cell is set in the control condition ( except for $[K]_e=34.8mM$), and only neighbor cells establish gap junctions. As indicated by green (spontaneous oscillating mode) and red (agonist-dependent oscillating mode) lines in Fig. (\ref{fig15}), the oscillation level gradually evolves into a spread band along with increasing cell numbers. This fact infers that a broader-range timing or synchronization among cells is acceptable for rhythmical oscillations in longer clusters. Conversely, properties of red curves in Fig. (\ref{fig15}a) suggest that, due to including more interactions among cells, more transient growing states (positive frequency values) are excited, resulting in the prolongation of duration periods of vasomotions. With an input of delta-function calcium pulse in this resonance medium, we observe functionalities of SMCs' rhythmic oscillations by time-domain calculations.

\begin{figure*}[h]
\centering
\includegraphics[scale=0.35]{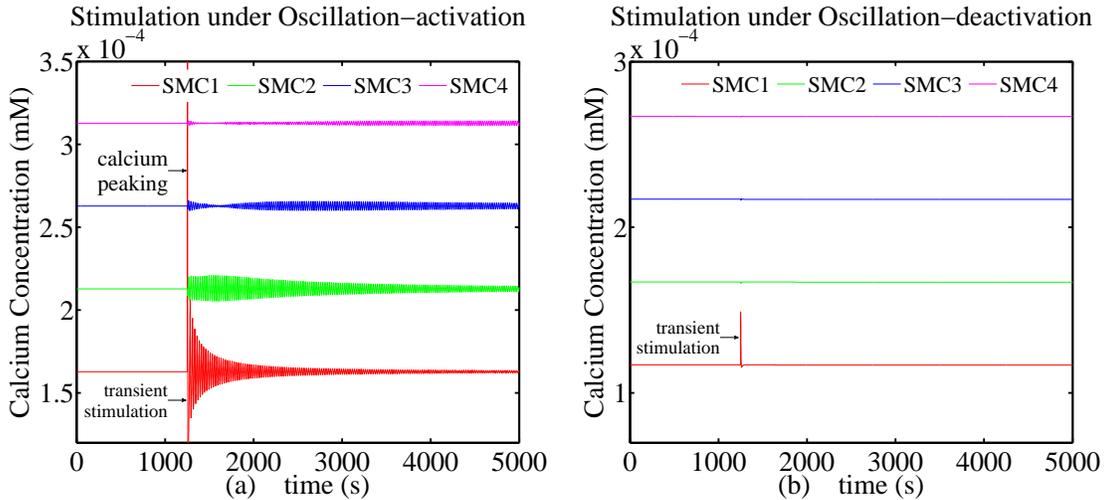}
\caption{ Calcium responses against delta-function calcium stimulation to SMC1 for homogeneous 6-SMCs:
(a) at $[K]_e=34.6$ (oscillation-activation, see region $\mathrm{\mathbf{II}}$ in Fig. \ref{fig4}), and (b) at $[K]_e=20.0$ (oscillation-deactivation, see region $\mathrm{\mathbf{I}}$ in Fig. \ref{fig4}). }\label{fig16}
\end{figure*}

As illustrated in Fig (\ref{fig16}a), with the same modeling parameters for Fig. (\ref{fig15}), the temporal changes of $[Ca^{2+}]_i$ of a 6-SMC cluster are evaluated. In this case, $[K]_e$ is reduced to be $34.6mM$ (oscillation-activation, see Fig. \ref{fig4}) so as to initially prepare in-equilibrium cells. Another 6-SMC cluster, with $[K]_e=20mM$ (oscillation-deactivation, see Fig. \ref{fig4}) is computed in Fig (\ref{fig16}b). Before calcium-pulse stimulation both 6-SMC clusters remain in equilibrium and are indistinguishable by observations. With the given stimulation (suddenly raising the calcium concentration on SMC1 by $35\%$), the cluster under the oscillation-activation first arouses significant calcium peaking by transient resonance in SMC1 \cite{spark1} (see Fig. \ref{fig16}a), and continuously brings calcium signalling toward other SMCs. Numerical results (not shown in the figure) indicate that calcium peaking arises from the activation of ryanodine receptors, which cause $[Ca^{2+}]_i$ to be released from the sarcoplasmic reticulum of cell $I$. For the cluster under oscillation-deactivation in Fig. (\ref{fig16}b), the stimulation from SMC1 dissipates fast and no signalling communications among cells occur, conforming with the observation in experiments of rat mesenteric arterioles \cite{wave1}.

\begin{figure*}[h]
\centering
\includegraphics[scale=0.35]{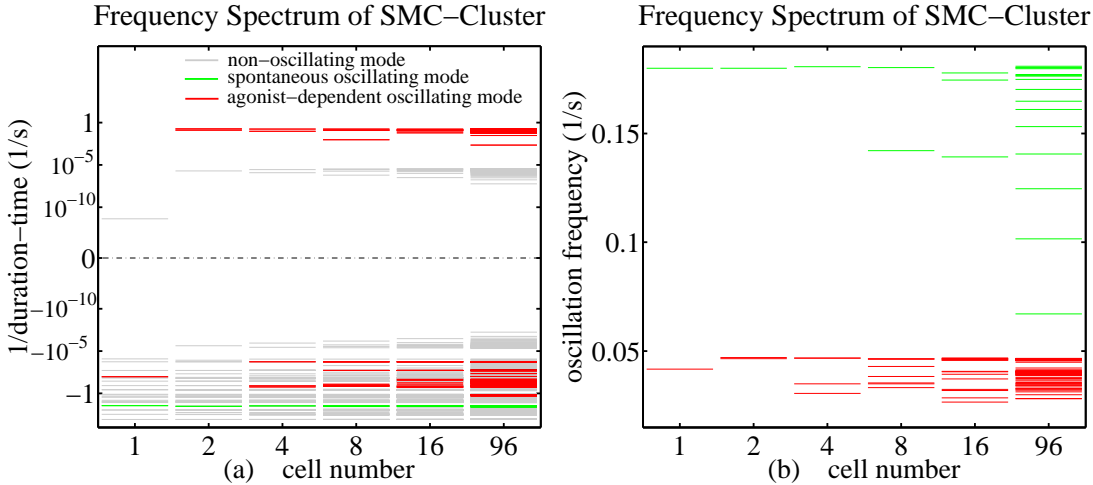}
\caption{ Frequency spectrum of inhomogeneous SMC clusters at different cell numbers: (a) real parts of eigenvalues related to growing or decaying time,
(b) imaginary parts of eigenvalues related to oscillation periods.}\label{fig17}
\end{figure*}

We next include the inhomogeneity of SMCS for biological complexity. Here, $15\%$ stochastic variations of SMC volumes \cite{sizedep1} are introduced. Figure (\ref{fig17}a) depicts similar properties on the agonist-dependent oscillating modes (red lines), although the growing mode seems to be more easily excited due to the fluctuation of cell volumes. It is noted that the spontaneous oscillating modes (green lines) remain relatively insensitive to cell number as well as cell uniformity. Figure (\ref{fig17}b) exhibits a broader but less dense level spectrum, versus Fig. (\ref{fig15}b). This fact could imply that the vasomotion in inhomogeneous clusters can decay faster than that in homogeneous ones due to less-overlapping oscillating levels.

\begin{figure*}[h]
\centering
\includegraphics[scale=0.35]{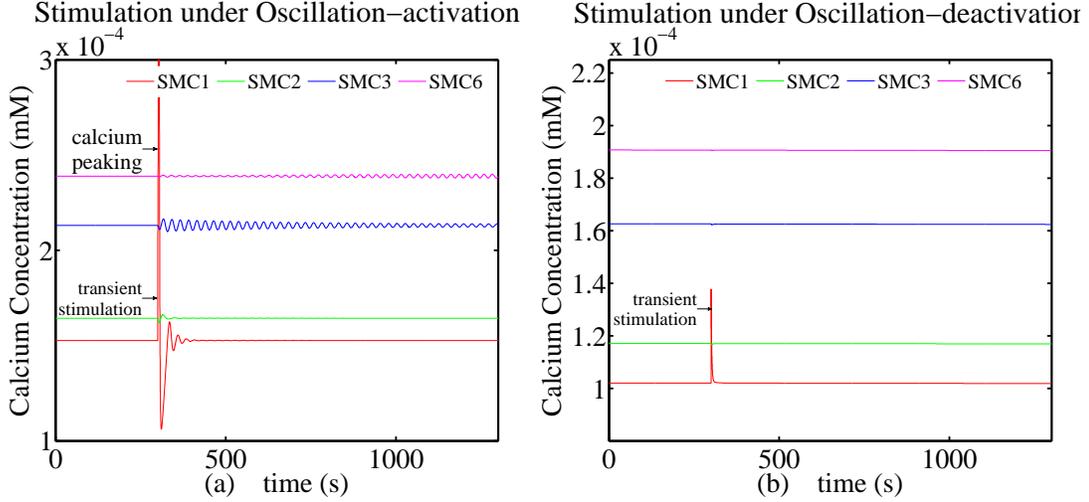}
\caption{ Calcium responses against delta-function calcium stimulation to SMC1 for homogeneous 6-SMCs:
(a) at $[K]_e=35.0$ (oscillation-activation, see region $\mathrm{\mathbf{II}}$ in Fig. \ref{fig4}), and (b) at $[K]_e=20.0$ (oscillation-deactivation, see region $\mathrm{\mathbf{I}}$ in Fig. \ref{fig4}). }\label{fig18}
\end{figure*}

We carry out time-domain analysis to study the influences of cell uniformity. Figure (\ref{fig18}a) presents one inhomogeneous 6-SMC cluster with oscillation-activation (region $\mathrm{\mathbf{II}}$ in Fig \ref{fig4}), while Fig. (\ref{fig18}b) helps analyze another inhomogeneous cluster with oscillation-deactivation (region $\mathrm{\mathbf{I}}$ in Fig \ref{fig4}). In the case of Fig. (\ref{fig18}a), $[K]_e$ is set to be $35.0mM$ in preparation for the initial in-equilibrium cells. In the case of Fig. (\ref{fig18}b), $[K]_e$ is reduced to be $20mM$ to achieve the oscillation-deactivation condition.
 For the oscillation-deactivation condition, the stimulation from SMC1 dissipates fast and no signal communications among cells occur.  For the oscillation-activation condition, we find that the signalling transference in inhomogeneous clusters decays relatively faster than that in homogeneous ones; otherwise, the signaling delivery among cells presents properties differing from the molecular diffusions and characterizes a frog-leap manner (from SMC1 to SMC3), depending on the specific inhomogeneity. Our calculation explain the observations in the literature\cite{ex3,ex5}.

\section{Conclusion}
We have developed herein a detailed biophysical algorithm that intuitionally investigate characteristics of rhythmicity and synchronization related calcium regulation in SMC. Implemented with frequency-domain and time-domain analyses for a single cell, this work recognizes the inherent properties of rhythmical calcium oscillations and validates the utilizations of the eigensystem formulation. In the case of finite SMC clusters, we study the influences of synchronization and resonance conditions, and look at functionalities of cell rhythmicity, calcium peaking, and calcium waves. Relevant calculations offer information underlying the present experimental observations found in the literature. In the future, accompanied by abundant pathological data, this approach could pave an alternate avenue toward physiological and pathological determinations.

\clearpage

\appendix

\section{Common variables} \label{appa}
\subsection{Standard parameter values and definitions} \label{appa1}
\begin{eqnarray*}
z_{K} &=&1,\text{ }z_{Na}=1,\text{ }z_{Ca}=2,\text{ } \\
z_{Cl} &=&-1,\text{ }N_{Av}=6.022\cdot 10^{23}, \\
R &=&8341.47mJ/molK,\text{ }F=96485.34C/mol, \\
T &=&293.0K,\text{ }C_{m}=25pF,\text{ }A_{m}=10^{-6}C_{m}\text{ }cm^{2}, \\
\left[ Ca\right] _{e} &=&2.0mM,\text{ }\left[ N_{a}\right] _{e}=140.0mM, \\
\left[ K\right] _{e} &=&5.0mM,\text{ }\left[ Cl\right] _{e}=129.0mM, \\
vol_{i} &=&1\text{ }p\ell ,\text{ }vol_{Ca}=0.7\text{ }p\ell ,\text{ } \\
vol_{SRu} &=&0.07\text{ }p\ell ,\text{ }vol_{SRr}=0.007\text{ }p\ell
\end{eqnarray*}

On the quasi-equilibrium condition, the ionic concentrations that we are interested
in can be expressed by a time-independent constant term plus a
time-dependent fluctuation:
\begin{eqnarray*}
\left[ Ca\right] _{i} &=&\left[ \overline{Ca}\right] _{i}+\left[ \delta _{Ca}%
\right] _{i},\text{ }\left[ Ca\right] _{r}=\left[ \overline{Ca}\right] _{r}+%
\left[ \delta _{Ca}\right] _{r}, \\
\left[ Ca\right] _{u} &=&\left[ \overline{Ca}\right] _{u}+\left[ \delta _{Ca}%
\right] _{u},\text{ }\left[ Na\right] _{i}=\left[ \overline{Na}\right] _{i}+%
\left[ \delta _{Na}\right] _{i}, \\
\left[ K\right] _{i} &=&\left[ \overline{K}\right] _{i}+\left[ \delta _{K}%
\right] _{i},\text{ }\left[ Cl\right] _{i}=\left[ \overline{Cl}\right] _{i}+%
\left[ \delta _{Cl}\right] _{i}, \\
V_{m} &=&\overline{V}_{m}+\delta _{V_{m}}
\end{eqnarray*}%
Herein, $\overline{X}$ represents the equilibrium average constant of $X$
and $\delta _{X}$ represents the deviation from the equilibrium. Similar
representations are adopted for variables involved in transient processes,
including $d_{L}$, $f_{L}$, $p_{f}$, $p_{s}$, $p_{K}$, $q_{1}$, $q_{2}$, $%
P_{SOC}$, $R_{10}$, $R_{11}$, $R_{01}$, $h_{IP3}$, $\left[ R_{G}^{S}\right] $%
, $\left[ R_{P,G}^{S}\right] $, $\left[ G\right] $, $\left[ IP_{3}\right] $,
$\left[ PIP_{2}\right] $, $V_{cGMP}$, and $\left[ cGMP\right] $, and they are
introduced in the relevant paragraphs below. Relevant mathematical equations
\cite{mathmodel1}, e.g. reversal potentials for ion $X$, can hence be
obtained by using the Taylor series expansion to the first order of fluctuation $%
\delta _{X}$ at equilibrium:
\begin{eqnarray}
E_{X} &\simeq &\frac{RT}{z_{X}F}\ln \left( \left[ X\right] _{e}/\left[
\overline{X}\right] _{i}\right) -\frac{RT}{z_{X}F}\frac{\left[ \delta _{X}%
\right] _{i}}{\left[ \overline{X}\right] _{i}}  \nonumber \\
&\equiv &\overline{E}_{X}+\delta E_{X}
\end{eqnarray}%
Herein, we take $X\in \left\{ Ca,Na,K,Cl\right\} $ for example.

\subsection{Initial values of variables} \label{appa2}

\begin{eqnarray*}
\left[ Ca\right] _{i} &=&68.0\times 10^{-6}mM \\
\left[ Ca\right] _{r} &=&0.57mM,\text{ \ }\left[ Ca\right] _{u}=0.66mM \\
\left[ Na\right] _{i} &=&8.4mM,\text{ \ }\left[ K\right] _{i}=140.0mM \\
\left[ Cl\right] _{i} &=&59.4mM,\text{ \ }V_{m}=-59.4mV \\
\lbrack cGMP] &=&0.0mM,\text{ \ }[IP_{3}]=0.0mM \\
C_{m} &=&25pF,\text{ \ }A_{m}=C_{m}\times 10^{-6}cm^{2}
\end{eqnarray*}%
\begin{eqnarray*}
\lbrack NO] &=&10^{-5}mM,\text{ \ constant} \\
\lbrack NE] &=&2\times 10^{-4}mM,\text{ \ constant} \\
d_{L} &=&d_{L0},\text{ \ }f_{L}=f_{L0}
\end{eqnarray*}%
\begin{eqnarray*}
R_{cGMP} &=&\frac{\left[ cGMP\right] ^{2}}{\left[ cGMP\right] ^{2}+\left(
0.55\cdot 10^{-3}\text{ }mM\right) ^{2}} \\
p_{f} &=&p_{s}=\overline{p}_{o} \\
p_{K} &=&p_{Ko},\text{ \ }q_{1}=q_{2}=q_{0}
\end{eqnarray*}%
\begin{eqnarray*}
p_{SOC} &=&0.0,\text{ \ }R_{10}=0.0033 \\
R_{11} &=&0.000004,\text{ \ }R_{01}=0.9955 \\
h_{IP3} &=&K_{inh,IP3}/\left( \left[ Ca\right] _{i}+K_{inh,IP3}\right)
\end{eqnarray*}%
\begin{eqnarray*}
\left[ R_{P,G}^{S}\right]  &=&0,\text{ \ }\left[ R_{G}^{S}\right] =\left[
R_{T,G}\right] \xi _{G} \\
\left[ PIP_{2}\right]  &=&\left[ PIP_{2,T}\right] -\left(
1+k_{deg,G}/r_{r,G}\right) \gamma _{G}[IP_{3}]
\end{eqnarray*}%
\begin{eqnarray*}
r_{h,G0} &=&k_{deg,G}\gamma _{G}[IP_{3}]/\left[ PIP_{2}\right]  \\
\left[ G\right]  &=&r_{h,G0}\left( K_{c,G}+\left[ Ca\right] _{i}\right)
/\left( \alpha _{G}\left[ Ca\right] _{i}\right)  \\
V_{cGMP} &=&0.0
\end{eqnarray*}

\bigskip

\section{ Mathematical model equations for membrane electrophysiology} \label{appb}

\subsection{L-type voltage-operated $C_{a}^{2+}$ channels} \label{appb1}

\begin{eqnarray}
I_{VOCC} &=&10^{6}A_{m}P_{VOCC}d_{L}f_{L}V_{m}\frac{z_{Ca}^{2}F^{2}}{RT}
\nonumber \\
&&\times \frac{\left[ Ca\right] _{e}-\left[ Ca\right] _{i}e^{V_{m}z_{Ca}F/%
\left( RT\right) }}{1-e^{V_{m}z_{Ca}F/(RT)}}\text{ \ }\mathtt{[pA]}
\nonumber \\
&\simeq &\overline{I}_{VOCC}+\Delta I_{VOCC}^{V_{m}}\delta _{V_{m}}+\Delta
I_{VOCC}^{Ca_{i}}\left[ \delta _{Ca}\right] _{i}+  \nonumber \\
&&\Delta I_{VOCC}^{d_{L}}\delta _{d_{L}}+\Delta I_{VOCC}^{f_{L}}\delta
_{f_{L}}
\end{eqnarray}%
\begin{eqnarray}
\frac{dd_{L}}{dt} &=&\frac{d_{L0}-d_{L}}{\tau _{d_{L}}}\simeq \frac{d\delta
_{d_{L}}}{dt}\equiv \Delta _{d_{L}}^{V_{m}}\delta _{V_{m}}+\Delta
_{d_{L}}^{d_{L}}\delta _{d_{L}} \\
\frac{df_{L}}{dt} &=&\frac{f_{L0}-f_{L}}{\tau _{f_{L}}}\simeq \frac{d\delta
_{f_{L}}}{dt}\equiv \Delta _{f_{L}}^{V_{m}}\delta _{V_{m}}+\Delta
_{f_{L}}^{f_{L}}\delta _{f_{L}}
\end{eqnarray}%
on quasi-equilibrium conditions (i.e. $dd_{L}/dt=d\overline{d}%
_{L}/dt+d\delta _{d_{L}}/dt\simeq d\delta _{d_{L}}/dt$; $df_{L}/dt=d%
\overline{f}_{L}/dt+d\delta _{f_{L}}/dt\simeq d\delta _{f_{L}}/dt$) with
relevant variables%
\begin{eqnarray}
P_{VOCC} &=&1.88\cdot 10^{-5}\text{ }cm/s  \nonumber \\
d_{L0} &=&\left[ 1+e^{-V_{m}/8.3\text{ }mV}\right] ^{-1} \\
f_{L0} &=&\left[ 1+e^{(V_{m}+42.0\text{ }mV)/9.1\text{ }mV}\right] ^{-1} \\
\tau _{d_{L}} &=&2.5e^{-(V_{m}+40\text{ }mV)^{2}/(30\text{ }mV)^{2}}+1.15%
\text{ }ms \\
\tau _{f_{L}} &=&65e^{-(V_{m}+35\text{ }mV)^{2}/(25\text{ }mV)^{2}}+45\text{
}ms
\end{eqnarray}%
and the fluctuation terms
\begin{eqnarray}
\Delta I_{VOCC}^{V_{m}} &=&10^{6}\frac{A_{m}P_{VOCC}z_{Ca}^{2}F^{2}}{RT}%
\cdot \lbrack \frac{\alpha _{1}}{\beta _{1}}+  \nonumber \\
&&\frac{\left( \left[ Ca\right] _{e}-\left[ \overline{Ca}\right] _{i}\right)
\gamma _{1}\overline{V}_{m}z_{Ca}F}{\beta _{1}\left[ 1-\gamma _{1}\right] RT}%
] \\
\Delta I_{VOCC}^{Ca_{i}} &=&-10^{6}\frac{A_{m}P_{VOCC}z_{Ca}^{2}F^{2}}{RT}%
\frac{\overline{V}_{m}\gamma _{1}}{\beta _{1}}
\end{eqnarray}%
\begin{eqnarray}
\Delta I_{VOCC}^{d_{L}} &=&10^{6}\frac{A_{m}P_{VOCC}z_{Ca}^{2}F^{2}}{RT}%
\frac{\alpha _{1}\overline{V}_{m}}{\beta _{1}}  \nonumber \\
&&\times \left[ 1+e^{-\overline{V}_{m}/8.3}\right] \\
\Delta I_{VOCC}^{f_{L}} &=&10^{6}\frac{A_{m}P_{VOCC}z_{Ca}^{2}F^{2}}{RT}%
\frac{\alpha _{1}\overline{V}_{m}}{\beta _{1}}  \nonumber \\
&&\times \left[ 1+e^{(\overline{V}_{m}+42)/9.1}\right]
\end{eqnarray}%
\begin{eqnarray}
\Delta _{d_{L}}^{V_{m}} &=&\frac{e^{-\overline{V}_{m}/8.3}}{8.3\eta
_{1}\left( 1+e^{-\overline{V}_{m}/8.3}\right) ^{2}} \\
\Delta _{d_{L}}^{d_{L}} &=&-\frac{1}{\eta _{1}} \\
\Delta _{f_{L}}^{V_{m}} &=&-\frac{e^{(\overline{V}_{m}+42)/9.1}}{9.1\kappa
_{1}\left[ 1+e^{(\overline{V}_{m}+42)/9.1}\right] ^{2}} \\
\Delta _{f_{L}}^{f_{L}} &=&-\frac{1}{\kappa _{1}}
\end{eqnarray}%
Herein,%
\begin{eqnarray}
\alpha _{1} &=&\left[ Ca\right] _{e}-\left[ \overline{Ca}\right] _{i}e^{%
\overline{V}_{m}z_{Ca}F/(RT)} \\
\beta _{1} &=&\left[ 1+e^{-\overline{V}_{m}/8.3}\right] \left[ 1+e^{(%
\overline{V}_{m}+42)/9.1}\right]  \nonumber \\
&&\times \left[ 1-e^{\overline{V}_{m}z_{Ca}F/(RT)}\right]
\end{eqnarray}%
\begin{eqnarray}
\gamma _{1} &=&e^{\overline{V}_{m}z_{Ca}F/(RT)} \\
\eta _{1} &=&2.5e^{-(\overline{V}_{m}+40)^{2}/(30)^{2}}+1.15\text{ } \\
\kappa _{1} &=&65e^{-(\overline{V}_{m}+35)^{2}/(25)^{2}}+45
\end{eqnarray}

\bigskip

\subsection{Large conductance $C_{a}^{2+}$-activated $K^{+}$ channels} \label{appb2}

\begin{eqnarray}
I_{BKCa} &=&A_{m}N_{BKCa}P_{KCa}i_{KCa}\text{ \ }\mathtt{[pA]}  \nonumber \\
&\simeq &\overline{I}_{BKCa}+\Delta I_{BKCa}^{V_{m}}\delta _{V_{m}}+\Delta
I_{BKCa}^{K_{i}}\left[ \delta _{K}\right] _{i}+  \nonumber \\
&&\Delta I_{BKCa}^{p_{f}}\delta _{p_{f}}+\Delta I_{BKCa}^{p_{s}}\delta
_{p_{s}}
\end{eqnarray}%
\begin{eqnarray}
\frac{dp_{f}}{dt} &=&\frac{\overline{p}_{o}-p_{f}}{\tau _{p_{f}}}\simeq
\frac{d\delta _{p_{f}}}{dt}\equiv \Delta _{p_{f}}^{cGMP}\left[ \delta _{cGMP}%
\right]  \nonumber \\
&&+\Delta _{p_{f}}^{V_{m}}\delta _{V_{m}}+\Delta _{p_{f}}^{p_{f}}\delta
_{p_{f}}+\Delta _{p_{f}}^{Cai}\left[ \delta _{Ca}\right] _{i} \\
\frac{dp_{s}}{dt} &=&\frac{\overline{p}_{o}-p_{s}}{\tau _{p_{s}}}\simeq
\frac{d\delta _{p_{s}}}{dt}\equiv \Delta _{p_{s}}^{cGMP}\left[ \delta _{cGMP}%
\right]  \nonumber \\
&&+\Delta _{p_{s}}^{V_{m}}\delta _{V_{m}}+\Delta _{p_{s}}^{p_{s}}\delta
_{p_{s}}+\Delta _{p_{s}}^{Cai}\left[ \delta _{Ca}\right] _{i}
\end{eqnarray}%
on quasi-equilibrium conditions (i.e. $dp_{f}/dt=d\overline{p}%
_{f}/dt+d\delta _{p_{f}}/dt\simeq d\delta _{p_{f}}/dt$; $dp_{s}/dt=d%
\overline{p}_{s}/dt+d\delta _{p_{s}}/dt\simeq d\delta _{p_{s}}/dt$) with
relevant variables%
\begin{eqnarray*}
N_{BKCa} &=&6.6\cdot 10^{6}\text{ }1/cm^{2}\text{ } \\
P_{BKCa} &=&3.9\cdot 10^{-13}\text{ }cm^{3}/s \\
\tau _{p_{f}} &=&0.84\text{ }ms,\text{ }\tau _{p_{s}}=35.9\text{ }ms \\
dV_{1/2KCaNO} &=&46.3\text{ }mV,\text{ }dV_{1/2KCacGMP}=76\text{ }mV
\end{eqnarray*}%
\begin{eqnarray}
\overline{p}_{o} &=&\left[ 1+e^{-\frac{V_{m}-V_{1/2KCa}}{18.25\text{ }mV}}%
\right] ^{-1} \\
P_{KCa} &=&0.17p_{f}+0.83p_{s} \\
V_{1/2KCa} &=&-41.7\log _{10}\left( \left[ Ca\right] _{i}\right)
-dV_{1/2KCaNO}R_{NO}  \nonumber \\
&&-dV_{1/2KCacGMP}R_{cGMP}-128.2
\end{eqnarray}%
\begin{eqnarray}
R_{NO} &=&\frac{\left[ NO\right] }{\left[ NO\right] +2\cdot 10^{-4}\text{ }mM%
} \\
R_{cGMP} &=&\frac{\left[ cGMP\right] ^{2}}{\left[ cGMP\right] ^{2}+\left(
1.5\cdot 10^{-3}\text{ }mM\right) ^{2}} \\
i_{KCa} &=&10^{6}P_{BKCa}V_{m}\frac{F^{2}}{RT} \\
&&\times \frac{\left[ K\right] _{e}-\left[ K\right] _{i}e^{V_{m}F/(RT)}}{%
1-e^{V_{m}F/(RT)}}  \nonumber
\end{eqnarray}%
and the fluctuation terms%
\begin{eqnarray}
\Delta I_{BKCa}^{V_{m}} &=&10^{6}\frac{A_{m}N_{BKCa}P_{BKCa}F^{2}}{RT}\cdot
\lbrack  \nonumber \\
&&\frac{\gamma _{2}}{\left( 1-\alpha _{2}\right) \left( 1+\beta _{2}\right) }
\nonumber \\
&&+\frac{\alpha _{2}F\overline{V}_{m}\left( \left[ K\right] _{e}-\left[
\overline{K}\right] _{i}\right) }{\left( \alpha _{2}-1\right) ^{2}\left(
1+\beta _{2}\right) RT}] \\
\Delta I_{BKCa}^{K_{i}} &=&10^{6}\frac{A_{m}N_{BKCa}P_{BKCa}F^{2}}{RT}
\nonumber \\
&&\times \frac{\alpha _{2}\overline{V}_{m}}{\left( \alpha _{2}-1\right)
\left( 1+\beta _{2}\right) }
\end{eqnarray}%
\begin{eqnarray}
\Delta I_{BKCa}^{p_{f}} &=&1.7\cdot 10^{5}\frac{A_{m}N_{BKCa}P_{BKCa}F^{2}}{%
RT}\frac{\gamma _{2}\overline{V}_{m}}{1-\alpha _{2}} \\
\Delta I_{BKCa}^{p_{s}} &=&8.3\cdot 10^{5}\frac{A_{m}N_{BKCa}P_{BKCa}F^{2}}{%
RT}\frac{\gamma _{2}\overline{V}_{m}}{1-\alpha _{2}}
\end{eqnarray}%
\begin{eqnarray}
\Delta _{p_{f}}^{V_{m}} &=&\frac{4}{73}\frac{\beta _{2}}{\left( 1+\beta
_{2}\right) ^{2}\tau _{p_{f}}} \\
\Delta _{p_{f}}^{p_{f}} &=&\frac{-1}{\tau _{p_{f}}} \\
\Delta _{p_{f}}^{Cai} &=&\frac{834}{365\ln (10)}\frac{\beta _{2}}{\left(
1+\beta _{2}\right) ^{2}\tau _{p_{f}}\left[ \overline{Ca}\right] _{i}}
\end{eqnarray}%
\begin{eqnarray}
\Delta _{p_{f}}^{cGMP} &=&\frac{9}{3.65\cdot 10^{7}}\frac{\beta _{2}}{\left(
1+\beta _{2}\right) ^{2}\tau _{p_{f}}}\times  \nonumber \\
&&\frac{dV_{1/2KCacGMP}\left[ \overline{cGMP}\right] }{\left( \left[
\overline{cGMP}\right] ^{2}+2.25\cdot 10^{-6}\right) ^{2}}
\end{eqnarray}%
\begin{eqnarray}
\Delta _{p_{s}}^{V_{m}} &=&\frac{4}{73}\frac{\beta _{2}}{\left( 1+\beta
_{2}\right) ^{2}\tau _{p_{s}}} \\
\Delta _{p_{s}}^{p_{s}} &=&\frac{-1}{\tau _{p_{s}}} \\
\Delta _{p_{s}}^{Cai} &=&\frac{834}{365\ln (10)}\frac{\beta _{2}}{\left(
1+\beta _{2}\right) ^{2}\tau _{p_{s}}\left[ \overline{Ca}\right] _{i}}
\end{eqnarray}%
\begin{eqnarray}
\Delta _{p_{s}}^{cGMP} &=&\frac{9}{3.65\cdot 10^{7}}\frac{\beta _{2}}{\left(
1+\beta _{2}\right) ^{2}\tau _{p_{s}}}\times  \nonumber \\
&&\frac{dV_{1/2KCacGMP}\left[ \overline{cGMP}\right] }{\left( \left[
\overline{cGMP}\right] ^{2}+2.25\cdot 10^{-6}\right) ^{2}}
\end{eqnarray}%
Herein,%
\begin{eqnarray}
\alpha _{2} &=&e^{\overline{V}_{m}F/(RT)} \\
\beta _{2} &=&\exp \left[ -\frac{4}{73}\overline{V}_{m}-\frac{166.8}{73}\log
_{10}\left( \left[ \overline{Ca}\right] _{i}\right) \right]  \nonumber \\
&&\times \exp \left[ -\frac{4}{73}\left( 128.2+dV_{1/2KCaNO}R_{NO}\right) %
\right]  \nonumber \\
&&\times \exp \left[ -\frac{4}{73}dV_{1/2KCacGMP}\cdot \kappa _{2}\right]
\end{eqnarray}%
\begin{eqnarray}
\kappa _{2} &=&\frac{\left[ \overline{cGMP}\right] ^{2}}{\left[ \overline{%
cGMP}\right] ^{2}+\left( 1.5\cdot 10^{-3}\text{ }mM\right) ^{2}} \\
\gamma _{2} &=&\left[ K\right] _{e}-\left[ \overline{K}\right] _{i}e^{%
\overline{V}_{m}F/(RT)}
\end{eqnarray}

\bigskip

\subsection{Voltage-dependent $K^{+}$ channels} \label{appb3}

\begin{eqnarray}
I_{Kv} &=&g_{Kv}p_{K}\left( 0.45q_{1}+0.55q_{2}\right) \left(
V_{m}-E_{K}\right) \text{ \ }\mathtt{[pA]}  \nonumber \\
&\simeq &\overline{I}_{Kv}+\Delta I_{Kv}^{V_{m}}\delta _{V_{m}}+\Delta
I_{Kv}^{K_{i}}\left[ \delta _{K}\right] _{i}+  \nonumber \\
&&\Delta I_{Kv}^{p_{K}}\delta _{p_{K}}+\Delta I_{Kv}^{q_{1}}\delta
_{q_{1}}+\Delta I_{Kv}^{q_{2}}\delta _{q_{2}}
\end{eqnarray}%
\begin{eqnarray}
\frac{dp_{K}}{dt} &=&\frac{p_{ko}-p_{K}}{\tau _{p_{K}}}\simeq \frac{d\delta
_{p_{K}}}{dt}\equiv \Delta _{p_{K}}^{V_{m}}\delta _{V_{m}}+\Delta
_{p_{K}}^{p_{K}}\delta _{p_{K}} \\
\frac{dq_{1}}{dt} &=&\frac{q_{o}-q_{1}}{\tau _{q_{1}}}\simeq \frac{d\delta
_{q_{1}}}{dt}\equiv \Delta _{q_{1}}^{V_{m}}\delta _{V_{m}}+\Delta
_{q_{1}}^{q_{1}}\delta _{q_{1}} \\
\frac{dq_{2}}{dt} &=&\frac{q_{o}-q_{2}}{\tau _{q_{2}}}\simeq \frac{d\delta
_{q_{2}}}{dt}\equiv \Delta _{q_{2}}^{V_{m}}\delta _{V_{m}}+\Delta
_{q_{2}}^{q_{2}}\delta _{q_{2}}
\end{eqnarray}%
on quasi-equilibrium conditions with relevant variables%
\begin{eqnarray}
g_{Kv} &=&1.35\text{ }nS  \nonumber \\
\tau _{q_{1}} &=&371.0\text{ }ms,\text{ }\tau _{q_{2}}=2884.0\text{ }ms
\nonumber \\
\tau _{p_{K}} &=&61.5e^{-0.027V_{m}}\text{ }ms  \nonumber \\
p_{Ko} &=&\left[ 1+e^{-\frac{V_{m}+11\text{ }mV}{15\text{ }mV}}\right] ^{-1}
\\
q_{o} &=&\left[ 1+e^{\frac{V_{m}+40\text{ }mV}{14\text{ }mV}}\right] ^{-1}
\end{eqnarray}%
and the fluctuation terms%
\begin{eqnarray}
\Delta I_{Kv}^{V_{m}} &=&\frac{g_{Kv}}{\left( 1+\alpha _{3}\right) \left(
1+\beta _{3}\right) } \\
\Delta I_{Kv}^{K_{i}} &=&\frac{g_{Kv}RT}{\left[ \overline{K}\right]
_{i}z_{K}F\left( 1+\alpha _{3}\right) \left( 1+\beta _{3}\right) } \\
\Delta I_{Kv}^{p_{K}} &=&\frac{g_{Kv}\left( \overline{V}_{m}-\overline{E}%
_{K}\right) }{\left( 1+\beta _{3}\right) }
\end{eqnarray}%
\begin{eqnarray}
\Delta I_{Kv}^{q_{1}} &=&\frac{9g_{Kv}\left( \overline{V}_{m}-\overline{E}%
_{K}\right) }{20\left( 1+\alpha _{3}\right) } \\
\Delta I_{Kv}^{q_{2}} &=&\frac{11g_{Kv}\left( \overline{V}_{m}-\overline{E}%
_{K}\right) }{20\left( 1+\alpha _{3}\right) }
\end{eqnarray}%
\begin{eqnarray}
\Delta _{p_{K}}^{V_{m}} &=&\frac{2e^{-\frac{11}{15}-\frac{119}{3000}%
\overline{V}_{m}}}{1845\left( 1+\alpha _{3}\right) ^{2}},\text{ }\Delta
_{p_{K}}^{p_{K}}=-\frac{2e^{0.027\overline{V}_{m}}}{123} \\
\Delta _{q_{1}}^{V_{m}} &=&\frac{-\beta _{3}}{14\tau _{q_{1}}\left( 1+\beta
_{3}\right) ^{2}},\text{ }\Delta _{q_{1}}^{q_{1}}=\frac{-1}{\tau _{q_{1}}} \\
\Delta _{q_{2}}^{V_{m}} &=&\frac{-\beta _{3}}{14\tau _{q_{2}}\left( 1+\beta
_{3}\right) ^{2}},\text{ }\Delta _{q_{2}}^{q_{2}}=\frac{-1}{\tau _{q_{2}}}
\end{eqnarray}%
Herein,%
\begin{equation}
\alpha _{3}=e^{\frac{-1}{15}\left( \overline{V}_{m}+11\right) },\text{ }%
\beta _{3}=e^{\frac{1}{14}\left( \overline{V}_{m}+40\right) }
\end{equation}

\bigskip

\subsection{Unspecified $K^{+}$ leak channels} \label{appb4}

\begin{eqnarray}
I_{Kleak} &=&g_{Kleak}\left( V_{m}-E_{K}\right) \text{ \ }\mathtt{[pA]}
\nonumber \\
&\simeq &\overline{I}_{Kleak}+\Delta I_{Kleak}^{V_{m}}\delta _{V_{m}}+\Delta
I_{Kleak}^{K_{i}}\left[ \delta _{K}\right] _{i}
\end{eqnarray}%
on quasi-equilibrium conditions with relevant variables%
\[
g_{Kleak}=0.067\text{ }nS
\]%
and the fluctuation terms%
\begin{eqnarray}
\Delta I_{Kleak}^{V_{m}} &=&g_{Kleak} \\
\Delta I_{Kleak}^{K_{i}} &=&\frac{g_{Kleak}RT}{\left[ \overline{K}\right]
_{i}z_{K}F}
\end{eqnarray}

\bigskip

\subsection{Non-selective cation channels} \label{appb5}

\begin{eqnarray}
I_{CaNSC} &=&10^{6}A_{m}d_{NSC}P_{oNSC}P_{CaNSC}V_{m}\frac{z_{Ca}^{2}F^{2}}{%
RT}  \nonumber \\
&&\times \frac{\left[ Ca\right] _{e}-\left[ Ca\right] _{i}e^{V_{m}z_{Ca}F/%
\left( RT\right) }}{1-e^{V_{m}z_{Ca}F/(RT)}}  \nonumber \\
&\simeq &\overline{I}_{CaNSC}+\Delta I_{CaNSC}^{V_{m}}\delta _{V_{m}}
\nonumber \\
&&+\Delta I_{CaNSC}^{Ca_{i}}\left[ \delta _{Ca}\right] _{i}
\end{eqnarray}%
\begin{eqnarray}
I_{NaNSC} &=&10^{6}A_{m}\left[ \frac{\left[ DAG\right] }{\left[ DAG\right]
+K_{NSC}}+d_{NSC}\right] V_{m}  \nonumber \\
&&\frac{P_{oNSC}P_{NaNSC}F^{2}}{RT}\frac{\left[ Na\right] _{e}-\left[ Na%
\right] _{i}e^{V_{m}F/\left( RT\right) }}{1-e^{V_{m}F/(RT)}}  \nonumber \\
&\simeq &\overline{I}_{NaNSC}+\Delta I_{NaNSC}^{V_{m}}\delta _{V_{m}}
\nonumber \\
&&+\Delta I_{NaNSC}^{Na_{i}}\left[ \delta _{Na}\right] _{i}
\end{eqnarray}%
\begin{eqnarray}
I_{KNSC} &=&10^{6}A_{m}\left[ \frac{\left[ DAG\right] }{\left[ DAG\right]
+K_{NSC}}+d_{NSC}\right] V_{m}  \nonumber \\
&&\frac{P_{oNSC}P_{KNSC}F^{2}}{RT}\frac{\left[ K\right] _{e}-\left[ K\right]
_{i}e^{V_{m}F/\left( RT\right) }}{1-e^{V_{m}F/(RT)}}  \nonumber \\
&\simeq &\overline{I}_{KNSC}+\Delta I_{KNSC}^{V_{m}}\delta _{V_{m}}
\nonumber \\
&&+\Delta I_{KNSC}^{K_{i}}\left[ \delta _{K}\right] _{i}
\end{eqnarray}%
on quasi-equilibrium conditions with relevant variables%
\begin{eqnarray}
d_{NSC} &=&0.0244,\text{ }K_{NSC}=3\cdot 10^{-3}\text{ }mM,  \nonumber \\
P_{NaNSC} &=&5.11\cdot 10^{-7}\text{ }cm/s,  \nonumber \\
\lbrack DAG] &=&[\overline{IP}_{3}] \\
P_{KNSC} &=&1.06P_{NaNSC},P_{CaNSC}=4.54P_{NaNSC}  \nonumber \\
P_{oNSC} &=&\frac{0.57}{1+e^{-\frac{V_{m}-47.12\text{ }mV}{24.24\text{ }mV}}}%
+0.43
\end{eqnarray}%
and the fluctuation terms%
\begin{eqnarray}
\Delta I_{CaNSC}^{V_{m}} &=&\frac{0.43\eta _{5}\beta _{5}P_{CaNSC}}{1-\gamma
_{5}^{z_{Ca}}}+  \nonumber \\
&&\frac{\left( \left[ Ca\right] _{e}-\left[ \overline{Ca}\right] _{i}\right)
\gamma _{5}^{z_{Ca}}P_{CaNSC}\beta _{5}F\overline{V}_{m}z_{Ca}}{100\left(
1-\gamma _{5}^{z_{Ca}}\right) ^{2}RT/43}  \nonumber \\
&&+\frac{0.57\eta _{5}\beta _{5}P_{CaNSC}}{\left( 1-\gamma
_{5}^{z_{Ca}}\right) \left( 1+\zeta _{5}\right) }+  \nonumber \\
&&\frac{\left( \left[ Ca\right] _{e}-\left[ \overline{Ca}\right] _{i}\right)
\gamma _{5}^{z_{Ca}}P_{CaNSC}\beta _{5}F\overline{V}_{m}z_{Ca}}{100\left(
1-\gamma _{5}^{z_{Ca}}\right) ^{2}\left( 1+\zeta _{5}\right) RT/57}
\nonumber \\
&&+\frac{19\zeta _{5}\beta _{5}P_{CaNSC}\overline{V}_{m}\eta _{5}}{808\left(
1+\zeta _{5}\right) ^{2}\left( 1-\gamma _{5}^{z_{Ca}}\right) }
\end{eqnarray}%
\begin{equation}
\Delta I_{CaNSC}^{Ca_{i}}=\frac{\gamma _{5}^{z_{Ca}}P_{CaNSC}\beta _{5}%
\overline{V}_{m}}{\gamma _{5}^{z_{Ca}}-1}\left( 0.43+\frac{0.57}{1+\zeta _{5}%
}\right)
\end{equation}%
\begin{eqnarray}
\Delta I_{NaNSC}^{V_{m}} &=&\frac{0.43\lambda _{5}\alpha _{5}P_{NaNSC}}{%
1-\gamma _{5}}+  \nonumber \\
&&\frac{\left( \left[ Na\right] _{e}-\left[ \overline{Na}\right] _{i}\right)
\gamma _{5}P_{NaNSC}\alpha _{5}F\overline{V}_{m}}{100\left( 1-\gamma
_{5}\right) ^{2}RT/43}  \nonumber \\
&&+\frac{0.57\lambda _{5}\alpha _{5}P_{NaNSC}}{\left( 1-\gamma _{5}\right)
\left( 1+\zeta _{5}\right) }+  \nonumber \\
&&\frac{\left( \left[ Na\right] _{e}-\left[ \overline{Na}\right] _{i}\right)
\gamma _{5}P_{NaNSC}\alpha _{5}F\overline{V}_{m}}{100\left( 1-\gamma
_{5}\right) ^{2}\left( 1+\zeta _{5}\right) RT/57}  \nonumber \\
&&+\frac{19\zeta _{5}\alpha _{5}P_{NaNSC}\overline{V}_{m}\lambda _{5}}{%
808\left( 1+\zeta _{5}\right) ^{2}\left( 1-\gamma _{5}\right) }
\end{eqnarray}%
\begin{equation}
\Delta I_{NaNSC}^{Na_{i}}=\frac{\gamma _{5}P_{NaNSC}\alpha _{5}\overline{V}%
_{m}}{\gamma _{5}-1}\left( 0.43+\frac{0.57}{1+\zeta _{5}}\right)
\end{equation}%
\begin{eqnarray}
\Delta I_{KNSC}^{V_{m}} &=&\frac{0.43\kappa _{5}\alpha _{5}P_{KNSC}}{%
1-\gamma _{5}}+  \nonumber \\
&&\frac{\left( \left[ K\right] _{e}-\left[ \overline{K}\right] _{i}\right)
\gamma _{5}P_{KNSC}\alpha _{5}F\overline{V}_{m}}{100\left( 1-\gamma
_{5}\right) ^{2}RT/43}  \nonumber \\
&&+\frac{0.57\kappa _{5}\alpha _{5}P_{KNSC}}{\left( 1-\gamma _{5}\right)
\left( 1+\zeta _{5}\right) }+  \nonumber \\
&&\frac{\left( \left[ K\right] _{e}-\left[ \overline{K}\right] _{i}\right)
\gamma _{5}P_{KNSC}\alpha _{5}F\overline{V}_{m}}{100\left( 1-\gamma
_{5}\right) ^{2}\left( 1+\zeta _{5}\right) RT/57}  \nonumber \\
&&+\frac{19\zeta _{5}\alpha _{5}P_{KNSC}\overline{V}_{m}\kappa _{5}}{%
808\left( 1+\zeta _{5}\right) ^{2}\left( 1-\gamma _{5}\right) }
\end{eqnarray}%
\begin{equation}
\Delta I_{KNSC}^{K_{i}}=\frac{\gamma _{5}P_{KNSC}\alpha _{5}\overline{V}_{m}%
}{\gamma _{5}-1}\left( 0.43+\frac{0.57}{1+\zeta _{5}}\right)
\end{equation}%
Herein,%
\begin{eqnarray}
\alpha _{5} &=&\frac{10^{6}A_{m}F^{2}}{RT}\left[ \frac{\left[ DAG\right] }{%
\left[ DAG\right] +K_{NSC}}+d_{NSC}\right] \\
\beta _{5} &=&10^{6}A_{m}d_{NSC}\frac{z_{Ca}^{2}F^{2}}{RT} \\
\gamma _{5} &=&e^{\frac{\overline{V}_{m}F}{RT}},\text{ }\zeta _{5}=e^{-\frac{%
\overline{V}_{m}-47.12}{24.24}} \\
\eta _{5} &=&\left[ Ca\right] _{e}-\left[ \overline{Ca}\right] _{i}e^{%
\overline{V}_{m}z_{Ca}F/\left( RT\right) } \\
\lambda _{5} &=&\left[ Na\right] _{e}-\left[ \overline{Na}\right] _{i}e^{%
\overline{V}_{m}F/\left( RT\right) } \\
\kappa _{5} &=&\left[ K\right] _{e}-\left[ \overline{K}\right] _{i}e^{%
\overline{V}_{m}F/\left( RT\right) }
\end{eqnarray}

\bigskip

\subsection{Store-operated non-selective cation channels} \label{appb6}

\begin{eqnarray}
I_{SOCCa} &=&g_{SOCCa}P_{SOC}\left( V_{m}-E_{Ca}\right)  \nonumber \\
&\simeq &\overline{I}_{SOCCa}+\Delta I_{SOCCa}^{P_{SOC}}\delta _{P_{SOC}}+
\nonumber \\
&&\Delta I_{SOCCa}^{V_{m}}\delta _{V_{m}}+\Delta I_{SOCCa}^{Ca_{i}}\left[
\delta _{Ca}\right] _{i} \\
I_{SOCNa} &=&g_{SOCNa}P_{SOC}\left( V_{m}-E_{Na}\right)  \nonumber \\
&\simeq &\overline{I}_{SOCNa}+\Delta I_{SOCNa}^{P_{SOC}}\delta _{P_{SOC}}+
\nonumber \\
&&\Delta I_{SOCNa}^{V_{m}}\delta _{V_{m}}+\Delta I_{SOCNa}^{Na_{i}}\left[
\delta _{Na}\right] _{i}
\end{eqnarray}%
\begin{eqnarray}
\frac{dp_{SOC}}{dt} &=&\frac{p_{SOC,o}-p_{SOC}}{\tau _{SOC}}\simeq \frac{%
d\delta _{p_{SOC}}}{dt}  \nonumber \\
&\equiv &\Delta _{p_{SOC}}^{Ca_{u}}\delta _{Ca_{u}}+\Delta
_{p_{SOC}}^{p_{SOC}}\delta _{p_{SOC}}
\end{eqnarray}%
on quasi-equilibrium conditions with relevant variables%
\begin{eqnarray}
g_{SOCCa} &=&0.0083\text{ }nS,\text{ }g_{SOCNa}=0.0575\text{ }nS  \nonumber
\\
\tau _{SOC} &=&100.0\text{ }ms  \nonumber \\
P_{SOC,o} &=&\left( 1+\frac{\left[ Ca\right] _{u}}{10^{-4}\text{ }mM}\right)
^{-1}
\end{eqnarray}%
and the fluctuation terms%
\begin{eqnarray}
\Delta I_{SOCCa}^{V_{m}} &=&\frac{g_{SOCCa}}{1+10^{4}\left[ \overline{Ca}%
\right] _{u}} \\
\Delta I_{SOCCa}^{Ca_{i}} &=&\frac{g_{SOCCa}RT}{\left[ \overline{Ca}\right]
_{i}\left( 1+10^{4}\left[ \overline{Ca}\right] _{u}\right) Fz_{Ca}} \\
\Delta I_{SOCCa}^{P_{SOC}} &=&g_{SOCCa}\left( \overline{V}_{m}-\overline{E}%
_{Ca}\right) \\
\Delta I_{SOCNa}^{V_{m}} &=&\frac{g_{SOCNa}}{1+10^{4}\left[ \overline{Ca}%
\right] _{u}} \\
\Delta I_{SOCNa}^{Na_{i}} &=&\frac{g_{SOCNa}RT}{\left[ \overline{Na}\right]
_{i}\left( 1+10^{4}\left[ \overline{Ca}\right] _{u}\right) F} \\
\Delta I_{SOCNa}^{P_{SOC}} &=&g_{SOCNa}\left( \overline{V}_{m}-\overline{E}%
_{Na}\right)
\end{eqnarray}%
\begin{equation}
\Delta _{p_{SOC}}^{Ca_{u}}=\frac{-10^{4}\tau _{SOC}^{-1}}{\left( 1+10^{4}%
\left[ \overline{Ca}\right] _{u}\right) ^{2}},\Delta _{p_{SOC}}^{p_{SOC}}=%
\frac{-1}{\tau _{SOC}}
\end{equation}

\bigskip

\subsection{Calcium-activated chloride channels} \label{appb7}

\begin{eqnarray}
I_{ClCa} &=&C_{m}g_{ClCa}P_{Cl}\left( V_{m}-E_{Cl}\right) \text{ }\mathtt{%
[pA]}  \nonumber \\
&\simeq &\overline{I}_{ClCa}+\Delta I_{ClCa}^{V_{m}}\delta _{V_{m}}+\Delta
I_{ClCa}^{Ca_{i}}\left[ \delta _{Ca}\right] _{i}+  \nonumber \\
&&\Delta I_{ClCa}^{Cl_{i}}\left[ \delta _{Cl}\right] _{i}+\Delta
I_{ClCa}^{cGMP}\left[ \delta _{cGMP}\right]
\end{eqnarray}%
on quasi-equilibrium conditions with relevant variables%
\begin{eqnarray*}
g_{ClCa} &=&0.23\text{ }nS/pF,\text{ }n_{ClcGMP}=3.3 \\
K_{ClCa} &=&3.65\cdot 10^{-4}\text{ }mM,\text{ }n_{ClCa}=2 \\
R_{ClcGMP} &=&0.0132 \\
K_{ClcGMP} &=&6.4\cdot 10^{-3}\text{ }mM
\end{eqnarray*}%
\begin{eqnarray}
P_{Cl} &=&R_{ClcGMP}\frac{\left( \left[ Ca\right] _{i}\right) ^{n_{ClCa}}}{%
\left( \left[ Ca\right] _{i}\right) ^{n_{ClCa}}+\left( K_{ClCa}\right)
^{n_{ClCa}}}+  \nonumber \\
&&\alpha _{Cl}\frac{\left( \left[ Ca\right] _{i}\right) ^{n_{ClCa}}}{\left( %
\left[ Ca\right] _{i}\right) ^{n_{ClCa}}+\left( K_{ClCa,cGMP}\right)
^{n_{ClCa}}} \\
\alpha _{Cl} &=&\frac{\left( [cGMP]\right) ^{n_{ClcGMP}}}{\left(
[cGMP]\right) ^{n_{ClcGMP}}+\left( K_{ClcGMP}\right) ^{n_{ClcGMP}}}
\end{eqnarray}%
\begin{equation}
K_{ClCa,cGMP}=(1-0.9\alpha _{Cl})\cdot 4\cdot 10^{-4}\text{ }mM
\end{equation}%
and the fluctuation terms%
\begin{eqnarray}
\Delta I_{ClCa}^{V_{m}} &=&C_{m}g_{ClCa}[\alpha _{7}R_{ClcGMP}+\beta
_{7}\gamma _{7}] \\
\Delta I_{ClCa}^{Ca_{i}} &=&C_{m}g_{ClCa}n_{ClCa}\left( \overline{V}_{m}-%
\overline{E}_{Cl}\right) \left[ \overline{Ca}\right] _{i}^{-n_{ClCa}-1}
\nonumber \\
&&\times \lbrack \alpha _{7}^{2}R_{ClcGMP}\left( K_{ClCa}\right) ^{n_{ClCa}}
\nonumber \\
&&+\beta _{7}^{2}\gamma _{7}\left( \frac{1-0.9\gamma _{7}}{2500}\right)
^{n_{ClCa}}] \\
\Delta I_{ClCa}^{Cl_{i}} &=&\frac{-C_{m}g_{ClCa}RT\left( \alpha
_{7}R_{ClcGMP}+\beta _{7}\gamma _{7}\right) }{\left[ \overline{Cl}\right]
_{i}F}
\end{eqnarray}%
\begin{eqnarray}
\Delta I_{ClCa}^{cGMP} &=&C_{m}g_{ClCa}\beta _{7}^{2}n_{ClcGMP}\left(
\overline{V}_{m}-\overline{E}_{Cl}\right) \times \lbrack  \nonumber \\
&&\frac{\gamma _{7}^{2}\left( K_{ClcGMP}\right) ^{n_{ClcGMP}}}{\left( [%
\overline{cGMP}]\right) ^{1+n_{ClcGMP}}}+  \nonumber \\
&&\frac{\left( K_{ClcGMP}\right) ^{n_{ClcGMP}}\gamma _{7}^{2}\left( \frac{%
1-0.9\gamma _{7}}{2500}\right) ^{n_{ClCa}}}{\left( \left[ \overline{Ca}%
\right] _{i}\right) ^{n_{ClCa}}\left( [\overline{cGMP}]\right)
^{1+n_{ClcGMP}}}+  \nonumber \\
&&\frac{9n_{ClCa}\gamma _{7}^{2}\kappa _{7}\left( \frac{1-0.9\gamma _{7}}{%
2500}\right) ^{n_{ClCa}}}{\left( \left[ \overline{Ca}\right] _{i}\right)
^{n_{ClCa}}[\overline{cGMP}]}]
\end{eqnarray}%
Herein,%
\begin{eqnarray}
\alpha _{7} &=&\frac{\left( \left[ \overline{Ca}\right] _{i}\right)
^{n_{ClCa}}}{\left( \left[ \overline{Ca}\right] _{i}\right)
^{n_{ClCa}}+\left( K_{ClCa}\right) ^{n_{ClCa}}} \\
\beta _{7} &=&\frac{\left( \left[ \overline{Ca}\right] _{i}\right)
^{n_{ClCa}}}{\left( \left[ \overline{Ca}\right] _{i}\right)
^{n_{ClCa}}+\left( \frac{1-0.9\gamma _{7}}{2500}\right) ^{n_{ClCa}}} \\
\gamma _{7} &=&\frac{\left( [\overline{cGMP}]\right) ^{n_{ClcGMP}}}{\left( [%
\overline{cGMP}]\right) ^{n_{ClcGMP}}+\left( K_{ClcGMP}\right) ^{n_{ClcGMP}}}
\\
\kappa _{7} &=&\frac{\left( K_{ClcGMP}\right) ^{n_{ClcGMP}}}{\left( [%
\overline{cGMP}]\right) ^{n_{ClcGMP}}+10\left( K_{ClcGMP}\right)
^{n_{ClcGMP}}}
\end{eqnarray}

\bigskip

\subsection{Plasma membrane $Ca^{2+}$ pump} \label{appb8}

\begin{eqnarray}
I_{PMCA} &=&I_{PMCA,0}\frac{\left[ Ca\right] _{i}}{\left[ Ca\right]
_{i}+K_{m,PMCA}}  \nonumber \\
&\simeq &\overline{I}_{PMCA}+\Delta I_{PMCA}^{Ca_{i}}\left[ \delta _{Ca}%
\right] _{i}
\end{eqnarray}%
on quasi-equilibrium conditions with relevant variables%
\begin{eqnarray*}
I_{PMCA,0} &=&5.37\text{ }pA \\
K_{m,PMCA} &=&1.7\cdot 10^{-4}\text{ }mM
\end{eqnarray*}%
and the fluctuation terms%
\begin{equation}
\Delta I_{PMCA}^{Ca_{i}}=I_{PMCA,0}\frac{K_{m,PMCA}}{\left( \left[ \overline{%
Ca}\right] _{i}+K_{m,PMCA}\right) ^{2}}
\end{equation}

\bigskip

\subsection{Plasma membrane $Na^{+}-Ca^{2+}$ exchange} \label{appb9}

\begin{eqnarray}
I_{NCX} &=&g_{NCX}R_{NCX,cGMP}\times  \nonumber \\
&&\frac{\text{ }\left[ Na\right] _{i}^{3}\text{ }\left[ Ca\right] _{e}\phi
_{F}-\left[ Na\right] _{e}^{3}\text{ }\left[ Ca\right] _{i}\phi _{R}}{%
1+d_{NCX}\left( \left[ Na\right] _{e}^{3}\text{ }\left[ Ca\right] _{i}+\left[
Na\right] _{i}^{3}\text{ }\left[ Ca\right] _{e}\right) }  \nonumber \\
&\simeq &\overline{I}_{NCX}+\Delta I_{NCX}^{V_{m}}\delta _{V_{m}}+\Delta
I_{NCX}^{Ca_{i}}\left[ \delta _{Ca}\right] _{i}+  \nonumber \\
&&\Delta I_{NCX}^{Na_{i}}\left[ \delta _{Na}\right] _{i}+\Delta
I_{NCX}^{cGMP}\left[ \delta _{cGMP}\right]
\end{eqnarray}%
on quasi-equilibrium conditions with relevant variables%
\begin{eqnarray}
g_{NCX} &=&0.000487\text{ }pA  \nonumber \\
d_{NCX} &=&0.0003,\text{ }\gamma _{NCX}=0.45  \nonumber \\
R_{NCX,cGMP} &=&1+\frac{0.55[cGMP]}{[cGMP]+0.045\text{ }mM} \\
\phi _{F} &=&\exp \left[ \frac{\gamma _{NXC}V_{m}F}{RT}\right] \\
\phi _{R} &=&\exp \left[ \frac{\left( \gamma _{NXC}-1\right) V_{m}F}{RT}%
\right]
\end{eqnarray}%
and the fluctuation terms%
\begin{eqnarray}
\Delta I_{NCX}^{V_{m}} &=&\frac{\gamma _{9}F}{\alpha _{9}RT}\{\beta _{9}^{-1}%
\left[ \overline{Ca}\right] _{i}\left[ Na\right] _{e}^{3}(1-\gamma _{NCX})
\nonumber \\
&&+\left[ Ca\right] _{e}\left[ \overline{Na}\right] _{i}^{3}\gamma _{NCX}\}
\\
\Delta I_{NCX}^{Ca_{i}} &=&\frac{-\gamma _{9}}{\alpha _{9}^{2}}\{\beta
_{9}^{-1}\left[ Na\right] _{e}^{3}+d_{NCX}\left[ Na\right] _{e}^{3}\left[ Ca%
\right] _{e}\left[ \overline{Na}\right] _{i}^{3}  \nonumber \\
&&+d_{NCX}\beta _{9}^{-1}\left[ Na\right] _{e}^{3}\left[ Ca\right] _{e}\left[
\overline{Na}\right] _{i}^{3}\}
\end{eqnarray}%
\begin{eqnarray}
\Delta I_{NCX}^{Na_{i}} &=&\frac{3\gamma _{9}}{\alpha _{9}^{2}}\{d_{NCX}%
\left[ Ca\right] _{e}\left[ \overline{Ca}\right] _{i}\left[ Na\right]
_{e}^{3}\left[ \overline{Na}\right] _{i}^{2}  \nonumber \\
&&+\beta _{9}^{-1}d_{NCX}\left[ Ca\right] _{e}\left[ \overline{Ca}\right]
_{i}\left[ Na\right] _{e}^{3}\left[ \overline{Na}\right] _{i}^{2}  \nonumber
\\
&&+\left[ Ca\right] _{e}\left[ \overline{Na}\right] _{i}^{2}\}
\end{eqnarray}%
\begin{eqnarray}
\Delta I_{NCX}^{cGMP} &=&\frac{990g_{NCX}\cdot \beta _{9}^{\gamma _{NCX}}}{%
\alpha _{9}\left( 9+200[\overline{cGMP}]\right) ^{2}}\{\left[ Ca\right] _{e}%
\left[ \overline{Na}\right] _{i}^{3}-  \nonumber \\
&&\beta _{9}^{-1}\left[ \overline{Ca}\right] _{i}\left[ Na\right] _{e}^{3}\}
\end{eqnarray}%
Herein,%
\begin{eqnarray}
\alpha _{9} &=&1+d_{NCX}\left( \left[ Na\right] _{e}^{3}\text{ }\left[
\overline{Ca}\right] _{i}+\left[ \overline{Na}\right] _{i}^{3}\text{ }\left[
Ca\right] _{e}\right) \\
\beta _{9} &=&\exp \left[ \overline{V}_{m}F/(RT)\right] \\
\gamma _{9} &=&g_{NCX}\cdot \kappa _{9}\exp \left[ \gamma _{NCX}\overline{V}%
_{m}F/(RT)\right] \\
\kappa _{9} &=&1+\frac{0.55[\overline{cGMP}]}{[\overline{cGMP}]+0.045}
\end{eqnarray}

\bigskip

\subsection{Sodium-potassium pump} \label{appb10}

\begin{eqnarray}
I_{NaK} &=&C_{m}I_{NaK,0}Q\frac{\left[ Na\right] _{i}^{n_{HNai}}}{\left[ Na%
\right] _{i}^{n_{HNai}}+Na_{dNai}^{n_{HNai}}}\times  \nonumber \\
&&\frac{\left[ K\right] _{e}^{n_{HKe}}}{\left[ K\right]
_{e}^{n_{HKe}}+K_{dKe}^{n_{HKe}}}\frac{V_{m}+150\text{ }mV}{V_{m}+200\text{ }%
mV}  \nonumber \\
&\simeq &\overline{I}_{NaK}+\Delta I_{NaK}^{V_{m}}\delta _{V_{m}}+\Delta
I_{NaK}^{Na_{i}}\left[ \delta _{Na}\right] _{i}
\end{eqnarray}%
on quasi-equilibrium conditions with relevant variables%
\begin{eqnarray}
n_{HKe} &=&1.1,\text{ }n_{HNai}=1.7  \nonumber \\
K_{dKe} &=&1.6\text{ }mM,\text{ }Na_{dNai}=22\text{ }mM  \nonumber \\
I_{NaK,0} &=&2.3083\text{ }pA/pF  \nonumber \\
Q &=&Q_{10}^{(T-309.15\text{ }K)/(10\text{ }K)},\text{ }Q_{10}=1.87
\end{eqnarray}%
and the fluctuation terms%
\begin{eqnarray}
\Delta I_{NaK}^{V_{m}} &=&\frac{50C_{m}I_{NaK,0}Q\left[ K\right]
_{e}^{n_{HKe}}\left[ \overline{Na}\right] _{i}^{n_{HNai}}}{\alpha _{10}\beta
_{10}\gamma _{10}^{2}} \\
\Delta I_{NaK}^{Na_{i}} &=&\frac{C_{m}I_{NaK,0}Q}{\alpha _{10}\beta
_{10}^{2}\gamma _{10}}n_{HNai}\left( \overline{V}_{m}+150\right)  \nonumber
\\
&&\times \left[ K\right] _{e}^{n_{HKe}}Na_{dNai}^{n_{HNai}}\left[ \overline{%
Na}\right] _{i}^{n_{HNai}-1}
\end{eqnarray}%
Herein,%
\begin{eqnarray}
\alpha _{10} &=&\left[ K\right] _{e}^{n_{HKe}}+K_{dKe}^{n_{HKe}} \\
\beta _{10} &=&\left[ \overline{Na}\right]
_{i}^{n_{HNai}}+Na_{dNai}^{n_{HNai}} \\
\gamma _{10} &=&\left( \overline{V}_{m}+200\right)
\end{eqnarray}

\bigskip

\subsection{Sodium-potassium-chloride cotransport} \label{appb11}

\begin{eqnarray}
I_{NaKCl}^{Cl} &=&-10^{9}z_{Cl}R_{NaKCl,cGMP}A_{m}L_{NaKCl}RFT  \nonumber \\
&&\times \ln \left( \frac{\left[ N_{a}\right] _{e}}{\left[ N_{a}\right] _{i}%
}\frac{\left[ K\right] _{e}}{\left[ K\right] _{i}}\frac{\left[ Cl\right]
_{e}^{2}}{\left[ Cl\right] _{i}^{2}}\right) \text{ }\mathtt{[pA]}  \nonumber
\\
&\simeq &\overline{I}_{NaKCl}^{Cl}+\Delta I_{NaKCl}^{Cl,Na_{i}}\left[ \delta
_{Na}\right] _{i}+  \nonumber \\
&&\Delta I_{NaKCl}^{Cl,K_{i}}\left[ \delta _{K}\right] _{i}+\Delta
I_{NaKCl}^{Cl,Cl_{i}}\left[ \delta _{Cl}\right] _{i}  \nonumber \\
&&+\Delta I_{NaKCl}^{Cl,cGMP}\left[ \delta _{cGMP}\right]
\end{eqnarray}%
\begin{eqnarray}
I_{NaKCl}^{Na} &\equiv &-\frac{1}{2}I_{NaKCl}^{Cl}  \nonumber \\
&\simeq &\overline{I}_{NaKCl}^{Na}+\Delta I_{NaKCl}^{Na,Na_{i}}\left[ \delta
_{Na}\right] _{i}+  \nonumber \\
&&\Delta I_{NaKCl}^{Na,K_{i}}\left[ \delta _{K}\right] _{i}+\Delta
I_{NaKCl}^{Na,Cl_{i}}\left[ \delta _{Cl}\right] _{i}  \nonumber \\
&&+\Delta I_{NaKCl}^{Na,cGMP}\left[ \delta _{cGMP}\right]
\end{eqnarray}%
\begin{eqnarray}
I_{NaKCl}^{K} &\equiv &-\frac{1}{2}I_{NaKCl}^{Cl}  \nonumber \\
&\simeq &\overline{I}_{NaKCl}^{K}+\Delta I_{NaKCl}^{K,Na_{i}}\left[ \delta
_{Na}\right] _{i}+  \nonumber \\
&&\Delta I_{NaKCl}^{K,K_{i}}\left[ \delta _{K}\right] _{i}+\Delta
I_{NaKCl}^{K,Cl_{i}}\left[ \delta _{Cl}\right] _{i}  \nonumber \\
&&+\Delta I_{NaKCl}^{K,cGMP}\left[ \delta _{cGMP}\right]
\end{eqnarray}%
on quasi-equilibrium conditions with relevant variables%
\begin{eqnarray}
L_{NaKCL} &=&1.79\cdot 10^{-17}\text{ }mole^{2}/(sJcm^{2})  \nonumber \\
R_{NaKCl,cGMP} &=&1+\frac{3.5\left[ cGMP\right] }{\left[ cGMP\right]
+6.4\cdot 10^{-3}mM}
\end{eqnarray}%
and the fluctuation terms%
\begin{eqnarray}
\Delta I_{NaKCl}^{Cl,Na_{i}} &=&\frac{\alpha _{11}\beta _{11}}{\left[
\overline{Na}\right] _{i}} \\
\Delta I_{NaKCl}^{Cl,K_{i}} &=&\frac{\alpha _{11}\beta _{11}}{\left[
\overline{K}\right] _{i}} \\
\Delta I_{NaKCl}^{Cl,Cl_{i}} &=&\frac{2\alpha _{11}\beta _{11}}{\left[
\overline{Cl}\right] _{i}} \\
\Delta I_{NaKCl}^{Cl,cGMP} &=&\frac{-8750\alpha _{11}\gamma _{11}}{\left(
4+625\left[ \overline{cGMP}\right] \right) ^{2}}
\end{eqnarray}%
\begin{eqnarray}
\Delta I_{NaKCl}^{Na,Na_{i}} &=&\Delta I_{NaKCl}^{K,Na_{i}}=\frac{-\Delta
I_{NaKCl}^{Cl,Na_{i}}}{2} \\
\Delta I_{NaKCl}^{Na,K_{i}} &=&\Delta I_{NaKCl}^{K,K_{i}}=\frac{-\Delta
I_{NaKCl}^{Cl,K_{i}}}{2} \\
\Delta I_{NaKCl}^{Na,Cl_{i}} &=&\Delta I_{NaKCl}^{K,Cl_{i}}=\frac{-\Delta
I_{NaKCl}^{Cl,Cl_{i}}}{2} \\
\Delta I_{NaKCl}^{Na,cGMP} &=&\Delta I_{NaKCl}^{K,cGMP}=\frac{-\Delta
I_{NaKCl}^{Cl,cGMP}}{2}
\end{eqnarray}%
Herein,%
\begin{eqnarray}
\alpha _{11} &=&10^{9}z_{Cl}A_{m}L_{NaKCl}RFT \\
\beta _{11} &=&1+\frac{3.5\left[ \overline{cGMP}\right] }{\left[ \overline{%
cGMP}\right] +6.4\cdot 10^{-3}} \\
\gamma _{11} &=&\ln \left( \frac{\left[ N_{a}\right] _{e}}{\left[ \overline{%
N_{a}}\right] _{i}}\frac{\left[ K\right] _{e}}{\left[ \overline{K}\right]
_{i}}\frac{\left[ Cl\right] _{e}^{2}}{\left[ \overline{Cl}\right] _{i}^{2}}%
\right)
\end{eqnarray}

\bigskip

\subsection{Intercellular ionic communication} \label{appb12}

\begin{eqnarray}
I_{S,GJ} &=&-Pz_{S}^{2}\frac{V_{GJ}F^{2}}{RT}\frac{%
[S]_{C}-[S]_{i}e^{-z_{S}V_{GJ}F/RT}}{1-e^{-z_{S}V_{GJ}F/RT}}  \nonumber \\
&\simeq &\overline{I}_{S,GJ}+\Delta I_{S,GJ}^{SC}\left[ \delta _{S}\right]
_{C}+\Delta I_{S,GJ}^{S_{i}}\left[ \delta _{S}\right] _{i}+  \nonumber \\
&&\Delta I_{S,GJ}^{V_{mC}}\delta _{V_{mC}}+\Delta I_{S,GJ}^{V_{m}}\delta
_{V_{m}}+  \nonumber \\
&&\Delta I_{S,GJ}^{Ca_{i}}\left[ \delta _{Ca}\right] _{i}+\Delta
I_{S,GJ}^{Na_{i}}\left[ \delta _{Na}\right] _{i}+  \nonumber \\
&&\Delta I_{S,GJ}^{K_{i}}\left[ \delta _{K}\right] _{i}+\Delta
I_{S,GJ}^{Cl_{i}}\left[ \delta _{Cl}\right] _{i}
\end{eqnarray}%
on quasi-equilibrium conditions with relevant variables%
\begin{eqnarray}
G_{GJ} &=&2\text{ }nS  \nonumber \\
V_{GJ} &=&V_{mC}-V_{m} \\
P &=&\frac{G_{GJ}RT}{F^{2}\sum_{S^{\prime }}\left( z_{S^{\prime
}}^{2}[S^{\prime }]_{i}\right) }
\end{eqnarray}%
where $S$ and $S^{\prime }$ represent all accessible ions: $Ca^{2+}$, $%
Na^{+} $, $K^{+}$, $Cl^{-}$. The suffix $C$ denotes the variables of nearby
SMCs coupled to the local one. The fluctuation terms are given by%
\begin{eqnarray}
\Delta I_{S,GJ}^{SC} &=&\frac{-G_{GJ}\gamma _{12}z_{S}^{2}}{\left( 1-\alpha
_{12}^{z_{S}}\right) \beta _{12}} \\
\Delta I_{S,GJ}^{S} &=&\frac{G_{GJ}\gamma _{12}z_{S}^{2}\alpha _{12}^{z_{S}}%
}{\left( 1-\alpha _{12}^{z_{S}}\right) \beta _{12}} \\
\Delta I_{S,GJ}^{V_{mC}} &=&\frac{G_{GJ}z_{S}^{3}\alpha _{12}^{z_{S}}\gamma
_{12}F\left( \left[ \overline{S}\right] _{C}-\left[ \overline{S}\right]
_{i}\right) }{\left( 1-\alpha _{12}^{z_{S}}\right) ^{2}\beta _{12}RT}
\nonumber \\
&&-\frac{G_{GJ}z_{S}^{2}\left( \left[ \overline{S}\right] _{C}-\left[
\overline{S}\right] _{i}\alpha _{12}^{z_{S}}\right) }{\left( 1-\alpha
_{12}^{z_{S}}\right) \beta _{12}} \\
\Delta I_{S,GJ}^{V_{m}} &=&-\Delta I_{S,GJ}^{V_{mC}}
\end{eqnarray}%
\begin{eqnarray}
\Delta I_{S,GJ}^{Ca_{i}} &=&\frac{G_{GJ}\gamma
_{12}z_{Ca}^{2}z_{S}^{2}\left( \left[ \overline{S}\right] _{C}-\left[
\overline{S}\right] _{i}\alpha _{12}^{z_{S}}\right) }{\left( 1-\alpha
_{12}^{z_{S}}\right) \beta _{12}^{2}} \\
\Delta I_{S,GJ}^{Na_{i}} &=&\frac{G_{GJ}\gamma
_{12}z_{Na}^{2}z_{S}^{2}\left( \left[ \overline{S}\right] _{C}-\left[
\overline{S}\right] _{i}\alpha _{12}^{z_{S}}\right) }{\left( 1-\alpha
_{12}^{z_{S}}\right) \beta _{12}^{2}} \\
\Delta I_{S,GJ}^{K_{i}} &=&\frac{G_{GJ}\gamma _{12}z_{K}^{2}z_{S}^{2}\left( %
\left[ \overline{S}\right] _{C}-\left[ \overline{S}\right] _{i}\alpha
_{12}^{z_{S}}\right) }{\left( 1-\alpha _{12}^{z_{S}}\right) \beta _{12}^{2}}
\\
\Delta I_{S,GJ}^{Cl_{i}} &=&\frac{G_{GJ}\gamma
_{12}z_{Cl}^{2}z_{S}^{2}\left( \left[ \overline{S}\right] _{C}-\left[
\overline{S}\right] _{i}\alpha _{12}^{z_{S}}\right) }{\left( 1-\alpha
_{12}^{z_{S}}\right) \beta _{12}^{2}}
\end{eqnarray}%
Herein,%
\begin{eqnarray}
\alpha _{12} &=&\exp \left[ \frac{-F\left( \overline{V}_{mC}-\overline{V}%
_{m}\right) }{RT}\right] \\
\beta _{12} &=&z_{Ca}^{2}\left[ \overline{Ca}\right] _{i}+z_{Na}^{2}\left[
\overline{Na}\right] _{i}+  \nonumber \\
&&z_{K}^{2}\left[ \overline{K}\right] _{i}+z_{Cl}^{2}\left[ \overline{Cl}%
\right] _{i} \\
\gamma _{12} &=&\overline{V}_{mC}-\overline{V}_{m}
\end{eqnarray}

\bigskip

\section{Mathematical model equations for sarcoplasmic reticulum} \label{appc}

\subsection{Calcium-induced Calcium-release (CICR) mechanism of sarcoplasmic
reticulum}  \label{appc1}

\begin{eqnarray}
I_{SERCA} &=&I_{SERCA,0}\frac{\left[ Ca\right] _{i}}{\left[ Ca\right]
_{i}+K_{m,up}}  \nonumber \\
&\simeq &\overline{I}_{SERCA}+\Delta I_{SERCA}^{Ca_{i}}\left[ \delta _{Ca}%
\right] _{i} \\
I_{tr} &=&\left( \left[ Ca\right] _{u}-\left[ Ca\right] _{r}\right) \frac{%
z_{Ca}vol_{u}F}{\tau _{tr}}  \nonumber \\
&\simeq &\overline{I}_{tr}+\Delta I_{tr}^{Ca_{u}}\left[ \delta _{Ca}\right]
_{u}+\Delta I_{tr}^{Ca_{r}}\left[ \delta _{Ca}\right] _{r}
\end{eqnarray}%
\begin{eqnarray}
I_{rel} &=&\left( \left[ Ca\right] _{r}-\left[ Ca\right] _{i}\right) \frac{%
\left( R_{10}^{2}+R_{leak}\right) z_{Ca}vol_{r}F}{\tau _{rel}}  \nonumber \\
&\simeq &\overline{I}_{rel}+\Delta I_{rel}^{Ca_{i}}\left[ \delta _{Ca}\right]
_{i}+\Delta I_{rel}^{Ca_{r}}\left[ \delta _{Ca}\right] _{r}  \nonumber \\
&&+\Delta I_{rel}^{R_{10}}\delta _{R_{10}}
\end{eqnarray}%
on quasi-equilibrium conditions with relevant variables%
\begin{eqnarray*}
I_{SERCA,0} &=&6.68\text{ }pA,\text{ }K_{m,up}=10^{-3}\text{ }mM \\
\tau _{tr} &=&1000.0\text{ }ms,\text{ }\tau _{rel}=0.0333\text{ }ms \\
R_{leak} &=&1.07\cdot 10^{-5}
\end{eqnarray*}%
and the fluctuation terms%
\begin{eqnarray}
\Delta I_{SERCA}^{Ca_{i}} &=&\frac{I_{SERCA,0}K_{m,up}}{\left( \left[
\overline{Ca}\right] _{i}+K_{m,up}\right) ^{2}} \\
\Delta I_{tr}^{Ca_{r}} &=&\frac{-z_{Ca}vol_{u}F}{\tau _{tr}} \\
\Delta I_{tr}^{Ca_{u}} &=&\frac{z_{Ca}vol_{u}F}{\tau _{tr}}
\end{eqnarray}%
\begin{eqnarray}
\Delta I_{rel}^{Ca_{r}} &=&\frac{z_{Ca}vol_{r}F}{\tau _{rel}}\left(
\overline{R}_{10}^{2}+R_{leak}\right) \\
\Delta I_{rel}^{Ca_{i}} &=&\frac{-z_{Ca}vol_{r}F}{\tau _{rel}}\left(
\overline{R}_{10}^{2}+R_{leak}\right) \\
\Delta I_{rel}^{R_{10}} &=&\frac{2z_{Ca}vol_{r}F\overline{R}_{10}}{\tau
_{rel}}\left( \left[ \overline{Ca}\right] _{r}-\left[ \overline{Ca}\right]
_{i}\right)
\end{eqnarray}

\bigskip

\subsection{Ryanodine receptor}  \label{appc2}

\begin{eqnarray}
\frac{dR_{10}}{dt} &=&K_{r1}\left[ Ca\right] _{i}^{2}R_{00}-\left(
K_{-r1}+K_{r2}\left[ Ca\right] _{i}\right) R_{10}  \nonumber \\
&&+K_{-r2}R_{11} \\
\frac{dR_{11}}{dt} &=&K_{r2}\left[ Ca\right] _{i}R_{10}-\left(
K_{-r1}+K_{-r2}\right) R_{11}  \nonumber \\
&&+K_{r1}\left[ Ca\right] _{i}^{2}R_{01} \\
\frac{dR_{01}}{dt} &=&K_{r2}\left[ Ca\right] _{i}R_{00}-\left( K_{-r2}+K_{r1}%
\left[ Ca\right] _{i}^{2}\right) R_{01}  \nonumber \\
&&+K_{-r1}R_{11}
\end{eqnarray}%
\begin{equation}
R_{00}=1-R_{01}-R_{10}-R_{11}
\end{equation}%
On the quasi-equilibrium conditions, the equations can be re-arranged by
defining $R_{ij}=\overline{R}_{ij}+\delta _{R_{ij}}$ as the time-independent
constant term $(d\overline{R}_{ij}/dt=0)$ plus the time-dependent
fluctuation $\delta _{R_{ij}}$:%
\begin{eqnarray}
\frac{d\delta _{R_{10}}}{dt} &\simeq &\Delta _{R_{10}}^{Ca_{i}}\left[ \delta
_{Ca}\right] _{i}+\Delta _{R_{10}}^{R_{10}}\delta _{R_{10}}+\Delta
_{R_{10}}^{R_{11}}\delta _{R_{11}}  \nonumber \\
&&+\Delta _{R_{10}}^{R_{01}}\delta _{R_{01}} \\
\frac{d\delta _{R_{11}}}{dt} &\simeq &\Delta _{R_{11}}^{Ca_{i}}\left[ \delta
_{Ca}\right] _{i}+\Delta _{R_{11}}^{R_{10}}\delta _{R_{10}}+\Delta
_{R_{11}}^{R_{11}}\delta _{R_{11}}  \nonumber \\
&&+\Delta _{R_{11}}^{R_{01}}\delta _{R_{01}} \\
\frac{d\delta _{R_{01}}}{dt} &\simeq &\Delta _{R_{01}}^{Ca_{i}}\left[ \delta
_{Ca}\right] _{i}+\Delta _{R_{01}}^{R_{10}}\delta _{R_{10}}+\Delta
_{R_{01}}^{R_{11}}\delta _{R_{11}}  \nonumber \\
&&+\Delta _{R_{01}}^{R_{01}}\delta _{R_{01}}
\end{eqnarray}%
with relevant variables%
\begin{eqnarray*}
K_{r1} &=&2500.0\text{ }mM^{-2}ms^{-1} \\
K_{r2} &=&1.05\text{ }mM^{-1}ms^{-1} \\
K_{-r1} &=&0.0076\text{ }ms^{-1},\text{ }K_{-r2}=0.084\text{ }ms^{-1}
\end{eqnarray*}%
and the fluctuation terms%
\begin{eqnarray}
\Delta _{R_{10}}^{R_{10}} &=&-K_{-r1}-\left[ \overline{Ca}\right]
_{i}^{2}K_{r1}-\left[ \overline{Ca}\right] _{i}K_{r2} \\
\Delta _{R_{10}}^{R_{11}} &=&K_{-r2}-\left[ \overline{Ca}\right]
_{i}^{2}K_{r1}\text{ } \\
\Delta _{R_{10}}^{R_{01}} &=&-\left[ \overline{Ca}\right] _{i}^{2}K_{r1} \\
\Delta _{R_{10}}^{Ca_{i}} &=&2\left[ \overline{Ca}\right] _{i}K_{r1}-2\left[
\overline{Ca}\right] _{i}K_{r1}\overline{R}_{01}-K_{r2}\overline{R}_{10}
\nonumber \\
&&-2\left[ \overline{Ca}\right] _{i}K_{r1}\overline{R}_{10}-2\left[
\overline{Ca}\right] _{i}K_{r1}\overline{R}_{11}
\end{eqnarray}%
\begin{eqnarray}
\Delta _{R_{11}}^{R_{10}} &=&\left[ \overline{Ca}\right] _{i}K_{r2},\text{ }%
\Delta _{R_{11}}^{R_{01}}=\left[ \overline{Ca}\right] _{i}^{2}K_{r1} \\
\Delta _{R_{11}}^{R_{11}} &=&-K_{-r1}-K_{-r2} \\
\Delta _{R_{11}}^{Ca_{i}} &=&2\left[ \overline{Ca}\right] _{i}K_{r1}%
\overline{R}_{01}+K_{r2}\overline{R}_{10}
\end{eqnarray}%
\begin{eqnarray}
\Delta _{R_{01}}^{R_{10}} &=&-\left[ \overline{Ca}\right] _{i}K_{r2} \\
\Delta _{R_{01}}^{R_{11}} &=&K_{-r1}-\left[ \overline{Ca}\right] _{i}K_{r2}
\\
\Delta _{R_{01}}^{R_{01}} &=&-K_{-r2}-\left[ \overline{Ca}\right]
_{i}^{2}K_{r1}-\left[ \overline{Ca}\right] _{i}K_{r2} \\
\Delta _{R_{01}}^{Ca_{i}} &=&-2\left[ \overline{Ca}\right] _{i}K_{r1}%
\overline{R}_{01}-K_{r2}\overline{R}_{01}-K_{r2}\overline{R}_{10}\text{ \ }
\nonumber \\
&&-K_{r2}\overline{R}_{11}+K_{r2}
\end{eqnarray}

\bigskip

\subsection{$IP_{3}$ receptor}  \label{appc3}

\begin{eqnarray}
I_{IP3} &=&I_{IP3,0}z_{Ca}vol_{Ca}F\left( \left[ Ca\right] _{u}-\left[ Ca%
\right] _{i}\right) \times  \nonumber \\
&&\left( \frac{\left[ IP_{3}\right] }{\left[ IP_{3}\right] +K_{IP3}}\frac{%
\left[ Ca\right] _{i}h_{IP3}}{\left[ Ca\right] _{i}+K_{act,IP3}}\right) ^{3}
\nonumber \\
&\simeq &\overline{I}_{IP3}+\Delta I_{IP3}^{Ca_{i}}\left[ \delta _{Ca}\right]
_{i}+\Delta I_{IP3}^{Ca_{u}}\left[ \delta _{Ca}\right] _{u}  \nonumber \\
&&+\Delta I_{IP3}^{IP3}\left[ \delta _{IP3}\right] +\Delta
I_{IP3}^{h_{IP3}}\delta _{h_{IP3}} \\
\frac{dh_{IP3}}{dt} &=&K_{on,IP3}\left[ K_{inh,IP3}-\left( \left[ Ca\right]
_{i}+K_{inh,IP3}\right) h_{IP3}\right]  \nonumber \\
&\simeq &\frac{d\delta _{h_{IP3}}}{dt}=\Delta _{h_{IP3}}^{Ca_{i}}\left[
\delta _{Ca}\right] _{i}+\Delta I_{h_{IP3}}^{h_{IP3}}\delta _{h_{IP3}}
\end{eqnarray}%
on quasi-equilibrium conditions with relevant variables%
\begin{eqnarray*}
I_{IP3,0} &=&0.00288\text{ }ms^{-1},\text{ }K_{IP3}=1.2\cdot 10^{-4}\text{ }%
mM \\
K_{act,IP3} &=&1.7\cdot 10^{-4}\text{ }mM,\text{ }K_{inh,IP3}=10^{-4}\text{ }%
mM \\
K_{on,IP3} &=&1.4\text{ }ms^{-1}mM^{-1}
\end{eqnarray*}%
and the fluctuation terms
\begin{eqnarray}
\Delta I_{IP3}^{h_{IP3}} &=&\frac{3\alpha _{14}\overline{h}_{IP3}^{2}\left[
\overline{Ca}\right] _{i}^{3}\left[ \overline{IP}_{3}\right] ^{3}}{\beta
_{14}^{3}\gamma _{14}^{3}}  \nonumber \\
&&\times \left( \left[ \overline{Ca}\right] _{u}-\left[ \overline{Ca}\right]
_{i}\right) \\
\Delta I_{IP3}^{Ca_{u}} &=&\frac{\alpha _{14}\overline{h}_{IP3}^{3}\left[
\overline{Ca}\right] _{i}^{3}\left[ \overline{IP}_{3}\right] ^{3}}{\beta
_{14}^{3}\gamma _{14}^{3}}
\end{eqnarray}%
\begin{eqnarray}
\Delta I_{IP3}^{IP3} &=&\frac{3\alpha _{14}\overline{h}_{IP3}^{3}\left[
\overline{Ca}\right] _{i}^{3}\left[ \overline{IP}_{3}\right] ^{2}K_{IP3}}{%
\beta _{14}^{4}\gamma _{14}^{3}}  \nonumber \\
&&\times \left( \left[ \overline{Ca}\right] _{u}-\left[ \overline{Ca}\right]
_{i}\right) \\
\Delta I_{IP3}^{Ca_{i}} &=&\frac{\alpha _{14}\overline{h}_{IP3}^{3}\left[
\overline{IP}_{3}\right] ^{3}}{\beta _{14}^{3}\gamma _{14}^{4}}\{-4\left[
\overline{Ca}\right] _{i}^{3}K_{act,IP3}  \nonumber \\
&&-\left[ \overline{Ca}\right] _{i}^{4}+3\left[ \overline{Ca}\right] _{i}^{2}%
\left[ \overline{Ca}\right] _{u}K_{act,IP3}\}
\end{eqnarray}%
\begin{eqnarray}
\Delta _{h_{IP3}}^{Ca_{i}} &=&-\overline{h}_{IP3}K_{on,IP3} \\
\Delta I_{h_{IP3}}^{h_{IP3}} &=&-K_{on,IP3}\left( K_{inh,IP3}+\left[
\overline{Ca}\right] _{i}\right)
\end{eqnarray}%
Herein,%
\begin{eqnarray}
\alpha _{14} &=&I_{IP3,0}\cdot z_{Ca}\cdot vol_{Ca}\cdot F \\
\beta _{14} &=&\left[ \overline{IP}_{3}\right] +K_{IP3} \\
\gamma _{14} &=&\left[ \overline{Ca}\right] _{i}+K_{act,IP3}
\end{eqnarray}

\bigskip

\section{$\protect\alpha _{1}$-Adrenoceptor activation aand $IP_{3}$
formation}  \label{appd}

\begin{eqnarray}
\frac{d\left[ G\right] }{dt} &=&k_{a,G}\left( \delta _{G}+\rho _{r,G}\right)
\left( \left[ G_{T,G}\right] -\left[ G\right] \right) -k_{d,G}\left[ G\right]
\nonumber \\
&\simeq &\frac{d\left[ \delta _{G}\right] }{dt}=\Delta
_{G}^{R_{G}^{S}}[\delta _{R_{G}^{S}}]+\Delta _{G}^{G}\left[ \delta _{G}%
\right]
\end{eqnarray}%
\begin{eqnarray}
\frac{d\left[ IP_{3}\right] }{dt} &=&\frac{r_{h,G}}{\gamma _{G}}\left[
PIP_{2}\right] -k_{deg,G}\left[ IP_{3}\right]  \nonumber \\
&&+P_{IP3}\sum_{C}\left( \left[ IP_{3}\right] _{C}-\left[ IP_{3}\right]
\right)  \nonumber \\
&\simeq &\frac{d\left[ \delta _{IP3}\right] }{dt}=\Delta _{IP3}^{Ca_{i}}%
\left[ \delta _{Ca}\right] _{i}+\Delta _{IP3}^{G}\left[ \delta _{G}\right]
\nonumber \\
&&+\Delta _{IP3}^{PIP_{2}}\left[ \delta _{PIP_{2}}\right] +\Delta
_{IP3}^{IP3}\left[ \delta _{IP3}\right]  \nonumber \\
&&+\sum_{C}\Delta _{IP3}^{IP3C}\left[ \delta _{IP3}\right] _{C}
\end{eqnarray}%
\begin{eqnarray}
\frac{d\left[ R_{P,G}^{S}\right] }{dt} &=&\frac{\left[ NE\right] k_{p,G}%
\left[ R_{G}^{S}\right] }{K_{1,G}+\left[ NE\right] }-\frac{\left[ NE\right]
k_{e,G}\left[ R_{P,G}^{S}\right] }{K_{2,G}+\left[ NE\right] }  \nonumber \\
&\simeq &\frac{d[\delta _{R_{P,G}^{S}}]}{dt}  \nonumber \\
&=&\Delta _{R_{P,G}^{S}}^{R_{G}^{S}}[\delta _{R_{G}^{S}}]+\Delta
_{R_{P,G}^{S}}^{R_{P,G}^{S}}[\delta _{R_{P,G}^{S}}]
\end{eqnarray}%
\begin{eqnarray}
\frac{d\left[ R_{G}^{S}\right] }{dt} &=&k_{r,G}\xi _{G}\left[ R_{T,G}\right]
-k_{r,G}\left[ R_{P,G}^{S}\right] -  \nonumber \\
&&\left( k_{r,G}+\frac{k_{p,G}\left[ NE\right] }{K_{1,G}+\left[ NE\right] }%
\right) \left[ R_{G}^{S}\right]  \nonumber \\
&\simeq &\frac{d[\delta _{R_{G}^{S}}]}{dt}=\Delta
_{R_{G}^{S}}^{R_{G}^{S}}[\delta _{R_{G}^{S}}]+\Delta
_{R_{G}^{S}}^{R_{P,G}^{S}}[\delta _{R_{P,G}^{S}}]
\end{eqnarray}%
\begin{eqnarray}
\frac{d\left[ PIP_{2}\right] }{dt} &=&-\left( r_{h,G}+r_{r,G}\right) \left[
PIP_{2}\right] -r_{r,G}\gamma _{G}\left[ IP_{3}\right]  \nonumber \\
&&+r_{r,G}\left[ PIP_{2,T}\right]  \nonumber \\
&\simeq &\frac{d[\delta _{PIP_{2}}]}{dt}=\Delta _{PIP_{2}}^{Ca_{i}}\left[
\delta _{Ca}\right] _{i}+\Delta _{PIP_{2}}^{G}\left[ \delta _{G}\right]
\nonumber \\
&&+\Delta _{PIP_{2}}^{PIP_{2}}\left[ \delta _{PIP_{2}}\right] +\Delta
_{PIP_{2}}^{IP3}\left[ \delta _{IP3}\right]
\end{eqnarray}%
on quasi-equilibrium conditions with relevant variables%
\begin{eqnarray*}
\left[ R_{T,G}\right] &=&2\cdot 10^{4},\text{ }K_{1,G}=10^{-2}\text{ }mM \\
K_{2,G} &=&0.2\text{ }mM,\text{ }k_{r,G}=1.75\cdot 10^{-7}ms^{-1} \\
k_{e,G} &=&6\cdot 10^{-6}ms^{-1},\text{ }k_{a,G}=0.17\cdot 10^{-3}ms^{-1} \\
k_{deg,G} &=&1.25\cdot 10^{-3}ms^{-1},\text{ }\xi _{G}=0.85 \\
k_{d,G} &=&1.5\cdot 10^{-3}ms^{-1},\text{ }\left[ PIP_{2,T}\right] =5\cdot
10^{7} \\
r_{r,G} &=&1.5\cdot 10^{-5}ms^{-1},\text{ }K_{c,G}=4\cdot 10^{-4}\text{ }mM
\end{eqnarray*}%
\begin{eqnarray*}
\alpha _{G} &=&2.781\cdot 10^{-8}ms^{-1} \\
\left[ G_{T,G}\right] &=&10^{5},\text{ }k_{p,G}=10^{-4}\text{ }ms^{-1} \\
P_{IP3} &=&0.53\cdot 10^{-3}ms^{-1} \\
\gamma _{G} &=&10^{-15}N_{AV}\cdot vol_{i} \\
\delta _{G} &=&\frac{k_{d,G}\left[ G\right] }{k_{a,G}\left( \left[ G_{T,G}%
\right] -\left[ G\right] \right) },\text{ fixed at initial}
\end{eqnarray*}%
\begin{eqnarray}
\rho _{r,G} &=&\frac{\left[ NE\right] \left[ R_{G}^{S}\right] }{\xi _{G}%
\left[ R_{T,G}\right] \left( K_{1,G}+\left[ NE\right] \right) } \\
r_{h,G} &=&\alpha _{G}\frac{\left[ Ca\right] _{i}}{\left[ Ca\right]
_{i}+K_{c,G}}\left[ G\right]
\end{eqnarray}%
and the fluctuation terms%
\begin{eqnarray}
\Delta _{G}^{R_{G}^{S}} &=&\frac{\left( \left[ G_{T,G}\right] -\left[
\overline{G}\right] \right) k_{a,G}\left[ NE\right] }{\left( K_{1,G}+\left[
NE\right] \right) R_{T,G}\xi _{G}} \\
\Delta _{G}^{G} &=&-k_{d,G}-\frac{\text{ }k_{a,G}\left[ NE\right] [\overline{%
R}_{G}^{S}]}{\left( K_{1,G}+\left[ NE\right] \right) \left[ R_{T,G}\right]
\xi _{G}}  \nonumber \\
&&-k_{a,G}\delta _{G}
\end{eqnarray}%
\begin{eqnarray}
\Delta _{IP3}^{Ca_{i}} &=&\frac{K_{c,G}\left[ \overline{G}\right] \left[
\overline{PIP}_{2}\right] \alpha _{G}}{\left( \left[ \overline{Ca}\right]
_{i}+K_{c,G}\right) ^{2}\gamma _{G}} \\
\Delta _{IP3}^{G} &=&\frac{\left[ \overline{Ca}\right] _{i}\left[ \overline{%
PIP}_{2}\right] \alpha _{G}}{\left( \left[ \overline{Ca}\right]
_{i}+K_{c,G}\right) \gamma _{G}} \\
\Delta _{IP3}^{PIP_{2}} &=&\frac{\left[ \overline{Ca}\right] _{i}\left[
\overline{G}\right] \alpha _{G}}{\left( \left[ \overline{Ca}\right]
_{i}+K_{c,G}\right) \gamma _{G}} \\
\Delta _{IP3}^{IP3} &=&-k_{deg,G}-\sum_{C}P_{IP3} \\
\Delta _{IP3}^{IP3C} &=&P_{IP3}
\end{eqnarray}%
\begin{eqnarray}
\Delta _{R_{P,G}^{S}}^{R_{G}^{S}} &=&\frac{k_{p,G}\left[ NE\right] }{K_{1,G}+%
\left[ NE\right] } \\
\Delta _{R_{P,G}^{S}}^{R_{P,G}^{S}} &=&\frac{-k_{e,G}\left[ NE\right] }{%
K_{2,G}+\left[ NE\right] } \\
\Delta _{R_{G}^{S}}^{R_{G}^{S}} &=&-k_{r,G}-\frac{k_{p,G}\left[ NE\right] }{%
K_{1,G}+\left[ NE\right] } \\
\Delta _{R_{G}^{S}}^{R_{P,G}^{S}} &=&-k_{r,G},\text{ \ \ }\Delta
_{PIP_{2}}^{IP3}=-r_{r,G}\gamma _{G}
\end{eqnarray}%
\begin{eqnarray}
\Delta _{PIP_{2}}^{Ca_{i}} &=&-\frac{K_{c,G}\left[ \overline{G}\right] \left[
\overline{PIP}_{2}\right] \alpha _{G}}{\left( \left[ \overline{Ca}\right]
_{i}+K_{c,G}\right) ^{2}} \\
\Delta _{PIP_{2}}^{G} &=&-\frac{\left[ \overline{Ca}\right] _{i}\left[
\overline{PIP}_{2}\right] \alpha _{G}}{\left[ \overline{Ca}\right]
_{i}+K_{c,G}} \\
\Delta _{PIP_{2}}^{PIP_{2}} &=&-r_{r,G}-\frac{\left[ \overline{Ca}\right]
_{i}\left[ \overline{G}\right] \alpha _{G}}{\left[ \overline{Ca}\right]
_{i}+K_{c,G}}
\end{eqnarray}%
Herein, the $IP_{3}$ flux through gap junctions is also included, and $\left[
IP_{3}\right] _{C}$ denotes the\ concentrations for nearby SMCs coupled to
the local one.

\bigskip

\section{sGC activation and cGMP formation}  \label{appe}

\begin{eqnarray}
\frac{dV_{cGMP}}{dt} &=&\frac{V_{cGMP,0}-V_{cGMP}}{\tau _{sGC}}  \nonumber \\
&\simeq &\frac{d\delta _{V_{cGMP}}}{dt}=\Delta _{V_{cGMP}}^{V_{cGMP}}\delta
_{V_{cGMP}} \\
\frac{d\left[ cGMP\right] }{dt} &=&V_{cGMP}-k_{pde,cGMP}\frac{\left[ cGMP%
\right] ^{2}}{\left[ cGMP\right] +K_{m,pde}}  \nonumber \\
&\simeq &\frac{d\delta _{cGMP}}{dt}  \nonumber \\
&=&\Delta _{cGMP}^{V_{cGMP}}\delta _{V_{cGMP}}+\Delta _{cGMP}^{cGMP}\delta
_{cGMP}
\end{eqnarray}%
on quasi-equilibrium conditions with relevant variables%
\begin{eqnarray*}
k_{1,sGC} &=&2\cdot 10^{3}mM^{-1}ms^{-1} \\
k_{-1,sGC} &=&15\cdot 10^{-3}ms^{-1} \\
k_{2,sGC} &=&0.64\cdot 10^{-5}ms^{-1} \\
k_{-2,sGC} &=&0.1\cdot 10^{-6}ms^{-1}
\end{eqnarray*}%
\begin{eqnarray*}
k_{3,sGC} &=&4.2mM^{-1}ms^{-1} \\
k_{D,sGC} &=&0.4\cdot 10^{-3}ms^{-1} \\
k_{D\tau ,sGC} &=&10^{-4}ms^{-1},\text{ }B5_{sGC}=k_{2,sGC}/k_{3,sGC} \\
k_{pde,cGMP} &=&6.95\cdot 10^{-5}ms^{-1},\text{ }k_{m,pde}=10^{-3}mM \\
V_{cGMP,max} &=&1.26\cdot 10^{-7}mM/ms
\end{eqnarray*}%
\begin{eqnarray}
V_{cGMP,0} &=&V_{cGMP,max}\times \\
&&\frac{B5_{sGC}\left[ NO\right] +\left[ NO\right] ^{2}}{A0_{sGC}+A1_{sGC}%
\left[ NO\right] +\left[ NO\right] ^{2}}  \nonumber
\end{eqnarray}%
\begin{eqnarray}
A0_{sGC} &=&\frac{\left( k_{-1,sGC}+k_{2,sGC}\right) k_{D,sGC}}{%
k_{1,sGC}k_{3,sGC}}  \nonumber \\
&&+\frac{k_{-1,sGC}k_{-2,sGC}}{k_{1,sGC}k_{3,sGC}} \\
A1_{sGC} &=&\frac{\left( k_{1,sGC}+k_{3,sGC}\right) k_{D,sGC}}{%
k_{1,sGC}k_{3,sGC}}+  \nonumber \\
&&\frac{\left( k_{2,sGC}+k_{-2,sGC}\right) k_{1,sGC}}{k_{1,sGC}k_{3,sGC}}
\end{eqnarray}%
\begin{eqnarray}
\tau _{msGC} &=&\frac{1}{k_{3,sGC}\left[ NO\right] +k_{D\tau ,sGC}} \\
\tau _{ssGC} &=&\frac{1}{k_{-2,sGC}+k_{D\tau ,sGC}}-\tau _{msGC} \\
\tau _{sGC} &=&\tau _{msGC}+\frac{\tau _{ssGC}}{1+e^{-10\frac{%
V_{cGMP}-V_{cGMP,0}}{V_{cGMP,max}}}}
\end{eqnarray}%
and the fluctuation terms%
\begin{eqnarray}
\Delta _{V_{cGMP}}^{V_{cGMP}} &=&\frac{10\alpha _{16}\left( \overline{V}%
_{cGMP}-V_{cGMP,0}\right) \tau _{ssGC}}{V_{cGMP,max}\left[ \tau
_{ssGC}+\left( 1+\alpha _{16}\right) \tau _{msGC}\right] ^{2}}  \nonumber \\
&&-\frac{1}{\tau _{msGC}+\frac{\tau _{ssGC}}{1+\alpha _{16}}} \\
\Delta _{cGMP}^{V_{cGMP}} &=&1 \\
\Delta _{cGMP}^{cGMP} &=&-\frac{k_{pde,cGMP}\left[ \overline{cGMP}\right] }{%
\left( \left[ \overline{cGMP}\right] +K_{m,pde}\right) ^{2}}\times  \nonumber
\\
&&\left( \left[ \overline{cGMP}\right] +2K_{m,pde}\right)
\end{eqnarray}%
Herein%
\begin{equation}
\alpha _{16}=\exp \left( -10\frac{\overline{V}_{cGMP}-V_{cGMP,0}}{%
V_{cGMP,max}}\right)
\end{equation}

\bigskip

\section{Ionic balances and membrane potential}  \label{appf}

\begin{eqnarray}
\frac{d[Ca]_{u}}{dt} &=&\frac{I_{SERCA}-I_{tr}-I_{IP3}}{z_{Ca}vol_{u}F} \\
&\simeq &\frac{d\left[ \delta _{Ca}\right] _{u}}{dt}=\frac{\sum_{\nu
_{i}}\Delta I_{SERCA}^{\nu _{i}}-\Delta I_{tr}^{\nu _{i}}-\Delta
I_{IP3}^{\nu _{i}}}{z_{Ca}vol_{u}F}
\end{eqnarray}%
\begin{eqnarray}
\frac{d[Ca]_{r}}{dt} &=&\frac{I_{tr}-I_{rel}}{z_{Ca}vol_{r}F}\left[ 1+\frac{%
\left[ \overline{CSQN}\right] K_{CSQN}}{\left( K_{CSQN}+[Ca]_{r}\right) ^{2}}%
\right] ^{-1}  \nonumber \\
&\simeq &\frac{d\left[ \delta _{Ca}\right] _{r}}{dt}=\frac{\sum_{\nu
_{i}}\Delta I_{tr}^{\nu _{i}}-\Delta I_{rel}^{\nu _{i}}}{z_{Ca}vol_{r}F}%
\times  \nonumber \\
&&\left[ 1+\frac{\left[ \overline{CSQN}\right] K_{CSQN}}{\left( K_{CSQN}+[%
\overline{Ca}]_{r}\right) ^{2}}\right] ^{-1}
\end{eqnarray}%
\begin{eqnarray}
\frac{d[Ca]_{i}}{dt} &=&-\frac{I_{Ca,tot}}{z_{Ca}vol_{Ca}F}\times  \nonumber
\\
&&\left[ 1+\frac{\left[ \overline{S}_{CM}\right] K_{d}}{\left(
K_{d}+[Ca]_{i}\right) ^{2}}+\frac{\left[ \overline{B}_{F}\right] K_{dB}}{%
\left( K_{dB}+[Ca]_{i}\right) ^{2}}\right] ^{-1}  \nonumber \\
&\simeq &\frac{d\left[ \delta _{Ca}\right] _{i}}{dt}=-\frac{\Delta I_{Ca,tot}%
}{z_{Ca}vol_{Ca}F}\times \\
&&\left[ 1+\frac{\left[ \overline{S}_{CM}\right] K_{d}}{\left( K_{d}+[%
\overline{Ca}]_{i}\right) ^{2}}+\frac{\left[ \overline{B}_{F}\right] K_{dB}}{%
\left( K_{dB}+[\overline{Ca}]_{i}\right) ^{2}}\right] ^{-1}  \nonumber
\end{eqnarray}%
\begin{eqnarray}
\frac{d[Na]_{i}}{dt} &=&-\frac{I_{Na,tot}}{z_{Na}vol_{i}F}  \nonumber \\
&\simeq &\frac{d\left[ \delta _{Na}\right] _{i}}{dt}=-\frac{\Delta I_{Na,tot}%
}{z_{Na}vol_{i}F}
\end{eqnarray}%
\begin{eqnarray}
\frac{d[K]_{i}}{dt} &=&-\frac{I_{K,tot}}{z_{K}vol_{i}F}  \nonumber \\
&\simeq &\frac{d\left[ \delta _{K}\right] _{i}}{dt}=-\frac{\Delta I_{K,tot}}{%
z_{K}vol_{i}F}
\end{eqnarray}%
\begin{eqnarray}
\frac{d[Cl]_{i}}{dt} &=&-\frac{I_{Cl,tot}}{z_{Cl}vol_{i}F}  \nonumber \\
&\simeq &\frac{d\left[ \delta _{Cl}\right] _{i}}{dt}=-\frac{\Delta I_{Cl,tot}%
}{z_{Cl}vol_{i}F}
\end{eqnarray}%
\begin{eqnarray}
\frac{dV_{m}}{dt} &=&\frac{-I_{V,tot}+I_{stim}}{C_{m}}  \nonumber \\
&\simeq &\frac{d\delta _{V_{m}}}{dt}=\frac{\sum_{\nu _{i}}\left(
-I_{V,tot}\right) +I_{stim}}{C_{m}}
\end{eqnarray}%
for quasi-equilibrium conditions with $\nu _{i}$ being the model component.
Relevant variables are%
\begin{eqnarray*}
\left[ \overline{CSQN}\right] &=&15.0mM,\text{ \ }K_{CSQN}=0.8mM \\
\left[ \overline{S}_{CM}\right] &=&0.1mM,\text{ \ }K_{d}=0.00026mM \\
\left[ \overline{B}_{F}\right] &=&0.1mM,\text{ \ }K_{dB}=0.0005298mM
\end{eqnarray*}

\begin{eqnarray}
\Delta I_{Ca,tot} &=&\sum_{\nu _{i},C}\Delta I_{SOCCa}^{\nu _{i}}+\Delta
I_{VOCC}^{\nu _{i}}-2\Delta I_{NCX}^{\nu _{i}}  \nonumber \\
&&+\Delta I_{PMCA}^{\nu _{i}}+\Delta I_{CaNSC}^{\nu _{i}}+\Delta
I_{SERCA}^{\nu _{i}}-\Delta I_{rel}^{\nu _{i}}  \nonumber \\
&&-\Delta I_{IP3}^{\nu _{i}}+\Delta I_{Ca,GJ}^{\nu _{i},C} \\
\Delta I_{Na,tot} &=&\sum_{\nu _{i},C}\Delta I_{NaKCl}^{Na,\nu _{i}}+\Delta
I_{SOCNa}^{\nu _{i}}+3\Delta I_{NaK}^{\nu _{i}}  \nonumber \\
&&+3\Delta I_{NCX}^{\nu _{i}}+\Delta I_{NaNSC}^{\nu _{i}}+\Delta
I_{Na,GJ}^{\nu _{i},C}
\end{eqnarray}%
\begin{eqnarray}
\Delta I_{K,tot} &=&\sum_{\nu _{i},C}\Delta I_{NaKCl}^{K,\nu _{i}}+\Delta
I_{Kv}^{\nu _{i}}+\Delta I_{BKCa}^{\nu _{i}}+\Delta I_{KNSC}^{\nu _{i}}
\nonumber \\
&&+\Delta I_{Kleak}^{\nu _{i}}-2\Delta I_{NaK}^{\nu _{i}}+\Delta
I_{K,GJ}^{\nu _{i},C} \\
\Delta I_{Cl,tot} &=&\sum_{\nu _{i},C}\Delta I_{NaKCl}^{Cl,\nu _{i}}+\Delta
I_{ClCa}^{\nu _{i}}+\Delta I_{Cl.GJ}^{\nu _{i},C}
\end{eqnarray}%
\begin{eqnarray}
I_{V,tot} &=&\sum_{\nu _{i},C}\Delta I_{VOCC}^{\nu _{i}}+\Delta I_{Kv}^{\nu
_{i}}+\Delta I_{BKCa}^{\nu _{i}}+\Delta I_{Kleak}^{\nu _{i}}  \nonumber \\
&&+\Delta I_{NSC}^{\nu _{i}}+\Delta I_{SOC}^{\nu _{i}}+\Delta I_{ClCa}^{\nu
_{i}}+\Delta I_{PMCA}^{\nu _{i}}  \nonumber \\
&&+\Delta I_{NaK}^{\nu _{i}}+\Delta I_{NCX}^{\nu _{i}}+\Delta I_{Ca,GJ}^{\nu
_{i},C}+\Delta I_{Na,GJ}^{\nu _{i},C}  \nonumber \\
&&+\Delta I_{K,GJ}^{\nu _{i},C}+\Delta I_{Cl.GJ}^{\nu _{i},C}
\end{eqnarray}%
Herein, $\Delta I^{C}$ denotes the variation of current transferring from nearby cells to the
local one.


\begin{thebibliography}{99}
\bibitem{rhy1}{T. Tomita, Smooth muscle: An assessment of current knowledge,
edited by E. Bulbring, A. F. Brading, A. W. Jones and T. Tomita, 127-156 (1981).}

\bibitem{rhy2}{D. F. Van Helden, Pacemaker potentials in lymphatic
smooth muscle of the guinea-pig mesentery. J Physiol 471,
465-479 (1993).}

\bibitem{rhy3}{H. Hashitani, D. F. Van Helden, and H. Suzuki, Properties of
spontaneous depolarizations in circular smooth muscle cells
of rabbit urethra. Br J Pharmacol 118, 1627¡V1632 (1996).}

\bibitem{vasoa1}{K. Shimamura, F. Sekiguchi, and S. Sunano, Tension
oscillation in arteries and its abnormality in hypertensive
animals. Clin Exp Pharmacol Physiol 26, 275-284 (1999).}

\bibitem{vasoa2}{H. Nilsson, and C. Aalkjaer, Vasomotion: mechanisms and
physiological importance. Mol Interv 3, 79¡V89 (2003).}

\bibitem{vasoa3}{R. E. Haddock and C. E. Hill, Rhythmicity in arterial smooth muscle, J Physiol 566.3, 645-656 (2005).}

\bibitem{vasoa4}{C. Aalkjaer and H. Nilsson, Vasomotion: cellular background for the oscillator
and for the synchronization of smooth muscle cells, Br. J. Pharmacol. 144, 605¡V616 (2005).}

\bibitem{an1}{K. Kawasaki, K. Seki, and S. Hosoda, Spontaneous
rhythmic contractions in isolated human coronary
arteries. Experientia 37(12), 1291-1292 (1981).}

\bibitem{an2}{N. I. Gokina, R. D. Bevan, C. L. Walters, and J. A. Bevan,
Electrical Activity Underlying Rhythmic Contraction in
Human Pial Arteries. Circulation Research 78, 148-153 (1996).}

\bibitem{an3}{M. Omote, N. Kajimoto, and H. Mizusawa, The
ionic mechanism of phenylephrine-induced rhythmic
contractions in rabbit mesenteric arteries treated with
ryanodine, Acta Physiologica Scandinavica 147(1), 9-13 (1993).}

\bibitem{an4}{Y. Masuda, K. Okui, and Y. Fukuda, Fine spontaneous
contractions of the arterial wall of the rat in vitro,
Japanese Journal of Physiology 32, 453-457 (1982).}

\bibitem{ex1}{K. A. Dora, J. Xia, and B. R. Duling, Endothelial
cell signaling during conducted vasomotor responses,
American journal of physiology. Heart and circulatory
physiology 285(1), H119-H126 (2003).}

\bibitem{ex2}{M. Lamboley, A. Schuster, J. Beny, and J. Meister,
Recruitment of smooth muscle cells and arterial
vasomotion, American journal of physiology. Heart and
circulatory physiology 285(2), H562-9 (2003).}

\bibitem{ex3}{D. Seppey, R. Sauser, M. Koenigsberger, J. Beny, and
J. Meister, Intercellular calcium waves are associated
with the propagation of vasomotion along arterial
strips, American journal of physiology. Heart and
circulatory physiology 298(2), H488-96 (2010).}

\bibitem{ex4}{H. Peng, V. Matchkov, A. Ivarsen, C. Aalkjar,
and H. Nilsson, Hypothesis for the Initiation of
Vasomotion, Circulation Research 88(8), 810-815 (2001).}

\bibitem{ex5}{B. R. Duling and R. M. Berne, Propagated Vasodilation
in the Microcirculation of the Hamster Cheek Pouch,
Circulation Research 26(2), 163-170 (1970).}

\bibitem{perf1}{A. G. Tsai and M. Intaglietta, Evidence of flowmotion
induced changes in local tissue oxygenation. International Journal of Microcirculation: Clinical and Experimental (Sponsored by the European Society for Microcirculation) 12, 75¡V88 (1993).}

\bibitem{perf2}{M. Rucker, O. Strobel, B. Vollmar, F. Roesken, and
M. D. Menger, Vasomotion in critically perfused muscle
protects adjacent tissues from capillary perfusion failure,
American journal of physiology. Heart and circulatory
physiology 279, H550¡XH558 (2000).}

\bibitem{perf3}{T. Sakurai and N. Terui, Effects of sympathetically
induced vasomotion on tissue-capillary fluid exchange,
American journal of physiology. Heart and circulatory
physiology 291(4), H1761¡VH1767 (2006). [Online]. Available: http://www.ncbi.nlm.nih.gov/pubmed/16731646}

\bibitem{res1}{W. Funk, B. Endrich, K. Messmer, and M. Intaglietta,
Spontaneous arteriolar vasomotion as a determinant of
peripheral vascular resistance, Int J Microcirc Clin Exp 2, 11¡V25 (1983).}

\bibitem{res2}{C. Meyer, G. De Vries, S. T. Davidge, and D. C. Mayes,
Reassessing the mathematical modeling of the contribution
of vasomotion to vascular resistance, J Appl Physiol 92, 888¡V889 (2002).}

\bibitem{mathmodel1} {A. Kapela, A. Bezerianos, N. M. Tsoukias, A mathematical model of Ca2+ dynamics in rat mesenteric smooth muscle
cell: Agonist and NO stimulation, Journal of Theoretical Biology 253, 238¡V260 (2008).}

\bibitem{mathmodel2}{A. Kapela, S. Nagaraja, and N. M. Tsoukias, A mathematical model of vasoreactivity in rat mesenteric arterioles. II.
Conducted vasoreactivity, Am J Physiol Heart Circ Physiol 298, H52-H65 (2010).}

\bibitem{study1}{D. Parthimos, R. E. Haddock, C. E. Hill, and T. M. Griffith, Dynamics of A Three-Variable Nonlinear Model of Vasomotion:
Comparison of Theory and Experiment, Biophysical Journal 93, 1534-1556 (2007).}
\bibitem{study2}{M. Koenigsberger, R. Sauser, D. Seppey, J. L. Beny, and J. J. Meister, Calcium Dynamics and Vasomotion in Arteries Subject to Isometric,
Isobaric, and Isotonic Conditions, Biophysical Journal 95, 2728-2738 (2008).}

\bibitem{study3}{A. Goldbeter, G. Dupont, and M. J. Berridge,
Minimal model for signal-induced Ca2+ oscillations
and for their frequency encoding through protein
phosphorylation. Proceedings of the National Academy
of Sciences of the United States of America 87(4), 1461-5 (1990).}

\bibitem{study4}{J. M. Gonzalez-Fernandez and B. Ermentrout, On the
origin and dynamics of the vasomotion of small arteries,
Mathematical Biosciences 119, 127-167, (1994).}

\bibitem{study5}{D. Parthimos, D. H. Edwards, and T. M. Griffith, Minimal
model of arterial chaos generated by coupled intracellular
and membrane Ca2+ oscillators, American
journal of physiology. Heart and circulatory physiology 277, H1119-H1144 (1999).}

\bibitem{study6}{M. Koenigsberger, R. Sauser, M. Lamboley, J.-L.
Beny, and J. Meister, Ca2+ dynamics in a population
of smooth muscle cells: modeling the recruitment
and synchronization. Biophysical journal 87(1), 92-104 (2004).}

\bibitem{study7}{J. P. Johny and T. David, A numerical study into minimal conditions of arterial vasomotion. Thrissur, India: Proceedings of World Congress on Research and Innovations (2013). [Online]. Available: http://ir.canterbury.ac.nz/handle/10092/10508}

\bibitem{study8}{A. Kapela, J. Parikh, and N. M. Tsoukias, Multiple Factors Influence Calcium Synchronization in Arterial Vasomotion, Biophysical Journal 102, 211-220 (2012).}

\bibitem{book_K}{L. Xiang and R. L. Hester, Cardiovascular Responses to Exercise, Morgan $\&$ Claypool Life Sciences (2012).}

\bibitem{heart1}{Y. Y. L. Wang, M. Y. Jan, C. S. Shyu, C. A. Chiang, and W. K. Wang, IEEE Trans. Biomed. Eng. 51(1), 193 (2004).}

\bibitem{KClexp1}{A. Schuster, M. Lamboley, C. Grange, H. Oishi, J. L. Beny, N.   Stergiopulos, J. J. Meister, Calcium dynamics and vasomotion in rat mesenteric arteries,
J. Cardiovasc. Pharmacol. 43, 539-548 (2004).}

\bibitem{dyn1}{W. F. Jackson, Oscillations in active tension in hamster
aortas: role of the endothelium, Blood Vessels 25, 144-156 (1988).}

\bibitem{dyn2}{K Fujii, D. D. Heistad, and F. M. Faraci, Vasomotion of basilar
arteries in vivo, Am J Physiol 258, H1829-H1834 (1990).}

\bibitem{dyn3}{H. Gustafsson, A. Bulow, and H Nilsson, Rhythmic
contractions of isolated, pressurized small arteries from rat,
Acta Physiol Scand 152, 145-152 (1994).}

\bibitem{dyn4}{K. A. Dora, J. M. Hinton, S. D. Walker, and C. J. Garland,
An indirect influence of phenylephrine on the release of
endothelium-derived vasodilators in rat small mesenteric
artery, Br J Pharmacol 129, 381-387 (2000).}

\bibitem{dyn5}{K. Okazaki, S. Seki, N. Kanaya, J. Hattori, N. Tohse, and A. Namiki, Role of endothelium-derived hyperpolarizing factor
in phenylephrine-induced oscillatory vasomotion in rat
small mesenteric artery, Anesthesiology 98, 1164-1171 (2003).}

\bibitem{dyn6}{J. R. Mauban, and W. G. Wier, Essential role of EDHF in the
initiation and maintenance of adrenergic vasomotion in rat
mesenteric arteries, Am J Physiol Heart Circ Physiol 287,
H608-H616 (2004).}

\bibitem{dyn7}{H. Peng, V. Matchkov, A. Ivarsen, C. Aalkjaer, and H. Nilsson,
Hypothesis for the initiation of vasomotion, Circ Res 88,
810-815 (2001).}

\bibitem{dyn8} {I. S. Bartlett, G. J. Crane, T. O. Neild, and S. S. Segal, Electrophysiological basis of arteriolar vasomotion in vivo,
J Vasc Res 37, 568-575 (2000).}

\bibitem{dyn9} {H. Oishi, A. Schuster, W. Lamboley, N. Stergiopulos, J. J. Meister, and J. L. Beny, Role of membrane potential in vasomotion
of isolated pressurized rat arteries, Life Sci 71,
2239-2248 (2002).}

\bibitem{gjvalue1}{L. K. Moore, and J. M. Burt, Gap junction function in vascular smooth muscle: influence
of serotonin, Am J Physiol 269, H1481-1489 (1995).}
\bibitem{sizedep1}{M. J. Mulvany, U. Baandrup, H. J. Gundersen, Evidence for hyperplasia in mesenteric resistance vessels of spontaneously hypertensive rats using a three-dimensional disector, Circ Res. 57(5), 794-800 (1985).}
\bibitem{spark1} {C. D. Benham and T. B. Bolton, Spontaneous transient
outward currents in single visceral and vascular smooth
muscle cells of the rabbit, J Physiol 381, 385-406 (1986).}

\bibitem{wave1}{M. Koenigsberger,D. Seppey, J.-L. Beny, and J.-J. Meister, Mechanisms of Propagation of Intercellular Calcium Waves in Arterial
Smooth Muscle Cells, Biophysical Journal 99, 333-343 (2010).}

\end{thebibliography}
\end{document}